\documentclass{lmcs}
\pdfoutput=1

\usepackage[utf8]{inputenc}

\usepackage{lastpage}
\lmcsdoi{16}{3}{10}
\lmcsheading{}{\pageref{LastPage}}{}{}%
{Aug.~07,~2018}{Aug.~19,~2020}{}


\setlist[description]{labelindent=0mm,labelwidth=*,labelsep=2mm,leftmargin=1em}

\usepackage{prooftree}
\usepackage{derivation}
\usepackage{amssymb}
\usepackage{stmaryrd}
\usepackage{mathtools}
\usepackage{multicol}
\usepackage{tikz}
\usepackage{tikz-cd}
\usepackage{enumitem}
\usepackage[lmcs]{optional}

\newcommand{\parfunc}{\rightharpoonup}
\newcommand{\setcomp}[2]{\left\{ \, #1 \; \middle| \; #2 \, \right\}} 
\newcommand{\defeq}{\stackrel{\mathclap{\mbox{\tiny def}}}{=}}
\newcommand{\sem}[1]{\left\llbracket #1 \right\rrbracket}
\newcommand{\transit}[2]{\xRightarrow{#1}}

\newcommand{\ctxt}[2]{#1 \mathbin{;} #2}
\newcommand{\ibox}[1]{\textsf{box}\ #1}
\newcommand{\letbox}[3]{\textsf{let box}\ #1 \Leftarrow #2\ \textsf{in}\ #3}
\newcommand{\fixbox}[2]{\textsf{fix}\ #1\ \textsf{in  box}\ #2}
\newcommand{\later}{\mathop{\blacktriangleright}}

\newcommand{\mathsc}[1]{\text{\normalfont\scshape#1}}
\newcommand{\fv}[1]{\mathsc{Fv}\left(#1\right)}
\newcommand{\ufv}[1]{\mathsc{Fv}_0\left(#1\right)}
\newcommand{\bfv}[1]{\mathsc{Fv}_{\geq 1}\left(#1\right)}
\newcommand{\vars}[1]{\mathsc{Vars}\left(#1\right)}

\newcommand{\red}{\mathrel{\longrightarrow}}
\newcommand{\redp}{\mathrel{\Longrightarrow}}
\newcommand{\redt}{\mathrel{\longrightarrow^\ast}}

\newcommand{\vct}[1]{\overrightarrow{#1}}
\newcommand{\PSH}[1]{\textbf{Psh}\left(#1\right)}

\newlist{indproof}{itemize}{5}
\setlist[indproof]{%
  itemsep=5pt,  
  font={\sc}, 
  label={}
}

\newcommand{\case}[1]{\item{\sc Case({\normalfont #1})}.}

\usetikzlibrary{matrix,arrows}
\usetikzlibrary{fit,positioning,calc}

\begin{document}

\title{Dual-context Calculi for Modal Logic}

\author{G. A. Kavvos}
\address{Institut for Datalogi, Aarhus University,
  \r{A}bogade 34, 8200 Aarhus N, Denmark}
\email{alex.kavvos@cs.au.dk}
\thanks{This is a revised and extended version of a paper
presented at LICS 2017 \cite{Kavvos2017b}.}

\begin{abstract}

  We present natural deduction systems and associated modal lambda
  calculi for the necessity fragments of the normal modal logics K, T,
  K4, GL and S4. These systems are in the dual-context style: they
  feature two distinct zones of assumptions, one of which can be
  thought as modal, and the other as intuitionistic. We show that
  these calculi have their roots in in sequent calculi. We then
  investigate their metatheory, equip them with a confluent and
  strongly normalizing notion of reduction, and show that they
  coincide with the usual Hilbert systems up to provability. Finally,
  we investigate a categorical semantics which interprets the modality
  as a product-preserving functor.

\end{abstract}

\maketitle

\section*{Introduction}
  \label{section:introduction}

The developments that have taken place over the past twenty years
have shown that \emph{constructive modalities}---broadly construed
as unary type operators---are an important and versatile tool for
both type theory and programming language theory. 

Modalities have been used for various purposes within
dependent type theory, e.g.
  to express proof irrelevance \cite{Pfenning2001, Awodey2004} and truncation in Homotopy Type Theory 
    \cite{Rijke2020},
  to formalise Cartan geometry \cite{Wellen2017} and quantum gauge field theory \cite{Schreiber2014},
  to internalise parametricity arguments in dependent type theory \cite{Nuyts2017, Nuyts2018}, 
  to reason about differential cohesive toposes \cite{Gross2017},
  to formally prove theorems that relate topology and homotopy \cite{Shulman2018},
  and
  to construct models of universes internal to a topos \cite{Licata2018}.

To name but a few occurrences in programming language theory,
modalities have been used 
  in staged metaprogramming \cite{Davies2001a, Tsukada2010, Davies2017},
  to control the complexity of typed programs \cite{Hofmann1999},
  to enable recursion over higher-order abstract syntax \cite{Schurmann2001} and 
    to provide categorical models for it \cite{Hofmann1999b}, 
  to control information flow \cite{Abadi1999, Shikuma2008, Kavvos2019a},
  to design a $\lambda$-calculus for distributed computing and mobile code \cite{MurphyVII2004},
  to build models of stateful languages with recursive types \cite{Birkedal2012},
  in coinductive programming and guarded recursion \cite{Atkey2013, Clouston2016},
  in functional reactive programming \cite{Krishnaswami2011, Krishnaswami2011a, Krishnaswami2013, Bahr2019},
  in modelling contextual computation \cite{Orchard2014},
  and, finally,
  in compartmentalising effects \cite{Moggi1991} and combining them with resources \cite{Curien2016}.
  
Despite this wide applicability, there have been very few
foundational studies on the \emph{Curry-Howard-Lambek
correspondence} that underlies modal types: we have surveyed
relevant work in \cite{Kavvos2016b}. The major impediment to
carrying out such work is that the methods of modal proof theory
are at best kaleidoscopic: while one type of calculus may work
well for a specific logic, it may readily fail to express a
closely related one. It is possible to develop a working intuition
for these patterns, but it is much harder to precisely explain the
root of these difficulties. 

This paper has two goals. The first is to explain why a particular
pattern of natural deduction for modal logics---namely that of
\emph{dual contexts}---is suited to the necessity fragments of the
most popular \emph{normal modal logics}. The key intuition is that the
separation of assumptions into a modal zone and an intuitionistic zone
allows us to mimic rules from cut-free sequent calculi for these
logics. The second goal is to extend the Curry-Howard-Lambek
correspondence and its usual triptych of logic, computation, and
categories to normal modal logics.  We show that the dual-context term
languages admit a categorical semantics in which the modality is
interpreted by a \emph{product-preserving endofunctor} equipped with
gadgets corresponding to the sundry axioms of normal modal logic. This
deviates from previous approaches (in particular that of Bierman and
de Paiva \cite{Bierman2000a}) but leads to a simpler syntax that
nonetheless encompasses a large number of categorical models.

\subsection*{A road map}

Our investigation is structured as follows. First, we define
constructive versions of the most basic normal modal logics,
namely $\textsf{K}$, $\textsf{T}$, $\textsf{K4}$, $\textsf{GL}$
and $\textsf{S4}$, and present a Hilbert system for each
(\S\ref{sec:logics}). We then briefly recount previous attempts at
presenting natural deduction systems for them. This discussion
leads us to a systematic method for deriving dual-context systems
(\S\ref{sec:deriving}).

Then, in \S\ref{sec:metatheory} we reformulate these systems as
modal $\lambda$-calculi, and study their basic metatheory. By
writing down terms of the appropriate type, we show that our
dual-context systems are equivalent to the Hilbert systems given
in \S\ref{sec:deriving} at the level of provability. Following
that, we endow these terms with a notion of reduction
(\S\ref{sec:reduction}). We prove that this has the usual good
properties: it is confluent, strongly normalizing, and eliminates
all cuts, i.e. the normal forms satisfy the subformula property.
We stop short of deciding equality.

Finally, we give a categorical semantics for these calculi. We first
introduce the relevant category theory in \S\ref{sec:modalcats}: this
consists of a self-contained account of strong monoidal functors
between cartesian categories, which we prove coincide with
product-preserving functors. We then define various other gadgets used
for interpreting the rules of our calculi, including coherent
comultiplications (for \textsf{K4}) and counits (for \textsf{T}). In
the case of \textsf{S4}, which contains both of these logics, these
gadgets are required to satisfy the usual coherence equations of a
product-preserving \emph{comonad}. Finally, \textsf{GL} requires a
novel notion of \emph{modal fixed point}. We define a sound and
complete interpretation into these structures in \S\ref{sec:catsem}.

\subsection*{The case of G\"odel and L\"ob}

Perhaps one of the most interesting and surprising aspects of our
investigation is that the general pattern described in
\S\ref{sec:deriving} can be used to derive a natural deduction system
for a constructive version of the \emph{logic of provability}. While
the classical version of the logic itself is comprehensively covered
in the book by Boolos \cite{Boolos1994a}, numerous interesting
historical and mathematical facts about its constructive and
intuitionistic variants are collected in a survey by Tadeusz Litak
\cite{Litak2014}.

\textsf{GL} is the least normal modal logic containing the
\emph{G\"odel-L\"ob axiom}
\begin{equation}
  \label{eq:godel-lob-axiom}
  \Box(\Box A \rightarrow A) \rightarrow \Box A
  \tag{$\star$}
\end{equation} 
for every formula $A$. If we erase the modal operators from this
axiom we obtain the type $(A \rightarrow A) \rightarrow A$ of
\emph{fixed point combinators} at $A$.  It is thus natural to
suspect that, from a computational point of view, \textsf{GL}
should lead to a calculus equipped with some unusual, modal kind
of \emph{recursion}. 

This observation, which is originally due to Nakano
\cite{Nakano2000}, led to a fruitful research programme on what is
variously known as \emph{productive coprogramming}
\cite{Atkey2013}, \emph{guarded recursion} \cite{Birkedal2012,
Birkedal2013, Clouston2016, Guatto2018}, or \emph{corecursion}
\cite{Severi2012}. However, the underlying logic of these systems
is essentially a variant of the \emph{strong G\"odel-L\"ob logic}
\textsf{SL}, which is the least normal modal logic closed under
the stronger axiom \[
  (\Box A \rightarrow A) \rightarrow A
\] 
This axiom, which implies \eqref{eq:godel-lob-axiom} as well as the
unusual formula $A \rightarrow \Box A$, is incompatible with classical
normal modal logic, as no interesting Kripke frames satisfy the latter
formula \cite[Remark 17]{Litak2014}. Nevertheless, it appears time and
again in various intutionistic settings.


Returning to the weaker logic \textsf{GL}, we find that our
methods apply to yield a new natural deduction system, which---we
argue in \S\ref{sec:perennial}---is significantly simpler to the
only other known system for \textsf{GL}, viz. that of Bellin
\cite{Bellin1985}. Given the intricate relationship between
constructive provability logics, guarded recursion, and
coinductively-defined infinite data, it may come as a surprise
that the detour-eliminating reduction for this system is strongly
normalizing. We will discuss that in more detail, but the basic
intuition is that the introduction term in our calculus has a
coinductive behaviour: it recursively unfolds only when forced to
do so by an occurrence of the elimination rule.

\subsection*{Related work on dual contexts}

Dual-context calculi were pioneed by Girard \cite{Girard1993},
Andreoli \cite{Andreoli1992}, Wadler \cite{Wadler1993,
Wadler1994}, Plotkin \cite{Plotkin1993}, and Barber
\cite{Barber1996}, in the setting of \emph{multiplicative
exponential linear logic} and its `of course' ($!$) modality. They
were then imported into the intuitionistic setting by Davies and
Pfenning \cite{Davies2001, Davies2001a}, who introduced the system
for \textsf{S4} that we study in this paper. With the exception of
\textsf{S4}, the systems discussed in this paper are largely new.
In some cases, glimpses of similar patterns have appeared before.
Despite the popularity of the \textsf{S4} system in the
programming language community, there has been no detailed study
of its proof-theoretic properties.

An approach that is similar to ours for \textsf{K} and \textsf{K4}
has been considered by Pfenning \cite{Pfenning2013, Pfenning2015}
in the context of \emph{linear sequent calculi}, which seems to be
closely related to the work of Danos and Joinet on
\emph{elementary linear logic} \cite{Danos2003}.  However, this
work of Pfenning remains unpublished, and the natural deduction
system for \textsf{K} in this paper is independently due to the
present author. The technical innovations needed in presenting a
term calculus for \textsf{K4} and \textsf{GL} are new. 

The study of reduction conducted by Davies and Pfenning for
dual-context \textsf{S4} \cite{Davies2001} was limited to an
evaluation strategy used as operational semantics. In
\cite{Davies2001a} a notion of conversion is introduced, but the
study of its properties was left to a future paper, which never
appeared. Neither of these papers discusses commuting conversions.
The categorical interpretation of dual-context \textsf{S4} in
terms of product-preserving comonads was briefly sketched by
Hofmann \cite[\S 2.6.12]{Hofmann1999}.

\subsection*{Related work on modal proof theory}

The history of modal proof theory and constructive modal logics is
long and tumultuous, so we shall try to avoid the subject as much
as possible. A more thorough discussion of modal $\lambda$-calculi
may be found in \cite{Kavvos2016b}. For a broader survey of the
proof theory of modal logic we recommend \cite{Negri2011}.

While the earliest work goes back to Prawitz \cite{Prawitz1965},
the first modal $\lambda$-calculus seems to be the Bierman-de
Paiva system for \textsf{S4} \cite{Bierman1992a, Bierman1996a,
Bierman2000a}. For reasons we discuss in \S\ref{sec:deriving},
this system is unsatisfying: its proof theory requires many
commuting conversions to eliminate all cuts, and its syntax is
counterintuitive as a programming language. The proof-theoretic
aspect is extensively discussed by Goubault-Larrecq \cite[\S
5.2]{Goubault-Larrecq1996}.  The programming-related issues are
mentioned by Clouston et al.  \cite{Clouston2016}, who use a
similar style of calculus in the context of guarded recursion.
Clouston et al. try to argue that the ``burden presented by the
explicit substitutions seems quite small,'' but the fact this is a
leitmotif in their paper significantly weakens their argument.
Furthermore, in \emph{op.  cit.} it is argued that this style does
not economically adapt to dependent types. Indeed, the
dual-context type theory of Shulman \cite{Shulman2018} seems to
vindicate this claim.

There is little previous work on natural deduction for
sub-\textsf{S4} systems of normal modal logic. Martini and Masini
\cite{Martini1996} presented a \emph{Fitch-style} system. This was
later adapted by Davies and Pfenning \cite[\S 5]{Davies2001a}, who
call it the `Kripke-style' formulation. Under various restrictions
on its syntax, the Martini and Masini system captures \textsf{K},
\textsf{K4}, \textsf{T}, and \textsf{S4}, at the price of having
one's terms annotated with indices. The Kripke-style formulation
simplifies some of this presentation, but does not dispense with
indices. A simpler Fitch-style system for \textsf{K} was
extensively studied by Clouston \cite{Clouston2018}.

There is also a calculus in the style of Bierman and de Paiva for
\textsf{K}, which was introduced by Bellin, de Paiva, and Ritter
\cite{Bellin2001}. This suffered from some technical issues that
were later mitigated by Kakutani \cite{Kakutani2007a}. Some of the
unsatisfactory aspects of this calculus are discussed by de Paiva
and Ritter \cite{dePaiva2011}: they trace its roots to the
aforementioned system of Bellin \cite{Bellin1985} for \textsf{GL},
who hints at (but does not study) systems for \textsf{K} and
\textsf{K4}.

We argue that the dual-context formulations introduced in the
present work lead to simpler calculi: our terms feature neither
delayed substitutions, nor are they littered with indices. This
simplifies the metatheory, and makes them more practicable. Unlike
Bierman-de Paiva style calculi, dual-context calculi are simple
enough to lead to implementations: see e.g. the experimental work
of Wickline, Lee and Pfenning \cite{Wickline1998a} on
metaprogramming with dual-context \textsf{S4}. Moreover, our
calculi are simple enough to enable large-scale pen-and-paper
proofs: Shulman \cite{Shulman2018} used a dual-context dependent
type theory to produce a formal proof of Brouwer's fixed point
theorem. Thus, the fact that dual-context style can extended to a
range of sub-\textsf{S4} modalities is of independent interest in
exploring applications in various sub-\textsf{S4} settings.

Nevertheless, we ought to stress one central detail: our systems
strongly preserve products. That is, the proof theory of dual-context
systems induces an isomorphism of types
$\Box(A \times B) \cong \Box A \times \Box B$. In contrast, systems in
the Bierman-de Paiva style prove a bi-implication that does \emph{not}
necessarily extend to an isomorphism. Indeed, the categorical
semantics of Bierman-de Paiva \textsf{S4} require only a \emph{lax
  monoidal comonad}, which comes with a natural transformation
$\Box A \times \Box B \rightarrow \Box(A \times B)$ that is not
necessarily invertible. This means our work is closer to the system of
Clouston \cite{Clouston2018}, who requires that the modality have a
left adjoint, and thus that it preserve products. All things
considered, if we want to avoid product preservation we must revert to
the system with delayed substitutions.


\section{The Logics in Question}
  \label{sec:logics}

We are concerned with $(\land\rightarrow\Box)$ fragment of five of
the most commonly encountered \emph{normal modal logics} \cite[\S
1.6]{Blackburn2001} \cite[\S\S 1.6--1.7]{Fitting1993} \cite[\S\S
2--3]{Hughes1996} \cite[\S\S 4--5]{Boolos1994a}: \textsf{K}
(abbrv.  \textsf{CK}), \textsf{K4} (abbrv.  \textsf{CK4}),
\textsf{T} (abbrv. \textsf{CT}), \textsf{GL} (abbrv.
\textsf{CGL}), and \textsf{S4} (abbrv.  \textsf{CS4}). In this
section we shall discuss their common characteristics, and present
a Hilbert system for each.

\subsection{Constructive modal logics} 

All of the above logics belong to the group of \emph{constructive
modal logics}. These are a family of intuitionistic modal logics
which have been cherry-picked to satisfy a specific desideratum,
namely to have a well-behaved, Gentzen-style proof theory, and
thereby an associated computational interpretation.

The special behaviour of these logics is even more appreciable
when the possibility modality ($\lozenge$) is taken into
consideration. First, the de Morgan duality between $\Box$ and
$\lozenge$ breaks down, rendering them logically independent. For
that reason we shall mostly refer to the $\Box$ as the \emph{box}
modality. Second, the principles $\lozenge (A \lor B) \rightarrow
\lozenge A \lor \lozenge B$ and $\lnot\lozenge \bot$ are not
provable. These two principles are tautologies if we employ
traditional Kripke semantics \cite{Kripke1963}. Thus, the way to a
computational interpretation seems to necessitates that we eschew
the Kripkean analysis. Even though $\lozenge$ is essential in
pinpointing the salient differences between constructive modal
logics and other forms of intuitionistic modal logic---e.g. those
studied by \cite{Simpson1994}---it seems that its computational
interpretation is not very crisp. Hence, we restrict our study to
the better-behaved, and seemingly more applicable box modality.

\subsection{Preliminaries}

All of our modal logics shall be inductively defined sets of
formul\ae---the \emph{theorems} of the logic. These formul\ae~are
generated by the Backus-Naur form \[
  A, B\ ::=\
           p_i
      \mid \top
      \mid A \land B
      \mid A \rightarrow B 
      \mid \Box A
\] where $p_i$ is drawn from a countable set of propositions. The
sets of theorems will be generated by \emph{axioms}, closed under
some \emph{inference rules}. The set of axioms will always contain
(a) all the instances the axioms of intuitionistic propositional
logic, but over modal formul\ae; and (b) all instances of the
\emph{normality} axiom, also known as axiom \textsf{K} (after
Kripke). The set of inference rules will contain rules for using
axioms and assumptions, modus ponens, and the modal rule of
\emph{necessitation}, namely \[ 
  \begin{prooftree}
  A \in \mathcal{L}
    \justifies
  \Box A \in \mathcal{L}
\end{prooftree} 
\] The only thing that will then vary between any two of our
logics $\mathcal{L}$ will be the set of axioms.

\subsection{Axioms}
  \label{sec:axioms}

We write $(A_1) \oplus \dots \oplus (A_n)$ to mean the set of
theorems that are derivable from all instances of the axioms
$(A_1), \dots, (A_n)$ under the aforementioned rules of axiom,
assumption, modus ponens, and necessitation. Furthermore, we write
$(\mathsf{IPL}_\Box)$ to mean the set of all instances of the
axiom schemata of intuitionistic propositional logic, but over
modal formul\ae. We will use the following modal axiom schemata: 
\begin{align*}
  (\mathsf{K)} \quad 
    &\Box (A \rightarrow B) \rightarrow (\Box A \rightarrow \Box B)
      &
  (\mathsf{4)} \quad 
    &\Box A \rightarrow \Box \Box A \\
  (\mathsf{T)} \quad 
    &\Box A \rightarrow A
      &
  (\mathsf{GL)} \quad 
    &\Box (\Box A \rightarrow A) \rightarrow \Box A
\end{align*} 
Constructive \textsf{K} is then defined to be the minimal
\emph{normal} constructive modal logic. Constructive \textsf{K4}
adds axiom \textsf{4} to that. Likewise, constructive \textsf{T}
is the result of adding axiom \textsf{T} to \textsf{CK}.
Constructive \textsf{S4} results from mixing all these axiom
schemas together. Finally, we obtain constructive \textsf{GL} from
\textsf{CK} by adding the \emph{G\"odel-L\"ob} axiom \textsf{GL}.
In summary:
\begin{gather*}
  \begin{align*}
    \textsf{CK} &\defeq (\textsf{IPL}_\Box) \oplus (\textsf{K}) 
      &
    \textsf{CK4} &\defeq (\textsf{IPL}_\Box) \oplus (\textsf{K}) \oplus (\textsf{4})
      \\
    \textsf{CT} &\defeq (\textsf{IPL}_\Box) \oplus (\textsf{K}) \oplus (\textsf{T}) 
      &
    \textsf{CGL} &\defeq (\textsf{IPL}_\Box) \oplus (\textsf{K}) \oplus (\textsf{GL})
  \end{align*} 
    \\
  \textsf{CS4} \defeq (\textsf{IPL}_\Box) \oplus (\textsf{K}) \oplus (\textsf{4}) \oplus (\textsf{T})
\end{gather*}

\subsection{Hilbert systems}

We introduce a judgment of the form \[
  \Gamma \vdash A
\] where $\Gamma$ is a \emph{context}, i.e. a list of formul\ae~
defined by the grammar $\Gamma ::=\ \cdot \mid \Gamma, A$ where
$A$ is a single formula. We shall use the comma to also denote
concatenation. For example, $\Gamma, A, \Delta$ shall stand for
the juxtaposition of three things: the context $\Gamma$, the
context consisting of the single formula $A$, and the context
$\Delta$.

\begin{figure}
  \centering
  \renewcommand{\arraystretch}{2.7}

\begin{small}
\begin{tabular}{cccc}
  $
    \begin{prooftree}
      \justifies
        \Gamma, A, \Delta \vdash A
      \using
        (\textsf{assn})
    \end{prooftree} 
  $
 
  &

  $ 
  \begin{prooftree}
      A \text{ is an axiom}
    \justifies
      \Gamma \vdash A
    \using
    (\textsf{ax})
  \end{prooftree}
  $

  &

  $ 
  \begin{prooftree}
      \Gamma \vdash A \rightarrow B
    \quad
      \Gamma \vdash A
    \justifies
      \Gamma \vdash B
    \using
      (\textsf{MP})
  \end{prooftree}
  $

  &

  $ 
  \begin{prooftree}
      \vdash A
    \justifies
      \Gamma \vdash \Box A
    \using
      (\textsf{nec})
    \end{prooftree}
  $
\end{tabular}
\end{small}


  \caption{Hilbert systems}
  \label{fig:hilbert}
\end{figure}

This judgment is generated inductively, and includes rules for
axioms and assumptions: see Figure \ref{fig:hilbert}. The main
rule concerning the modality is that of \emph{necessitation},
which we need state carefully. Otherwise, we risk invalidating the
deduction theorem, leading to a common point of confusion in early
work on the proof theory of modal logic: see \cite{Hakli2012} for
a historical and technical account. To approach this issue, we
need to recall that necessitation bears a likeness to universal
quantification: $\Box A$ is a theorem just if $A$ is a theorem,
and there is no reason that this should be so if we need any
assumptions to prove $A$.  Thus, we should be able to infer $\Box
A$ (under any assumptions) only if we can infer $A$ without any
assumptions at all. In symbols:
\[ 
  \begin{prooftree} 
      \vdash A 
    \justifies 
      \Gamma \vdash \Box A 
  \end{prooftree}
\] 

To indicate that we are using the Hilbert system for e.g.
\textsf{CK}, we annotate the turnstile an write $\Gamma
\vdash_{\textsf{CK}} A$. We write $\Gamma \vdash A$ when the
relevant statement pertains to all systems.

\subsection{Metatheory of Hilbert systems}

\subsubsection{Structural rules} 

We establish the following basic facts about all our Hilbert
systems by a straightforward induction on the derivation of each
premise.

\begin{thm}[Structural \& Cut] 
  \label{thm:hilbscut}
The following rules are admissible.\footnote{Recall that a rule is
\emph{admissible} just if the existence of a proof of the
antecedent implies the existence of a proof of the conclusion,
where that existence is determined in our metatheory. In contrast,
a rule is \emph{derivable} just if a proof of the antecedent can
be used verbatim as a constituent part of a proof of the
conclusion.}
  \begin{multicols}{2}
  \begin{enumerate}[itemsep=4pt]
    \item (Weakening) $
      \begin{prooftree}
        \Gamma \vdash C
          \justifies
        \Gamma, A \vdash C
      \end{prooftree}
    $ 
    \item (Exchange) $
      \begin{prooftree}
        \Gamma, A, B, \Delta \vdash C
          \justifies
        \Gamma, B, A, \Delta \vdash C
      \end{prooftree}
    $
    \item (Contraction) $
      \begin{prooftree}
        \Gamma, A, A, \Delta \vdash C
          \justifies
        \Gamma, A, \Delta \vdash C
      \end{prooftree}
    $
    \item (Cut) $
      \begin{prooftree}
        \Gamma \vdash A 
          \qquad
        \Gamma, A, \Delta \vdash C
          \justifies
        \Gamma, \Delta \vdash C
      \end{prooftree}
    $
  \end{enumerate}
  \end{multicols}
\end{thm}

\begin{thm}[Deduction Theorem]
  \label{thm:deduction}
  The rule $
    \begin{prooftree}
      \Gamma, A \vdash B
        \justifies
      \Gamma \vdash A \rightarrow B
    \end{prooftree}
  $ is admissible.
\end{thm}

\subsubsection{Admissible Modal Rules}
  \label{sec:admissible}

We now consider some admissible rules that refer to the box
modality. These will prove useful when we tackle the proof of
equivalence between Hilbert systems and dual-context systems.

The first one is \emph{Scott's rule}, which ensures that if we
`box' all our assumptions then we can `box' the conclusion.  We
will see that, in categorical terms, Scott's rule expresses the
fact that the box is a \emph{functor}, and in particular one that
\emph{preserves products}. We write $\Box \Gamma$ to mean the
context $\Gamma$ which each assumption occurring in it boxed, i.e.
$\Box(A_1, \dots, A_n) \defeq \Box A_1, \dots, \Box A_n$.

\begin{thm}[Admissibility of Scott's rule]
  \label{thm:scott}
  The following rule is admissible: 
  \[
    \begin{prooftree}
      \Gamma \vdash A	
        \justifies
      \Box \Gamma \vdash \Box A
    \end{prooftree}
  \] 
\end{thm}
\begin{proof} 
Straightforward induction on the derivation of $\Gamma \vdash A$.
We show the case for modus ponens.  If the last step in the
derivation of $\Gamma \vdash A$ is of the form
\[
  \begin{prooftree}
    \[ 
      \leadsto
	\justifies
      \Gamma \vdash B \rightarrow A
    \]
      \quad
    \[
      \leadsto
	\justifies
      \Gamma \vdash B
    \]
      \justifies
    \Gamma \vdash A
  \end{prooftree}
\] 
then, by applying the induction hypotheses to the two
subderivations, we obtain proofs of $\Box \Gamma \vdash \Box(B
\rightarrow A)$ and $\Box \Gamma \vdash \Box B$.  We can then use
axiom \textsf{K} and modus ponens twice to build the desired
proof: 
\[
  \begin{prooftree}
    \[
      \[
	  \justifies
	\Box \Gamma \vdash \Box (B \rightarrow A) \rightarrow
	  \Box B \rightarrow \Box A
      \]
	\quad
      \[
	\leadsto
	  \justifies
	\Box \Gamma \vdash \Box (B \rightarrow A)
      \]
	\justifies
      \Box \Gamma \vdash \Box B \rightarrow \Box A
    \]
      \quad
    \[
      \leadsto
	\justifies
      \Box \Gamma \vdash \Box B
    \]
      \justifies
    \Box \Gamma \vdash \Box A
  \end{prooftree}
\]
\end{proof}

The above theorem also follows by the deduction theorem. However,
this proof implicitly considers derivations of $\Gamma \vdash A$
as terms of an underlying \emph{modal combinatory logic}. It is an
old observation by Curry and Feys \cite{Curry1958} that
combinatory logics vaguely correspond to Hilbert systems, and
Pfenning \cite{Pfenning2010} has sketched such a system for
\textsf{CS4}.

Next, we deal with a rule that is only derivable if the system
contains the axiom \textsf{T}. The gist of the rule is that $\Box
A$ is stronger than $A$, as it implies it in any context.

\begin{thm}[Admissibility of Veridicality] 
  If $\mathcal{L} \in \{\textsf{CT}, \textsf{CS4}\}$, then the
  following rule is admissible: \[
    \begin{prooftree}
      \Gamma \vdash_\mathcal{L} A	
        \justifies
      \Box \Gamma \vdash_\mathcal{L} A
    \end{prooftree}
  \]
\end{thm}
\begin{proof} By induction on the derivation of $\Gamma \vdash A$.
All the cases are straightforward, except the assumption rule. If
$\Gamma \vdash A$ because $A$ occurs in $\Gamma$, then $\Box
\Gamma \vdash \Box A$, and using modus ponens along with an
instance of axiom $\textsf{T}$ yields the result. \end{proof}

Finally, we present a rule that we call the \emph{Four rule}.  As
its name suggests, the Four rule encapsulates the deductive
behaviour of axiom \textsf{4}. In a nutshell, it expresses the
fact that if something is derivable from $\Box \Box A$ then it is
derivable from $\Box A$ itself.

The Four rule only pertains to logics that include all instances
of \textsf{4}. One of these logics is \textsf{CGL}, but in that
case \textsf{4} is a theorem, so we begin by deriving it.
\begin{lem}
  $\vdash_{\textsf{CGL}} \Box A \rightarrow \Box\Box A$
\end{lem}
\begin{proof}
  We follow \cite{Boolos1994a}. By using one of the conjunction
  axioms of $(\textsf{IPL}_\Box)$ and Scott's rule, we have $\Box
  (\Box A \land A) \vdash \Box A$, and hence $ A, \Box(\Box A
  \land A) \vdash \Box A \wedge A$ by weakening, axiom, and
  the one of the conjunction axioms. Then, Scott's rule followed
  by the deduction theorem yield that $ \Box A \vdash \Box
  \left(\Box\left(\Box A \land A\right) \rightarrow \Box A \land
  A\right) $. The conclusion matches the premise of the
  G\"odel-L\"ob axiom, so using modus ponens gives yields $ \Box A
  \vdash \Box\left(\Box A \land A\right) $. Cutting this with
  $\Box \left(\Box A \land A\right) \vdash \Box\Box A$ and using
  the deduction theorem completes the proof.
\end{proof}

\begin{thm}[Admissibility of the Four Rule] 
  \label{thm:four}
  If $\mathcal{L}$ is a logic that includes \textsf{4} either as
  axiom or as theorem, i.e.  if $\mathcal{L} \in \{\textsf{CK4},
  \textsf{CGL}, \textsf{CS4}\}$, then the following rule is
  admissible: 
  \[
    \begin{prooftree}
      \Box \Gamma, \Gamma \vdash_\mathcal{L} A	
        \justifies
      \Box \Gamma \vdash_\mathcal{L} \Box A
    \end{prooftree}
  \]
\end{thm}
\begin{proof}
  Induction on the derivation of $\Box \Gamma, \Gamma \vdash A$.
  Most cases are straightforward.  If $\Box \Gamma, \Gamma \vdash
  A$ by the assumption rule, it follows that $A$ either occurs in
  $\Box \Gamma$, or it occurs in $\Gamma$. If it occurs in $\Box
  \Gamma$, then it is of the form $\Box A'$; thus $\Box \Gamma
  \vdash \Box A'$, and using modus ponens alongside an instance of
  axiom \textsf{4} yields $\Box \Gamma \vdash \Box\Box A' = \Box
  A$. If, on the other hand, $A$ occurs in $\Gamma$, then $\Box
  \Gamma \vdash \Box A$ by the assumption rule.
\end{proof}

A slightly weaker variant of the Four rule appears in
\cite{Bierman2000a}, and follows by weakening.

\begin{cor} \label{thm:four_cor}
  If $\mathcal{L} \in \{\textsf{CK4}, \textsf{CGL},
  \textsf{CS4}\}$, then the following rule is admissible:\[
    \begin{prooftree}
      \Box \Gamma \vdash_\mathcal{L} A	
       	\justifies
      \Box \Gamma \vdash_\mathcal{L} \Box A
    \end{prooftree}
\]
\end{cor} 

If veridicality is admissible as well---i.e. in the case of
\textsf{CS4}---we can derive the theorem from the corollary. If
$\Box \Gamma, \Gamma \vdash A$, then $\Box\Box \Gamma, \Box \Gamma
\vdash A$ by veridicality, and repeatedly cutting with instances
of $\Box B \vdash \Box\Box B$ yields $\Box \Gamma, \Box \Gamma
\vdash A$.  Repeated uses of exchange and contraction then show
$\Box \Gamma \vdash A$, to which we apply the corollary. 

Finally, we show that \emph{L\"ob's rule} is admissible in
\textsf{CGL}.

\begin{thm}[L\"ob's Rule]
\label{thm:lob} 
  The rule $
    \begin{prooftree}
      \Box \Gamma, \Gamma, \Box A \vdash A
        \justifies
      \Box \Gamma \vdash \Box A
    \end{prooftree}
  $ is admissible in \textsf{CGL}.
\end{thm}
\begin{proof} 
  By the deduction theorem we infer that $\Box \Gamma, \Gamma
  \vdash \Box A \rightarrow A$, and hence, by the Four rule, $\Box
  \Gamma \vdash \Box (\Box A \rightarrow A)$. We then use the
  G\"odel-L\"ob axiom and modus ponens.
\end{proof}

The following, which follows by weakening, is often quoted as
L\"ob's rule.

\begin{cor} The rule 
  $
  \begin{prooftree}
    \Box \Gamma, \Box A \vdash A
      \justifies
    \Box \Gamma \vdash \Box A
  \end{prooftree}
  $
  is admissible in \textsf{CGL}.
\end{cor}

\section{From sequent calculi to dual contexts}
  \label{sec:deriving}

We will now discuss the problems that one usually faces when
devising modal $\lambda$-calculi for box modalities. We then
demonstrate how the dual-context pattern decisively deals with
many of these, by importing patterns found in well-behaved sequent
calculi.

\subsection{The perennial issues}
  \label{sec:perennial}

  Most work on the subject is concentrated on essentially two kinds of
  calculi: those with \emph{delayed substitutions}, following a style
  that was popularised by Bierman and de Paiva \cite{Bierman2000a};
  and those employing \emph{dual contexts}, a pattern that was
  imported into modal type theory by Davies and Pfenning
  \cite{Davies2001a, Davies2001}.

\paragraph{Explicit substitutions \`{a} la Bierman \& de Paiva}
  \label{sec:depaiva} 

The calculus introduced by Bierman and de Paiva made use of a
trick that was previously employed in the context of
Intuitionistic Linear Logic by \cite{Benton1993b} to ensure that
substitution is admissible. The trick is simple: \emph{if cut is
not admissible, then we build it into the introduction rule}.

In the case of \textsf{CS4}, the resultant syntax is an extension
of the ordinary simply-typed $\lambda$-calculus. The extension is
obtained by adding the following introduction rule:
\[
  \begin{prooftree}
    \Gamma \vdash M_1 : \Box A_1 
      \quad \dots \quad
    \Gamma \vdash M_n : \Box A_n
      \quad\quad
    x_1 : \Box A_1, \dots, x_n : \Box A_n \vdash N : B
      \justifies
    \Gamma \vdash \textsf{box } N \textsf{ with } M_1, \dots, M_n
    \textsf{ for } x_1, \dots x_n : \Box B
  \end{prooftree}
\] 
In this example, $x_1, \dots x_n$ comprise all the free variables
that may occur in $N$. They must all be `modal,' in that their
type has to start with a box. We are allowed to place a box in
front of $B$, but we must provide a substitute $M_i$ for each of
these free variables. Of course, this $M_i$ must also be of modal
type. In short: \emph{all the data that goes into the making of
something of type $\Box B$ must be `boxed.'} The given substitutes
$M_i$ are `frozen' as part of the term of type $\Box B$: they
become a \emph{delayed substitution}\footnote{These are often
referred to as \emph{explicit substitutions}. The present author
reserves this term for those that are intentionally build into a
calculus, and are not an artifact of proof-theoretic desires.} in
the syntax. This is a combined introduction and cut rule: the
introduction part ensures that modal data depend only on modal
data, and the cut part ensures that substitution is admissible.

The elimination rule is simpler by comparison, and incorporates
axiom \textsf{T}: \[
  \begin{prooftree}
      \Gamma \vdash M : \Box A
        \justifies
      \Gamma \vdash \textsf{unbox } M : A
  \end{prooftree}
\] 
In order to ensure admissibility of cut and hence subject
reduction, the $\beta$-rule associated with these rules has the
effect of unrolling the delayed substitutions \emph{en masse}: 
\[
  \textsf{unbox }(\textsf{box } N \textsf{ with } M_1, \dots, M_n
    \textsf{ for } x_1, \dots x_n)
  \longrightarrow
    N[M_1/x_1, \dots, M_n/x_n]
\] 
Calculi of this sort are notorious for suffering from two kinds of
problems: the need for \emph{commuting conversions}, and the lack
of \emph{`good' symmetries}.

\begin{description}

\item[Commuting Conversions] 

In order to maintain the validity of vital proof-theoretic
results, calculi with delayed subsitutions often require a large
number of \emph{commuting conversions}. The r\^{o}le of these
conversions is to expose `hidden' redexes, the existence of which
spoil the so-called \emph{subformula property}, i.e. the property
that in normal proofs all detours have been eliminated. The issue
of commuting conversions usually arises from positive connectives,
such as disjunction and existence: see the book by Girard \cite[\S
10.1]{Girard1989} for a particularly perspicuous discussion.

In calculi such as the above, commuting conversions invariably
take the form of \emph{structural rules} that reshuffle the
delayed substitutions. Such rules are traditionally found in
sequent calculi, but not in natural deduction, where they are
often admissible. Their presence in a natural deduction system is
incompatible with the view that natural deduction proofs comprise
the ``real proof objects''---see \cite[\S 5.4]{Girard1989}. In the
context of Bierman and de Paiva's system for \textsf{CS4},
Goubault-Larrecq \cite{Goubault-Larrecq1996} argues that systems
with such rules obscure the computational meaning of modal proofs.

\item[`Good' symmetries] 

The calculus of Bierman and de Paiva for \textsf{CS4} exhibits
reasonable symmetries: if we forget about the delayed
substitutions for a moment, then we can see an introduction and an
elimination rule, the latter post-inverse to the former: there is
reasonable \emph{harmony}.

Things are not that simple when it comes to other calculi of this
sort. As a first example, consider the calculus of Bellin-de
Paiva-Ritter \cite{Bellin2001} for \textsf{CK}. Its introduction
rule is only slightly different to the one for \textsf{CS4}, in
that the free variables need not be of modal type. However, the
substitutes for these free variables must still be modal. To
wit:
\[
  \begin{prooftree}
      \Gamma \vdash M_1 : \Box A_1 
      \quad \dots \quad
      \Gamma \vdash M_n : \Box A_n
    \quad
      x_1 : A_1, \dots, x_n : A_n \vdash N : B
    \justifies
      \Gamma \vdash 
        \textsf{box}\ N\ \textsf{with}\ M_1, \dots, M_n\
                         \textsf{for}\ x_1, \dots x_n : \Box B
  \end{prooftree}
\] In this calculus there can be no harmony, for there is no
elimination rule at all. Indeed, the only plausible `$\beta$-rule'
is very similar to a commuting conversion for \textsf{CS4} that
was studied by Goubault-Larrecq \cite{Goubault-Larrecq1996}. Its
function is to unbox any `canonical' terms in the delayed
substitutions; e.g \[
  \textsf{box}\ yx\
    \textsf{with}\ y, (\textsf{box}\ M\ \textsf{with}\ z\ \textsf{for}\ z), N\
    \textsf{for}\  y, x, w\
    \longrightarrow\
  \textsf{box } yM \textsf{ with } y, z, N \textsf{ for } y, z, w
\] This reduction locates `boxed' delayed substitutions of `boxed'
proofs, and combines them into a single `boxed' proof. See
\cite{Kakutani2007a} for a calculus for \textsf{CK} with this
rule. 

Secondly, this pattern leads to significant complexity in more
complicated systems, e.g. when we need \emph{diagonal
assumptions}, which are natural in the case of \textsf{GL}. The
only natural deduction system for \textsf{CGL}, which is due to
Bellin \cite{Bellin1985}, is of this form. Translating the proof
tree formulation to terms terms, its single modal rule is
\[
  \begin{prooftree}
      \begin{array}{c}
        \Gamma \vdash M_1 : \Box B_1 
          \quad \dots \quad
        \Gamma \vdash M_m : \Box B_m
          \quad
        \Gamma \vdash N_1 : \Box C_1 
          \quad \dots \quad
        \Gamma \vdash N_n : \Box C_n \\
        x_1 : B_1, \dots, x_m : B_m,
        y_1 : \Box C_1, \dots, y_n : \Box C_n,
        z : \Box A \vdash N : A
      \end{array}
    \justifies
      \Gamma \vdash 
        \textsf{fix}\ z\ \textsf{in}\ 
        \textsf{box}\ N\ \textsf{with}\ M_1, \dots, M_n \mid N_1, \dots, N_n\
                        \textsf{for}\ x_1, \dots x_m \mid y_1, \dots y_n : \Box A
  \end{prooftree}
\] 
This calculus is virtually at the midpoint between the Bierman-de
Paiva calculus for \textsf{CS4}, and the Bellin-de Paiva-Ritter
calculus for \textsf{CK}: some the assumptions that are being
closed come are `boxed,' and some are not. Evidently, this pattern
is closely related to L\"ob's rule (Theorem \ref{thm:lob}) for
$\textsf{CGL}$. It also features a \emph{diagonal assumption} $z :
\Box A$, which is bound in the resulting term.

Normalization for the terms of Bellin's calculus is by no means
easy to describe. The only two small-step reductions that this
calculus admits are also akin to commuting conversions. Using
vector notation for succinctness, write $\textsf{box}\ N\
\textsf{with}\ \vec{M} \mid \vec{N}\ \textsf{for}\ \vec{x} \mid
\vec{y}$ to mean $\textsf{fix}\ z\ \textsf{in}\ \textsf{box}\ N\
\textsf{with}\ \vec{M} \mid \vec{N}\ \textsf{for}\ \vec{x} \mid
\vec{y}$, where $z$ is a fresh variable that is not free in $N$.
Following Bellin, we call this a \textsf{K4R} application. The
first small-step reduction (``\textsf{K4R} reduction'') is
essentially the one for \textsf{CK} given above, which here takes
the form
\begin{align*}
  &\textsf{fix}\ z\ \textsf{in}\
  \textsf{box}\ M\
    \textsf{with}\ 
      \vec{P}, 
      (\textsf{box}\ N\ \textsf{with}\ \vec{S} \mid \vec{T} \ \textsf{for}\ \vec{v}\mid\vec{v'}), 
      \vec{Q}
        \mid
      \vec{R}\
    \textsf{for}\ 
      \vec{x}, w, \vec{z} \mid \vec{y} \\
  \longrightarrow\
  &\textsf{fix}\ z\ \textsf{in}\
  \textsf{box}\ M[N/w]\
    \textsf{with}\ 
      \vec{P}, 
      \vec{S},
      \vec{Q}
        \mid
      \vec{T},
      \vec{R}\
    \textsf{for}\ 
      \vec{x}, \vec{v}, \vec{y} \mid \vec{v'}, \vec{y}
\end{align*} 
The second rule (``segment reduction step'') is closer to a
commuting conversion for \textsf{CS4}:
\begin{align*}
  &\textsf{fix}\ z\ \textsf{in}\
  \textsf{box}\ M\
    \textsf{with}\ 
      \vec{P} 
        \mid
      \vec{Q},
      (\textsf{fix}\ b\ \textsf{in}\ N\ 
        \textsf{with}\ \vec{S} \mid \vec{T}\ \textsf{for}\ \vec{v}\mid\vec{w}), 
      \vec{R}\
    \textsf{for}\ 
      \vec{x} \mid \vec{y}, z, \vec{y'} \\
  \longrightarrow\
  &\textsf{fix}\ z\ \textsf{in}\
  \textsf{box}\ M[
      (\textsf{fix}\ b\ \textsf{in}\ N\ 
        \textsf{with}\ \vec{c} \mid \vec{d}\ \textsf{for}\ \vec{v} \mid \vec{w})/z]\
    \textsf{with}\ 
      \vec{P}, 
      \vec{S}
        \mid
      \vec{Q},
      \vec{T},
      \vec{R}\
    \textsf{for}\ 
      \vec{x}, \vec{c} \mid \vec{y}, \vec{d}, \vec{y'}
\end{align*} 
Normalization of proofs for Bellin's calculus is contingent on an
auxiliary recursive algorithm, whose purpose is to turn every
modal rule into a \textsf{K4R} application by \emph{eliminating
diagonal assumptions}, so that the small-step reductions can then
simplify the remaining cuts. The details are far too complex to
reproduce here.

\end{description} 
It is thus evident that, once we step out of \textsf{CS4}, the use
of systems based on `mixed' introduction rules with delayed
subsitutions becomes less and less tenable: the resulting systems
lack harmony, and proof normalization becomes frighteningly
complicated.

In order to reach a better solution we must overcome two problems:
(a) we must `decouple' the two flavours---introduction and
cut---that together constitute the introduction or mixed modal
rules; and (b) we must minimize as much as possible the commuting
conversions---in particular, we should strive to free them from
any computational content. 

\subsection*{\bfseries Dual contexts}
  \label{sec:dual}

The right intuition for achieving this decoupling was introduced
by Girard \cite{Girard1993} in his attempt to combine classical,
intuitionistic, and linear logic in one system, and also
independently by Andreoli \cite{Andreoli1992} in the context of
linear logic programming. The idea is simple and can be turned
into a slogan: \emph{segregate assumptions}. This means that we
should divide our usual context of assumptions in two, or---even
better---think of it as consisting of \emph{two zones}. We should
think of one zone as the \emph{primary zone}, and the assumptions
occuring in it as the `ordinary' sort of assumptions. The other
zone is the \emph{secondary zone}, and the assumptions in it
normally have a different flavour. In this setting the
introduction rule explains the interaction between the two
contexts, whereas the elimination rule effects substitution for
the secondary context.

This idea has been most profitable in the case of the \emph{Dual
Intuitionistic Linear Logic} (\textsf{DILL}) of Plotkin and Barber
\cite{Plotkin1993, Barber1996} where the primary context consists
of \emph{linear} assumptions, and the secondary one consists of
\emph{intuitionstic} assumptions. The `of course' modality ($!$)
of Linear Logic is very much like a \textsf{S4} modality,
and---simply by lifting the linearity restrictions---Davies and
Pfenning \cite{Davies2001, Davies2001a} adapted this work to the
modal logic \textsf{CS4} with considerable success. In this
system, hereafter referred to as \emph{dual constructive
\textsf{S4}} (\textsf{DS4}), the primary context consists of
\emph{intuitionistic} assumptions, whereas the secondary context
consists of \emph{modal} assumptions.

The systems of Barber, Plotkin, Davies and Pfenning do not
immediately seem adaptable to other logics. Indeed, the pattern
may at first seem limited to modalities like `of course' and the
necessity of \textsf{S4}, which categorically correspond to
\emph{comonads}. As a comonad can be decomposed into an
adjunction, one might think that the dual-context pattern
implicitly makes use of the underlying universal property. In the
rest of this section we show that not only this is not so, but
that the dual-context style can be adapted to capture the
necessity fragments of all of the logics introduced in
\S\ref{sec:logics}.

\subsection{Deriving dual-context calculi}

Gentzen introduced the sequent calculus in the 1930s
\cite{Gentzen1935a, Gentzen1935b} in order to study normalization
of proofs, which we call \emph{cut elimination} in this context.
A proof in the sequent calculus consists of a tree of
\emph{sequents}, which take the form $\Gamma \vdash A$, where
$\Gamma$ is a context. Thus in our notation a sequent is a
different name for a judgment, like the ones in natural
deduction.\footnote{Fundamental differences arise in the classical
case, which features sequents of the form $\Gamma \vdash \Delta$
where $\Gamma$ and $\Delta$ are lists of formulae. In
intuitionistic logic $\Delta$ consists of at most one formula: see
\cite[\S 5.1.3]{Girard1989}.} The rules, however, are different:
they come in two flavours: \emph{left rules} and \emph{right
rules}.  Broadly speaking, right rules are exactly the
introduction rules of natural deduction, as they only concern the
conclusion $A$ of the sequent. The left rules play a r\^{o}le similar
to that of elimination rules, but they do so by `gerrymandering'
with the assumptions in $\Gamma$. See the in-depth discussion of
Girard \cite[\S 5.4]{Girard1989} on the correspondence between
natural deduction and sequent calculus.

The first attempts to forge sequent calculi for modal logics began
in the 1950s, with the formulation of a sequent calculus for
\textsf{S4} by Curry \cite{Curry1952} and Ohnisi and Matsumoto
\cite{Ohnisi1957, Ohnisi1959}. There was also some limited success
for other simple modal logics, mainly involving the axioms we
discuss here: see the surveys of Ono \cite{Ono1998} and Wansing
\cite{Wansing2002}.

\subsubsection{The Introduction Rules}

Let us consider the (intuitionistic) right rule for the logic
\textsf{S4}: \[
  \begin{prooftree}
    \Box \Gamma \vdash A
      \justifies
    \Box \Gamma \vdash \Box A
      \using
    (\Box\mathcal{R})
  \end{prooftree}
\] One cannot help but notice this rule has an intuitive
computational interpretation in terms of `flow of data.' We can
read it as follows: if only modal data are used in inferring $A$,
then we may obtain $\Box A$. Like in the Bierman-de Paiva
calculus, only `boxed' things may flow into something that is
`boxed.' 

Let us at the same time consider dual-context judgments. These
take the form
\[
  \ctxt{\Delta}{\Gamma} \vdash A
\] 
where both $\Delta$ and $\Gamma$ are contexts. The assumptions in
$\Delta$ are to be thought of as \emph{modal}, whereas the
assumptions in $\Gamma$ are run-of-the-mill intuitionistic
assumptions. A loose translation of such a a judgment to the
`ordinary' sort would be 
\[
  \ctxt{\Delta}{\Gamma} \vdash A 
    \qquad \rightsquigarrow \qquad 
  \Box \Delta, \Gamma \vdash A
\] 
Under this translation, if we `mimic' the right rule for
\textsf{S4} we would obtain the following: \[
  \begin{prooftree}
    \ctxt{\Delta}{\cdot} \vdash A
      \justifies
    \ctxt{\Delta}{\cdot} \vdash \Box A
  \end{prooftree}
\] 
where $\cdot$ denotes the empty context. However, natural
deduction systems do not have any structural rules, so we have to
include some kind of opportunity to weaken the context in the
above rule. If we do so, the result is 
\[
  \begin{prooftree}
    \ctxt{\Delta}{\cdot} \vdash A
      \justifies
    \ctxt{\Delta}{\Gamma} \vdash \Box A
  \end{prooftree}
\]

\noindent Under the translation described above, this is exactly
the right rule for \textsf{S4}, with some extra weakening
included. Incidentally, it is also exactly the introduction rule
of Davies-Pfenning dual-context system for \textsf{S4}
\cite{Davies2001}.

This pattern can be harvested to turn the right rules for the box
in sequent calculi into introduction rules in dual-context
systems. We tackle each case separately, except \textsf{T}, which
we discuss in \S \ref{sec:varrule}.

\paragraph{\textsf{K}}

The case for \textsf{K} is slightly harder to fathom at first
sight. This is because its sequent only has a single rule for the
modality, namely \emph{Scott's rule}: \[
  \begin{prooftree}
        \Gamma \vdash A
      \justifies
        \Box\Gamma \vdash \Box A
  \end{prooftree}
\] 
As Bellin et al. \cite{Bellin2001} discuss, this rule is
unsavoury: it is both a left and a right rule at the same time. It
cannot be split into two rules, which is the pattern that bestows
sequent calculus its fundamental symmetries.  Despite this,
Scott's rule is reasonably well-behaved. Leivant
\cite{Leivant1981} and Valentini \cite{Valentini1982} showed that
incorporating it yields a system which admits cut elimination.  It
also also appears in the sequent calculus for \textsf{CK} studied
by Wijesekera \cite{Wijesekera1990}.

According to the preceding translation, our introduction rule
should be 
\[ 
  \begin{prooftree}
      \ctxt{\cdot}{\Delta} \vdash A 
    \justifies
      \ctxt{\Delta}{\cdot} \vdash \Box A
  \end{prooftree}
\] 
Indeed, we emulate Scott's rule by ensuring that \emph{all the
intuitionistic assumptions must become modal, at once}. The final
form is reached again by adding opportunities for weakening: 
\[
  \begin{prooftree}
    \ctxt{\cdot}{\Delta} \vdash A
      \justifies
    \ctxt{\Delta}{\Gamma} \vdash \Box A
  \end{prooftree}
\]

At this point the reader may vehemently protest that this
introduction rule is not in the spirit of natural deduction, as we
are shamelessly messing with assumptions. So much is true. But it
is also true that even the most well-behaved fragments of natural
deduction are not really trees, but involve some `back edges,'
e.g. to record when and which assumptions are discharged: see
\cite[\S 2.1]{Girard1989}. The situation is even more involved
when it comes to the not-so-well-behaved positive fragment
($\lor\,\exists$): for example, the elimination rule for $\lor$,
namely 
\[
  \begin{prooftree}
    \Gamma \vdash A \lor B
      \quad
    \Gamma, A \vdash C
      \quad
    \Gamma, B \vdash C
      \justifies
    \Gamma \vdash C
  \end{prooftree} 
\] involves the silent elimination of two `temporary assumptions,'
$A$ and $B$. Rules involving such temporary assumptions are known
as rules in the style of Schroeder-Heister
\cite{Schroeder-Heister1984}. The sum of it all is this: \emph{the
proofs were never really trees}.

Consequently, the shameless shuffling of assumptions shall not be
a cause for concern. In fact, there is a simple way to think about
the `jump' that the context $\Delta$ makes from intuitionistic to
modal position. If we are writing down a deduction on the
blackboard, and we wish to introduce a box in front of the
conclusion, then all we have to do is to place a mark on all the
assumptions that are open at that point. This does not discharge
them, but it makes them modal: there shall be a fundamentally
different way of substituting for them, and it shall be a little
more complicated than the simple splicing of a proof tree at a
leaf.


\paragraph{\textsf{K4}}

The correct sequent calculus rule for the logic \textsf{K4}, as
well as the proof of cut elimination, is due to Sambin and
Valentini \cite{Sambin1982}.  Using elements from his joint work
with Sambin, as well some counterexamples found in the work of
Leivant \cite{Leivant1981} on \textsf{GL}, Valentini noticed that
the key property induced by axiom \textsf{4} is that anything
derivable from $\Box\Box A$ is derivable from $\Box A$. The
following (mixed left-and-right) rule for the encapsulates this
insight:
\[
  \begin{prooftree}
      \Box \Gamma, \Gamma \vdash A
    \justifies
      \Box \Gamma \vdash \Box A
  \end{prooftree}
\] Thus, to derive $\Box A$ from a bunch of boxed assumptions, it
suffices to derive $A$ from two copies of the same assumptions,
one boxed and one unboxed.\footnote{Indeed, this is the Four rule
we presented in \S\ref{sec:logics}.} This co-occurence of the
same assumptions in two forms will cause some mild technical
complications in the next section, but that will clarify the
structure of the `flow of data' in \textsf{K4}.

Following our previous recipe, a direct translation of this rule
yields \[
  \begin{prooftree}
      \ctxt{\Delta}{\Delta} \vdash A
    \justifies
      \ctxt{\Delta}{\Gamma} \vdash \Box A
  \end{prooftree}
\]

\paragraph{\textsf{GL}}

The correct formulation of sequent calculus for \textsf{GL} is a
difficult problem that has repeatedly received attention.  There
are simple solutions that guarantee that we can derive all and
only theorems of \textsf{GL}, but they fail to satisfy cut
elimination. 

The first attempt at a cut-free sequent calculus was that of
Leivant \cite{Leivant1981}. Soon thereafter, Valentini
\cite{Valentini1983} showed that Leivant's proof of cut
elimination was incorrect. Sambin and Valentini \cite{Sambin1980}
describe a procedure for building cut-free proofs for all provable
sequents, but their proof is semantic and goes through Kripke
structures, and hence does not constitute Gentzen-style cut
elimination. In \cite{Sambin1982}, the same authors collect and
describe in detail many early approaches, the reasons they do or
do not work, and all relevant results. Finally, Valentini
\cite{Valentini1983} shows that the same rule admits cut
elimination, but the proof is rather complicated, and derives from
the techniques of Bellin \cite{Bellin1985}. Recent progress on
clarifying that result may be found in Gor\'{e} and Ramanayke
\cite{Gore2012}. Another approach, this time based on infinitary
derivations, has been followed by Shamkanov \cite{Shamkanov2014}.

The Leivant-Valentini sequent calculus rule for \textsf{GL} is \[
  \begin{prooftree}
    \Box \Gamma, \Gamma, \Box A \vdash A
      \justifies
    \Box \Gamma \vdash \Box A
  \end{prooftree}
\] 
The only difference between this rule and the rule for \textsf{K4}
is the `diagonal assumption' $\Box A$. We can straightforwardly
use our translation to state it as an introduction rule:
\[
  \begin{prooftree}
    \ctxt{\Delta}{\Delta, \Box A} \vdash A
      \justifies
    \ctxt{\Delta}{\Gamma} \vdash \Box A
  \end{prooftree}
\]

\subsubsection{The Elimination Rule}
  \label{sec:elimrule}

As discussed in \S \ref{sec:dual}, in a dual-context calculus we
consider one context to be the \emph{primary zone}, and the other
to be the \emph{secondary zone}. Assumptions in the primary zone
are discharged by $\lambda$-abstraction. Thus, the function space
of \textsf{DILL} is linear, whereas the function space of
\textsf{DS4} is intuitionistic. 

In contrast, substituting for assumptions in the secondary zone is
the capacity of the \emph{elimination rule}. This is a customary
pattern for dual-context calculi: unlike primary assumptions,
substitution for secondary assumptions is essentially a cut rule.
In the term assignment system we will consider later, this takes
the form of a \emph{delayed substitution}, a type of `let
construct.' The rationale is this: the rest of the system controls
how secondary assumptions arise and are used, and the elimination
rule uniformly allows one to substitute for
them.\footnote{Alternative approaches have also been considered.
For example, one could introduce another abstraction operator,
i.e. a `modal $\lambda$.' This has been adopted by Pfenning
\cite{Pfenning2001} in a dependently-typed setting.} To wit: \[
  \begin{prooftree}
    \ctxt{\Delta}{\Gamma} \vdash \Box A
  \quad
    \ctxt{\Delta, A}{\Gamma} \vdash C
  \justifies
    \ctxt{\Delta}{\Gamma} \vdash C
  \using
    (\Box\mathcal{E})
  \end{prooftree}
\] 
The reader might protest that we are trying to pass a cut rule as
an elimination rule. Notwithstanding the hypocrisy, this is not
only common, but also the best presently known solution that
recovers the patterns of introduction and elimination in the
presence of modality. It is the core of our second slogan:
\emph{in dual-context systems, substitution is a cut rule for
secondary assumptions}. 

One cannot help but notice that such rules are also in the
infamous style of Schroeder-Heister \cite{Schroeder-Heister1984},
and also similar to the elimination rule for disjunction. As we
discussed in \S\ref{sec:perennial}, this kind of rule is known to
be problematic, as it automatically necessitates some commuting
conversions: unavoidably, the conclusion $C$ has no structural
relationship with anything else in sight. The pressing question
is whether this is an acceptable state of affairs. Unless we are
to engage in more complicated and radical schemes, the present
author is afraid that we must settle for it. Put simply, there is
no good way to do away with commuting conversions: they are
part-and-parcel of any sufficiently complicated type theory. All
we can hope for is to minimize their number, and state them
systematically.

\subsubsection{A second variable rule}
  \label{sec:varrule}

We have conveniently avoided discussing two things up to this
point: the left rule for $\Box$ in \textsf{S4}, which is the only
one of our logics that has both left and right rules, and the case
of \textsf{T}. These two are intimately related.

The left rule for necessity in \textsf{S4} is \[
  \begin{prooftree}
    \Gamma, A \vdash B
      \justifies
    \Gamma, \Box A \vdash B
      \using
    (\Box\mathcal{L})
  \end{prooftree}
\] 
We can intuitively read it as follows: if $A$ suffices to infer
$B$, then $\Box A$ is more than enough. This is a form of the
veridicality rule from \S\ref{sec:admissible}, and encapsulates
the axiom $\Box A \rightarrow A$. Together with Scott's rule, it
forms a sequent calculus where cut is admissible; this is
mentioned by Wansing \cite{Wansing2002}, and is attributed to
Ohnisi and Matsumoto \cite{Ohnisi1957}.

One way of emulating this rule in our framework would be to have a
construct that makes an assumption `jump' from one context to
another, but that is inelegant and leads to an unworkable
metatheory. The right way to imitate the left rule is to include a
rule that allows one to use a \emph{modal} assumption as if it
were merely intuitionistic. To wit: \[
  \begin{prooftree}
    \justifies
      \ctxt{\Delta, A, \Delta'}{\Gamma} \vdash A
  \end{prooftree}
\] This translates back to the sequent $\Box\Delta, \Box A, \Box
\Delta', \Gamma \vdash A$, which follows by $(\Box\mathcal{L})$.

A rule like this was introduced by Plotkin and Barber
\cite{Plotkin1993, Barber1996} for \emph{dereliction} in
\textsf{DILL}, and is also essential in Davies and Pfenning's
\textsf{DS4}. In our case, we use it in combination with the
introduction rule for \textsf{K} in order to make a system for
\textsf{T}.

\section{Types, terms, and metatheory}
  \label{sec:metatheory}

We now collect all the observations we have made in order to turn
our natural deduction systems into term assignment systems, i.e.
typed $\lambda$-calculi. First, we annotate each assumption $A$
with a variable, which we write as $x : A$. Then, we annotate each
judgment $\ctxt{\Delta}{\Gamma} \vdash A$ with a term $M$
representing the entire deduction that with that judgment as its
conclusion---see \cite[\S 3]{Girard1989} or \cite{Gallier1993,
Sorensen2006} for an introduction to term assignment.

\begin{figure}
  \centering
  \begin{small}
\begin{alignat*}{3}
  \textbf{Types}\qquad &
    A, B &&::=\   
        && p_i
      \mid A \times B 
      \mid A \rightarrow B 
      \mid \Box A
   \\
  \textbf{Typing Contexts}\qquad &
    \Gamma, \Delta &&::=\
        && \cdot 
      \mid \Gamma, x : A
   \\
  \textbf{Terms}\qquad &
    M, N &&::=\
        && x
      \mid \lambda x : A.\ M
      \mid M N
      \mid \langle M, N \rangle
      \mid \pi_i(M)
      \mid
    \\
      &
      &&
        && \ibox{M}
      \mid \letbox{u}{M}{N}
\end{alignat*}
  
\renewcommand{\arraystretch}{2.7}

\begin{tabular}{c c}

  $ 
  \begin{prooftree}
    \justifies
      \ctxt{\Delta}{\Gamma, x : A, \Gamma'} \vdash x:A
    \using
      (\textsf{var})
  \end{prooftree} 
  $

  &
  $ 
    \begin{prooftree}
      \justifies
        \ctxt{\Delta, u :  A, \Delta'}{\Gamma} \vdash u : A
      \using
        (\Box\textsf{var})
    \end{prooftree} 
  $

  \\


  $ 
  \begin{prooftree}
      \ctxt{\Delta}{\Gamma} \vdash M : A
    \quad
      \ctxt{\Delta}{\Gamma} \vdash N : B
    \justifies
      \ctxt{\Delta}{\Gamma} 
        \vdash \langle M, N \rangle : A \times B
    \using
      (\times\mathcal{I})
  \end{prooftree}
  $

  &

  $
  \begin{prooftree}
      \ctxt{\Delta}{\Gamma} \vdash M : A_1 \times A_2
    \justifies
      \ctxt{\Delta}{\Gamma} \vdash \pi_i\left(M\right) : A_i
    \using
      (\times\mathcal{E}_i)
  \end{prooftree}
  $

  \\


  $
  \begin{prooftree}
      \ctxt{\Delta}{\Gamma}, x : A \vdash M : B
    \justifies
      \ctxt{\Delta}{\Gamma} \vdash \lambda x : A. \; M : A \rightarrow B
    \using
      (\rightarrow\mathcal{I})
  \end{prooftree}
  $

  &

  $
  \begin{prooftree}
      \ctxt{\Delta}{\Gamma} \vdash M : A \rightarrow B
    \quad
      \ctxt{\Delta}{\Gamma} \vdash N : A
    \justifies
      \ctxt{\Delta}{\Gamma} \vdash M N : B
    \using
      (\rightarrow\mathcal{E})
  \end{prooftree}
  $

  \\


  $ 
    \begin{prooftree}
        \ctxt{\cdot}{\Delta} \vdash M : A
      \justifies
        \ctxt{\Delta}{\Gamma} \vdash \ibox{M} : \Box A
      \using
         (\Box\mathcal{I}_\textsf{K})
    \end{prooftree}
  $

  &

  $ 
    \begin{prooftree}
        \ctxt{\Delta}{\Delta^\bot} \vdash M^\bot : A
      \justifies
        \ctxt{\Delta}{\Gamma} \vdash \ibox{M} : \Box A
      \using
        (\Box\mathcal{I}_\textsf{K4})
    \end{prooftree}
  $

  \\

  $ 
    \begin{prooftree}
        \ctxt{\Delta}{\Delta^\bot, z^\bot : \Box A} \vdash M^\bot : A
      \justifies
        \ctxt{\Delta}{\Gamma} \vdash \fixbox{z}{M} : \Box A
      \using
        (\Box\mathcal{I}_\textsf{GL})
    \end{prooftree}
  $

  &

  $
    \begin{prooftree}
        \ctxt{\Delta}{\cdot} \vdash M : A
      \justifies
        \ctxt{\Delta}{\Gamma} \vdash \ibox{M} : \Box A
      \using
         (\Box\mathcal{I}_\textsf{S4})
    \end{prooftree}
  $

  \\
  
  \multicolumn{2}{c}{

  $
    \begin{prooftree}
        \ctxt{\Delta}{\Gamma} \vdash M : \Box A
      \quad\quad
        \ctxt{\Delta, u : A}{\Gamma} \vdash N : C
      \justifies
        \ctxt{\Delta}{\Gamma} \vdash \letbox{u}{M}{N} : C
      \using
        (\Box\mathcal{E})
    \end{prooftree} 
  $

  }

  \\

  \multicolumn{2}{l}{
  \textbf{Rules for \textsf{S4}:}

  $(\Box\mathcal{I}_\textsf{S4})$

  and 

  $(\Box\textsf{var})$

  }

  \\

  \multicolumn{2}{l}{
  \textbf{Rules for \textsf{T}}:

  $(\Box\mathcal{I}_\textsf{K})$ 

  and 

  $(\Box\textsf{var})$
  }

\end{tabular}
\end{small}


  \caption{Definition and typing judgments}
  \label{fig:types}
\end{figure}

The grammars defining types, terms and contexts, as well as the
typing rules for all our systems can be found in Figure
\ref{fig:types}. All of our systems contain the introduction and
elimination rules for products and functions, the variable rule
$(\textsf{var})$, and the box elimination rule
$(\Box\mathcal{E})$. Each of the systems for $\textsf{K}$,
$\textsf{K4}$ and $\textsf{GL}$ also contain the corresponding
introduction rule, e.g. $(\Box\mathcal{I}_\textsf{K})$. Finally,
the systems for $\textsf{T}$ and $\textsf{S4}$ each contain two
additional rules: the modal variable rule $(\Box\textsf{var})$,
and a modal introduction rule---$(\Box\mathcal{I}_\textsf{K})$ in
the first case, and $(\Box\mathcal{I}_\textsf{S4})$ in the second.
When we are at risk of confusion we annotate the turnstile with a
subscript to indicate which system we mean. 

From this point onwards, we assume Barendregt's conventions: terms
are equal up to $\alpha$-conversion, and bound variables are
silently renamed whenever necessary. In $\letbox{u}{M}{N}$, $u$ is
a bound variable in $N$. Finally, we write $N[M/x]$ to mean
capture-avoiding substitution of $M$ for $x$ in $N$. Furthermore,
we shall assume that whenever we write a judgment like
$\ctxt{\Delta}{\Gamma} \vdash M : A$, then $\Delta$ and $\Gamma$
are \emph{disjoint}, in the sense that $\vars{\Delta} \cap
\vars{\Gamma} = \emptyset$, where $ \vars{x_2 : A_1, \dots, x_n :
A_n} \defeq \{x_1, \dots, x_n\}$. This causes a mild technical
complication in the cases \textsf{K4} and \textsf{GL}.
Fortunately, the solution is relatively simple, and we explain it
now.

\subsection{Complementary variables}
  \label{sec:complementary}

Na\"{i}vely annotating the rule for \textsf{K4} would yield \[
  \begin{prooftree}
    \ctxt{\Delta}{\Delta} \vdash M : A
      \justifies
    \ctxt{\Delta}{\Gamma} \vdash \ibox{M} : \Box A
  \end{prooftree}
\] This, however, violates our convention that the two contexts
are disjoint: the same variables will appear both at modal and
intuitionistic positions.  To overcome this we introduce the
notion of \emph{complementary variables}. Let $\mathcal{V}$ be the
set of term variables for our calculi. A \emph{complementation
function} is an \emph{involution} on variables. That is, it is a
bijection $(-)^\bot : \mathcal{V} \xrightarrow{\cong} \mathcal{V}$
which happens to be its own inverse, i.e.
$\left(x^\bot\right)^\bot = x$.  The idea is that, if $u$ is the
modal variable representing some assumption in $\Delta$, we will
write $u^\bot$ to refer to a variable $x$, uniquely associated to
$u$, and representing the same assumption, but without a box in
front. For technical reasons, we would like that $x^\bot$ be the
same variable as $u$.

We extend the involution to contexts: \[
  (x_1: A_1, \dots, x_n : A_n)^\bot
    \defeq x_1^\bot : A_1, \dots, x_n^\bot : A_n
\] We also inductively extend $(-)^\bot$ to terms, with the
exception that it must not change anything inside a $\ibox{(-)}$
construct. It also need not change any bound modal variables, as
for \textsf{K4} and \textsf{GL} these shall only occur under
$\ibox{(-)}$ constructs: \begin{align*}
  (\lambda x : A. M)^\bot &\defeq \lambda x^\bot : A.\ M^\bot &
  (MN)^\bot &\defeq M^\bot N^\bot \\
  \langle M, N \rangle &\defeq \langle M^\bot, N^\bot \rangle &
  (\pi_i(M))^\bot &\defeq \pi_i(M^\bot) \\
  (\ibox M)^\bot &\defeq \ibox{M} & 
  (\letbox{u}{M}{N})^\bot &\defeq \letbox{u}{M^\bot}{N^\bot}
\end{align*} We use this machinery to modify the rule, so as to
maintain disjoint contexts. When we encounter an introduction rule
for the box and the context $\Delta$ gets `copied' to the
intuitionistic position, we will complement all variables in the
copy, as well as all variables occuring in $M$, but
not under any $\ibox{(-)}$ constructs, or bound by a
$\textsf{let}$:
\[
  \begin{prooftree}
    \ctxt{\Delta}{\Delta^\bot} \vdash M^\bot : A
      \justifies
    \ctxt{\Delta}{\Gamma} \vdash \ibox{M} : \Box A
  \end{prooftree}
\] 
As an example, here is a derivation of $\ctxt{\cdot}{\Box A}
\vdash \Box(A \wedge \Box A)$: 
\begin{small}
\[
  \begin{prooftree}
    \ctxt{\cdot}{x : \Box A} \vdash x : \Box A
      \quad
    \[
      \[
	\[
	    \justifies
	   \ctxt{u : A}{u^\bot : A} \vdash u^\bot : A
	\]
	\quad
	\[
	  \[
	      \justifies
	    \ctxt{u : A}{u^\bot : A} \vdash u^\bot : A
	  \]
	    \justifies
	   \ctxt{u : A}{u^\bot : A} \vdash \ibox{u} : \Box A
	\]
	  \justifies
	\ctxt{u : A}{u^\bot : A} \vdash
	        \langle u^\bot, \ibox{u}\rangle
		          : A \times \Box A
      \]
	\justifies
      \ctxt{u : A}{x : \Box A} \vdash
	\ibox{\langle u, \ibox{u}\rangle}
	  : \Box (A \times \Box A)
    \]
      \justifies
    \ctxt{\cdot}{x : \Box A} \vdash 
      \letbox{u}{x}{\ibox{\langle u, \ibox{u}\rangle}}
	: \Box (A \times \Box A)
  \end{prooftree}
\] \end{small}

We extend complementation to finite sets of variables, by setting
$\{x_1, \dots, x_n\} \defeq x_1^\bot, \dots, x_n^\bot$.  It is not
hard to see that the involutive behaviour of $(-)^\bot$ is
invariant inherited by these extensions, and that a number of
operations commute with $(-)^\bot$.

\begin{lem} \label{lem:involution} \hfill
  \begin{enumerate}
    \item For any context $\Delta$, $\left(\Delta^\bot\right)^\bot
    \equiv \Delta$.

    \item For any finite set of variables $S$,
    $\left(S^\bot\right)^\bot = S$.

    \item For any context $\Delta$, $\vars{\Delta^\bot} =
    \left(\vars{\Delta}\right)^\bot$.

    \item If $S, T$ are finite sets of variables, then $ S
    \subseteq T$ implies  $S^\bot \subseteq T^\bot$.
  \end{enumerate}
\end{lem}

There is a simple relationship between complementation and
substitution:

\begin{thm}
  \label{thm:bot}
  If $u^\bot$ is not free in $M$, then $\left(M[N/u]\right)^\bot
  \equiv M^\bot[N, N^\bot/u, u^\bot]$.
\end{thm}
\begin{proof} 
  By induction on $M$. Recall that $M \not\equiv u^\bot$ by
  assumption. The cases of $\lambda$-abstraction, application,
  pairing, projection, and $\letbox{u}{(-)}{(-)}$
  follow by the IH.
  \begin{indproof}
    \case{$u$}
      Then $
          \left(M[N/u]\right)^\bot
        \equiv 
          N^\bot
        \equiv
          u^\bot[N, N^\bot/u, u^\bot]
        \equiv
          M^\bot[N, N^\bot/u, u^\bot]
      $.
    \case{$v \not\equiv u$}
      Then $
          \left(M[N/u]\right)^\bot
        \equiv
          v^\bot
        \equiv
          v^\bot[N, N^\bot/u, u^\bot]
        \equiv
          M^\bot[N, N^\bot/u, u^\bot]
      $.

    \case{$\ibox{M'}$}
    We have: \[
        \left(\ibox{(M'[N/u])}\right)^\bot
      \equiv
        \ibox{(M'[N/u])}
      \equiv
        (\ibox{M'})^\bot[N, N^\bot/u, u^\bot]
    \] where the last step follows because $\ibox{M'} \equiv
    (\ibox{M'})^\bot$, and $u^\bot$ is not free in~$M'$.
\qedhere
  \end{indproof}
\end{proof}

To conclude our discussion of complementary variables, we
carefully define what it means for a pair of contexts to be
well-defined.

\begin{defi}[Well-defined contexts]
  A pair of contexts $\ctxt{\Delta}{\Gamma}$ is
  \emph{well-defined} just if
    \begin{enumerate}
      \item 
        They are \emph{disjoint}, i.e. $\vars{\Delta} \cap
        \vars{\Gamma} = \emptyset$.

      \item 
        In the cases of \textsf{K4} and \textsf{GL}, no two
        complementary variables occur in the same context; that
        is, $\vars{\Gamma} \cap \vars{\Gamma^\bot} = \emptyset$
        and $\vars{\Delta} \cap \vars{\Delta^\bot} = \emptyset$.
    \end{enumerate}
\end{defi} 
The second condition is easy to enforce, and will prove useful in
some technical proofs.

\subsection{Free variables: boxed and unboxed}

\begin{defi}[Free variables] \hfill
  \begin{enumerate}
    \item 
      The \emph{free variables} $\fv{M}$ of a term
      $M$ are defined by 
      \begin{gather*}
        \begin{align*}
          \fv{x}  &\defeq \{x\} &
          \fv{MN} &\defeq \fv{M} \cup \fv{N} \\
          \fv{\lambda x : A.\ M} &\defeq \fv{M} - \{x\} &
          \fv{\langle M , N \rangle} &\defeq \fv{M} \cup \fv{N} \\
          \fv{\pi_i(M)} &\defeq \fv{M} &
          \fv{\ibox{M}} &\defeq \fv{M}
        \end{align*} \\
        \fv{\letbox{u}{M}{N}} \defeq \fv{M} \cup \left(\fv{N} - \{u\}\right)
      \end{gather*}
      and for \textsf{GL} we replace the clause for
      $\ibox{(-)}$ with $\fv{\fixbox{z}{M}} \defeq \fv{M} -
      \{z\}$.

    \item 
      The \emph{unboxed free variables} $\ufv{M}$ of a term are
      those that \emph{do not} occur under the scope of a
      $\ibox{(-)}$ construct. The formal definition involves
      replacing the clause of $\fv{-}$ for $\ibox{(-)}$ with
      $\ufv{\ibox{M}} \defeq \emptyset$, and, for \textsf{GL},
      with $\ufv{\fixbox{z}{M}} \defeq \emptyset$.

    \item 
      The \emph{boxed free variables} $\bfv{M}$ of a term $M$ are
      those that \emph{do} occur under the scope of a $\ibox{(-)}$
      construct. The formal definition involves replacing the
      clauses of $\fv{-}$ for variables and for $\ibox{(-)}$ by
      $\bfv{x} \defeq \emptyset$ and $\bfv{\ibox{M}} \defeq
      \fv{M}$, and, for \textsf{GL}, by $\bfv{\fixbox{z}{M}}
      \defeq \fv{M} - \{z\}$.
  \end{enumerate}
\end{defi}

\noindent We can then prove the following theorem by a simple
induction on terms.

\begin{thm}[Free variables] \hfill
  \label{thm:freevar}
  \begin{enumerate}

    \item If $\ctxt{\Delta}{\Gamma, x : A, \Gamma'} \vdash M : A$
    and $x \not\in \fv{M}$, then $\ctxt{\Delta}{\Gamma, \Gamma'}
    \vdash M : A$.

    \item If $\ctxt{\Delta, u : A, \Delta'}{\Gamma} \vdash M : A$
    and $u \not\in \fv{M}$, then $\ctxt{\Delta, \Delta'}{\Gamma}
    \vdash M : A$.

    \item For every term $M$, $\fv{M} = \ufv{M} \cup \bfv{M}$.

    \item For every term $M$, $\ufv{M^\bot} = \ufv{M}^\bot$.

    \item For every term $M$, $\bfv{M^\bot} = \bfv{M}$.

    \item If $\mathcal{S} \in \{\textsf{DK}, \textsf{DK4},
      \textsf{DGL}\}$ and $\ctxt{\Delta}{\Gamma} \vdash_\mathcal{S}
      M : A$, then \begin{align*}
        \ufv{M} &\subseteq \vars{\Gamma}, &
        \bfv{M} &\subseteq \vars{\Delta}
      \end{align*}

    \item If $\mathcal{S} \in \{\textsf{DS4}, \textsf{DT}\}$
      and $\ctxt{\Delta}{\Gamma} \vdash_\mathcal{S} M : A$, then
      \begin{align*}
        \ufv{M} &\subseteq \vars{\Gamma} \cup \vars{\Delta}, &
        \bfv{M} &\subseteq \vars{\Delta}
      \end{align*}
  \end{enumerate}
\end{thm}

\opt{th}{
\begin{proof} 
  (1)-(5) follow by induction on $M$. We prove the rest.
  \begin{enumerate}
    \setcounter{enumi}{5}

    \item By induction on the derivation of $\ctxt{\Delta}{\Gamma}
    \vdash_\mathcal{S} M : A$. We show the cases for
    $(\Box\mathcal{I})$. The first statement follows trivially, as
    $\ufv{\ibox{M}} = \ufv{\fixbox{z}{M}} = \emptyset \subseteq
    \vars{\Gamma}$, so it remains to show the second statement. 

    For $(\Box\mathcal{I}_\textsf{K})$, we have
      \begin{derivation}
          \bfv{\ibox{M}}
        \since{definition}
          \fv{M}
        \since{(3)}
          \ufv{M} \cup \bfv{M}
        \since[\subseteq]{IH, twice}
          \vars{\Delta} \cup \vars{\cdot}
        \since{definition}
          \vars{\Delta}
      \end{derivation}

    For $(\Box\mathcal{I}_\textsf{K4})$, we have
      \begin{derivation}
        \bfv{\ibox{M}}
          \since{definition}
        \fv{M}
          \since{(3)}
        \ufv{M} \cup \bfv{M}
          \since{Lemma \ref{lem:involution}(2)}
        \left(\ufv{M}^\bot\right)^\bot \cup \bfv{M}
          \since[\subseteq]{(4), (5), and Lemma \ref{lem:involution}(4) }
        \left(\ufv{M^\bot}\right)^\bot \cup \bfv{M^\bot}
          \since[\subseteq]{IH twice, and Lemma \ref{lem:involution}(4)}
        \vars{\Delta^\bot}^\bot \cup \vars{\Delta}
          \since{Lemma \ref{lem:involution}(1) and (3)}
        \vars{\Delta}
      \end{derivation}

    For $(\Box\mathcal{I}_\textsf{GL})$, we have
    \begin{derivation}
        \bfv{\fixbox{z}{M}}
      \since{definition}
        \fv{M} - \{z\}
      \since{(3)}
        \ufv{M} \cup \bfv{M} - \{z\}
      \since{Lemma \ref{lem:involution}(2)}
        \left(\ufv{M}^\bot\right)^\bot \cup \bfv{M} - \{z\}
      \since[\subseteq]{(4), (5), Lemma \ref{lem:involution}(4),
      and monotonicity of subtraction.}
        \ufv{M^\bot}^\bot \cup \bfv{M^\bot} - \{z\}
      \since[\subseteq]{IH twice, and Lemma \ref{lem:involution}(4)
      and monotonicity of subtraction.}
        \left(\vars{\Delta^\bot} \cup \{z^\bot\}\right)^\bot 
          \cup \vars{\Delta} - \{z\}
      \since{Lemma \ref{lem:involution}(2, 3)}
        \left(\vars{\Delta} \cup \{z\} \cup \vars{\Delta}\right) - \{z\}
      \since{$z \not\in \vars{\Delta}$}
        \vars{\Delta}
    \end{derivation}

    \item By induction on the derivation of $\ctxt{\Delta}{\Gamma}
    \vdash_\mathcal{S} M : A$. We show the case for
    $(\Box\mathcal{I}_\mathsf{S4})$; the first statement is
    trivial, so we show the second:
    \begin{derivation}
      \bfv{\ibox{M}}
	\since{definition}
      \fv{M}
	\since{(1)}
      \ufv{M} \cup \bfv{M}
	\since[\subseteq]{IH, twice}
      \left(\vars{\Delta} \cup \vars{\cdot}\right) \cup \vars{\Delta}
	\since{definition}
      \vars{\Delta}
    \end{derivation}
  \end{enumerate}
\end{proof}
}

\subsection{Structural theorems}

As expected, our systems satisfy the standard menu of structural
results: weakening, contraction, exchange, and cut rules are
admissible.

\begin{thm}[Structural \& Cut] 
  \label{thm:scut}
  The following rules are admissible in all systems:
  \begin{multicols}{2}
  \begin{enumerate}
    \item (Weakening) \[
      \begin{prooftree}
          \ctxt{\Delta}{\Gamma, \Gamma'} \vdash M : C
        \justifies
          \ctxt{\Delta}{\Gamma, x : A, \Gamma'} \vdash M : C
      \end{prooftree}
    \]
    \item (Exchange) \[
      \begin{prooftree}
          \ctxt{\Delta}{\Gamma, x : A, y : B, \Gamma'} \vdash M : C
        \justifies
          \ctxt{\Delta}{\Gamma, y : B, x : A, \Gamma'} \vdash M : C
      \end{prooftree}
    \]
    \item (Contraction) \[
      \begin{prooftree}
          \ctxt{\Delta}{\Gamma, x : A, y : A, \Gamma'} \vdash M : A
        \justifies
          \ctxt{\Delta}{\Gamma, w : A, \Gamma'} \vdash M[w, w/x, y] : A
      \end{prooftree}
    \]
    \item (Cut) \[
      \begin{prooftree}
          \ctxt{\Delta}{\Gamma} \vdash N : A
        \qquad
          \ctxt{\Delta}{\Gamma, x : A, \Gamma'} \vdash M : C
        \justifies
          \ctxt{\Delta}{\Gamma, \Gamma'} \vdash M[N/x] : C
      \end{prooftree}
    \]
  \end{enumerate}
  \end{multicols}
\end{thm}
\begin{proof}
  By induction on the typing derivation of $M$. As an example, we
  show the case of $(\Box\mathcal{I}_\mathsf{K})$ for weakening.
  Suppose $\ctxt{\Delta}{\Gamma, \Gamma'} \vdash M : C$ by
  $(\Box\mathcal{I}_\mathsf{K})$. Then $M \equiv \ibox{M'}$ and $C
  \equiv \Box C'$ and $\ctxt{\cdot}{\Delta} \vdash M' : C'$.  A
  single use of $(\Box\mathcal{I}_\mathsf{K})$ then yields
  $\ctxt{\Delta}{\Gamma, x : A, \Gamma'} \vdash M : C$.
\end{proof}

\begin{thm}[Modal Structural]
  \label{thm:modalstruct}
  The following rules are admissible:
  \begin{multicols}{2}
  \begin{enumerate}
    \item (Modal Weakening) \[
      \begin{prooftree}
        \ctxt{\Delta, \Delta' }{\Gamma} \vdash M : C
          \justifies
        \ctxt{\Delta, u : A, \Delta'}{\Gamma} \vdash M : C
      \end{prooftree}
    \]
    \item (Modal Exchange) \[
      \begin{prooftree}
        \ctxt{\Delta, x : A, y : B, \Delta'}{\Gamma} \vdash M : C
          \justifies
        \ctxt{\Delta, y : B, x : A, \Delta'}{\Gamma} \vdash M : C
      \end{prooftree}
    \]
    \item (Modal Contraction) \[
      \begin{prooftree}
        \ctxt{\Delta, x : A, y : A, \Delta'}{\Gamma} \vdash M : C
          \justifies
        \ctxt{\Delta, w : A, \Delta'}{\Gamma} \vdash M[w, w/x, y] : C
      \end{prooftree}
    \]
  \end{enumerate}
  \end{multicols}
\end{thm}

\begin{proof} 
  Straightforward induction on the derivation of the premise.  As
  an example, we discuss a few cases of $(\Box\mathcal{I})$ for
  weakening, the rest being similar.

  If $\ctxt{\Delta, \Delta'}{\Gamma} \vdash M : A$ by
  $(\Box\mathcal{I}_\textsf{K})$, then $M \equiv \ibox{N}$ and $A
  \equiv \Box B$, with $\ctxt{\cdot}{\Delta, \Delta'} \vdash N :
  B$. We use Theorem \ref{thm:scut} to obtain
  $\ctxt{\cdot}{\Delta, x : A, \Delta'} \vdash N : B$, and then
  apply $(\Box\mathcal{I}_\textsf{K})$.

  If $\ctxt{\Delta, \Delta'}{\Gamma} \vdash M : A$ by
  $(\Box\mathcal{I}_\textsf{K4})$, then $M \equiv \ibox{N}$ and $A
  \equiv \Box B$, with $\ctxt{\Delta, \Delta'}{\Delta^\bot,
  \Delta'^\bot} \vdash N^\bot : B$. By the IH, we have that
  $\ctxt{\Delta, u : A, \Delta'}{\Delta^\bot, \Delta'^\bot} \vdash
  N^\bot : B$.  We use Theorem \ref{thm:scut} to deduce that
  $\ctxt{\Delta, u : A, \Delta'}{\Delta^\bot, u^\bot : A,
  \Delta'^\bot} \vdash N^\bot : B$, and then apply
  $(\Box\mathcal{I}_\textsf{K4})$.
\end{proof}

\begin{thm}[Modal Cut] 
  \label{thm:mcut}
  The following rules are admissible:
  \begin{enumerate}
    \item (Modal Cut for \textsf{DK}) \[
      \begin{prooftree}
          \ctxt{\cdot}{\Delta}
            \vdash_\textsf{DK} N : A
        \quad
          \ctxt{\Delta, u : A, \Delta'}{\Gamma}
            \vdash_\textsf{DK} M : C
        \justifies
          \ctxt{\Delta, \Delta'}{\Gamma}
            \vdash_\textsf{DK} M[N/u] : C
      \end{prooftree}
    \]
    \item (Modal Cut for \textsf{DK4}) \[
      \begin{prooftree}
          \ctxt{\Delta}{\Delta^\bot}
            \vdash_\textsf{DK4} N^\bot : A
        \quad
          \ctxt{\Delta, u : A, \Delta'}{\Gamma}
            \vdash_\textsf{DK4} M : C
        \justifies
          \ctxt{\Delta, \Delta'}{\Gamma} 
            \vdash_\textsf{DK4} M[N/u] : C
      \end{prooftree}
    \]
    \item (Modal Cut for \textsf{DGL}) \[
      \begin{prooftree}
          \ctxt{\Delta}{\Delta^\bot, z^\bot : \Box A}
            \vdash_\textsf{DGL} N^\bot : A
        \quad
          \ctxt{\Delta, u : A, \Delta'}{\Gamma}
            \vdash_\textsf{DGL} M : C
        \justifies
          \ctxt{\Delta, \Delta'}{\Gamma}
            \vdash_\textsf{DGL}
            M\left[N\left[\fixbox{z}{N}/z\right]/u\right] : C
      \end{prooftree}
    \]
    \item (Modal Cut for \textsf{DS4}) \[
      \begin{prooftree}
          \ctxt{\Delta}{\cdot} 
            \vdash_\textsf{DS4} N :  A
        \quad
          \ctxt{\Delta, u : A, \Delta'}{\Gamma}
            \vdash_\textsf{DS4} M : C
        \justifies
          \ctxt{\Delta, \Delta'}{\Gamma}
            \vdash_\textsf{DS4} M[N/u] : C
      \end{prooftree}
    \]
    \item (Modal Cut for \textsf{DT}) \[
      \begin{prooftree}
          \ctxt{\cdot}{\Delta}
            \vdash_\textsf{DT} N : A
        \quad
          \ctxt{\Delta, u : A, \Delta'}{\Gamma}
            \vdash_\textsf{DT} M : C
        \justifies
          \ctxt{\Delta, \Delta'}{\Gamma}
            \vdash_\textsf{DT} M[N/u] : C
      \end{prooftree}
    \]
  \end{enumerate}
\end{thm}
\begin{proof}
  By induction on the typing derivation of $M$. We show the case
  of $(\Box\mathcal{I})$, and---for \textsf{DS4} and
  \textsf{DT}---the case of modal variables $(\Box\textsf{var})$.
    \begin{enumerate}
      \item (\textsf{DK})
        If $\ctxt{\Delta, u : A, \Delta'}{\Gamma} \vdash M : C$ by
        $(\Box\mathcal{I}_\textsf{K})$, then $M \equiv \ibox{M'}$,
        $C \equiv \Box C'$, and \[
        \ctxt{\cdot}{\Delta, u : A, \Delta'}
          \vdash M' : C'
        \] By Theorem \ref{thm:scut}, we have \[
        \ctxt{\cdot}{\Delta, \Delta'}
          \vdash M'[N/u] : C
        \] and hence $\ctxt{\Delta, \Delta'}{\Gamma} \vdash
        \ibox{(M'[N/u])} : \Box C' \equiv C$ by an application of
        $(\Box\mathcal{I}_\textsf{K})$. But \[
        \ibox{\left(M'[N/u]\right)}
          \equiv
        \left(\ibox{M'}\right)[N/u]
          \equiv
        M[N/u]
        \] and hence we have the result.
	
    \item (\textsf{DK4})
      If $\ctxt{\Delta, u : A, \Delta'}{\Gamma} \vdash M : C$ by
      $(\Box\mathcal{I}_\textsf{K4})$, then $M \equiv \ibox{M'}$,
      $C \equiv \Box C'$, and \[
        \ctxt{\Delta, u : A, \Delta'}
        {\Delta^\bot, u^\bot : A, \Delta'^\bot}
            \vdash M'^\bot : C'
      \] By the IH, we have \[
        \ctxt{\Delta, \Delta'}
             {\Delta^\bot, u^\bot : A, \Delta'^\bot}
            \vdash M'^\bot[N/u] : C'
      \] and by Theorem \ref{thm:scut}, that yields \[
        \ctxt{\Delta, \Delta'}
             {\Delta^\bot, \Delta'^\bot}
            \vdash M'^\bot[N, N^\bot/u, u^\bot] : C'
      \] But, by Theorem \ref{thm:bot}, we have that $M'^\bot[N,
      N^\bot/u, u^\bot] \equiv \left(M'[N/u]\right)^\bot$, and
      hence by a use of $(\Box\mathcal{I}_\textsf{K4})$, we have \[
        \ctxt{\Delta, \Delta'}
        {\Gamma}
            \vdash \ibox{\left(M'[N/u]\right)} :
            \Box C' \equiv C
      \] and hence the result.

    \item (\textsf{DGL})
      If $\ctxt{\Delta, u : A, \Delta'}{\Gamma} \vdash M : C$ by
      $(\Box\mathcal{I}_\textsf{GL})$, then $M \equiv \fixbox{y}{M'}$,
      $C \equiv \Box C'$, and \[
        \ctxt{\Delta, u : A, \Delta'}
             {\Delta^\bot, u^\bot : A, \Delta'^\bot, y^\bot : \Box C'}
          \vdash M'^\bot : C'
      \] Write $N_\ast \defeq N[\fixbox{z}{N}/z]$. 
      By the first premise and the IH, we have that \[
        \ctxt{\Delta, \Delta'}
             {\Delta^\bot, u^\bot : A, \Delta'^\bot, y^\bot : \Box C'}
            \vdash M'^\bot\left[N_\ast/u\right] : C'
      \] We now need to substitute for $u^\bot$.  By an application of
      $(\Box\mathcal{I}_\textsf{GL})$ to the first premise we have \[
        \ctxt{\Delta}{\Delta^\bot} \vdash \fixbox{z}{N} : \Box A
      \] and hence by Theorem \ref{thm:scut} we substitute this into
      the first premise itself to get \[
        \ctxt{\Delta}{\Delta^\bot}
          \vdash N^\bot[\fixbox{z}{N}/z^\bot] : A
      \] But $N_\ast^\bot \equiv N^\bot[\fixbox{z}{N}/z^\bot]$, so
      by weakening and Theorem \ref{thm:scut}, we obtain \[
        \ctxt{\Delta, \Delta'}
             {\Delta^\bot, \Delta'^\bot, y^\bot : \Box C}
            \vdash M'^\bot[N_\ast, N_\ast^\bot / u, u^\bot] : C'
      \] But by well-definedness of contexts, $u^\bot \not\in \fv{M}$,
      so by Theorem \ref{thm:bot} we have that $M'^\bot[N_\ast,
      N_\ast^\bot/u, u^\bot] \equiv \left(M'[N_\ast/u]\right)^\bot$, and
      hence by a use of $(\Box\mathcal{I}_\textsf{GL})$, we have \[
        \ctxt{\Delta, \Delta'}
        {\Gamma}
            \vdash \fixbox{y}{\left(M'[N_\ast/u]\right)} :
            \Box C' \equiv C
      \] and hence the result.

    \item (\textsf{DS4}) 
	
    \begin{itemize}
      \item If $\ctxt{\Delta, u : A,
      \Delta'}{\Gamma} \vdash M : C$ by
      $(\Box\mathcal{I}_\textsf{S4})$ then $M \equiv \ibox{M'}$
      and $C \equiv \Box C'$ with \[
        \ctxt{\Delta, u : A, \Delta'}{\cdot} \vdash M' : C
      \] The IH then yields $\ctxt{\Delta, \Delta'}{\cdot} \vdash
      M'[N/u] : C$, and a single use of
      $(\Box\mathcal{I}_\textsf{S4})$ yields the result.

      \item If $\ctxt{\Delta, u : A, \Delta'}{\Gamma} \vdash M : C$ by
      $(\Box\textsf{var})$ then $M \equiv v$ for some $v$ such
      that $(v : C) \in \Delta, u : A, \Delta'$. There are two
      cases: \begin{itemize}
        \item $u \equiv v$: then $M[N/u] \equiv N$ and $A \equiv
        C$. The premise $\ctxt{\Delta}{\cdot} \vdash N : A$ along
        with weakening for both contexts yields the result.
      
        \item $u \not\equiv v$: then $M[N/u] \equiv M$, and $u$
        does not occur in $M$. It is easy to show that if
        $\ctxt{\Delta, u : A, \Delta'}{\Gamma} \vdash M : C$ and
        $u \not\in \bfv{M}$ then $\ctxt{\Delta,
        \Delta'}{\Gamma} \vdash M : C$.
      \end{itemize}
    \end{itemize}

	\item (\textsf{DT})

	\begin{itemize}
	  \item If $\ctxt{\Delta, u : A, \Delta'}{\Gamma} \vdash M :
	  C$ by $(\Box\mathcal{I}_\textsf{K})$ then we proceed as
	  in the case of \textsf{DK}.

	  \item If $\ctxt{\Delta, u : A, \Delta'}{\Gamma} \vdash M : C$ by
	  $(\Box\textsf{var})$ then $M \equiv v$ for some $v$ such
	  that $v : C \in \Delta, u : A, \Delta'$. There are two
	  cases: \begin{itemize}
	    \item $u \equiv v$: then $M[N/u] \equiv N$ and $A \equiv
	    C$. The premise $\ctxt{\cdot}{\Delta} \vdash N : A$
	    along with Theorem \ref{thm:dereliction} yields
	    $\ctxt{\Delta}{\cdot} \vdash N : A$. A series of
	    weakenings for both contexts then yields the result.
    
	    \item $u \not\equiv v$: then $M[N/u] \equiv M$, and $u$
	    does not occur in $M$. It is easy to show that if
	    $\ctxt{\Delta, u : A, \Delta'}{\Gamma} \vdash M : C$ and
	    $u \not\in \bfv{M}$ then $\ctxt{\Delta,
	    \Delta'}{\Gamma} \vdash M : C$.
	  \end{itemize}
	\end{itemize}
    \end{enumerate}
\end{proof}

Finally, in the cases where the \textsf{T} axiom is present, we
may move variables from the intuitionstic to the modal context:

\begin{thm}[Modal Dereliction]
  \label{thm:dereliction}
  If $\mathcal{S} \in \{\textsf{DS4}, \textsf{DT}\}$, then 
  $
    \begin{prooftree}
        \ctxt{\Delta}{\Gamma, \Gamma'} \vdash_\mathcal{S} M : A
      \justifies
        \ctxt{\Delta, \Gamma}{\Gamma'} \vdash_\mathcal{S} M : A
    \end{prooftree}
  $ 
  is admissible.
\end{thm}

\begin{proof}
  By induction on the derivation of $\ctxt{\Delta}{\Gamma,
  \Gamma'} \vdash M : A$. Most cases are straightfoward, except
  $(\textsf{var})$ and
  $(\Box\mathcal{I}_\mathsf{S4})$/$(\Box\mathcal{I}_\mathsf{K})$.
  If the judgment holds by $(\textsf{var})$, then $M \equiv x$ for
  some $(x : A) \in \Gamma, \Gamma'$. If $(x : A) \in \Gamma$, we
  use $(\Box\textsf{var})$ to conclude that $\ctxt{\Delta,
  \Gamma}{\Gamma'} \vdash x : A$. If $(x : A) \in \Gamma'$, then
  use of $(\textsf{var})$.  If the judgment holds by
  $(\Box\mathcal{I}_\mathsf{S4})$ then $M \equiv \ibox{M'}$ and $A
  \equiv \Box A'$ for some $M', A'$ with $\ctxt{\Delta}{\cdot}
  \vdash M' : A'$. Repeated use of weakening for the modal context
  followed by an application of $(\Box\mathcal{I}_\mathsf{S4})$
  yields the result.  The case of $(\Box\mathcal{I}_\mathsf{K})$
  is similar, but uses weakening for the intuitionistic context.
\end{proof}

\subsection{Equivalence between Hilbert and dual systems}
  \label{sec:equivalence}

In this section we prove that our dual-context $\lambda$-calculi
correspond to the Hilbert systems given in \S\ref{sec:logics}.
This ties the knot with respect to the Curry-Howard
correspondence.

Modulo the appearance of proof terms, the translation under which
this equivalence is shown is the same one that we used in
\S\ref{sec:deriving}: \[
  \ctxt{\Delta}{\Gamma} \vdash_{\textsf{D}\mathcal{L}} M : A 
    \quad \rightsquigarrow \quad 
  \Box \hat{\Delta}, \hat{\Gamma} \vdash_{\mathcal{L}} A
\] We write $\hat{\Gamma}$ to mean the context $\Gamma$
with all the variables removed: if $\Gamma \equiv x_1 : A_1,
\dots, x_n : A_n$, then $\hat{\Gamma} \defeq A_1, \dots, A_n$.

One direction of the proof involves showing that the axioms are
indeed derivable in the dual-context systems. The other direction
involves showing the admissibility of the dual-context rules in
the Hilbert systems.

First and foremost, we need to show that axiom $(\mathsf{K})$ is
derivable. It is easy to check that the term \[
  \mathsf{ax_K} \defeq
    \lambda f : \Box (A \rightarrow B). \;
    \lambda x : \Box A . \;
      \letbox{g}{f}{\letbox{y}{x}{\ibox{(g\,y)}}}
\] has type $\Box(A \rightarrow B) \rightarrow \Box A \rightarrow
\Box B$ in all our systems other than \textsf{GL}. For
\textsf{GL}, we instead use \[
  \mathsf{ax^\textsf{DGL}_K} \defeq
    \lambda f : \Box (A \rightarrow B). \;
    \lambda x : \Box A . \;
      \letbox{g}{f}{\letbox{y}{x}{\fixbox{z}{(g\,y)}}}
\] It is also not hard to see that in \textsf{DK4} and \textsf{DS4}
the terms \[
  \mathsf{ax_{4}} \defeq
    \lambda x : \Box A. \;
      \letbox{y}{x}{\ibox{\left(\ibox{y}\right)}}
\] have type $\Box A \rightarrow \Box\Box A$, which is exactly
axiom \textsf{4}. 

In the case of \textsf{DGL}, we need to show that the term \[
  \mathsf{ax_{GL}} \defeq
    \lambda x : \Box (\Box A \rightarrow A). \;
      \letbox{f}{x}{\left(\fixbox{z}{\left(f\,z\right)}\right)}
\] has type $\Box(\Box A \rightarrow A) \rightarrow \Box A$. The
most interesting part of the derivation can be found in Figure
\ref{fig:ax_w}.

\begin{figure}[h]
  \centering
  \input{ax_w}
  \caption{Derivation of the G\"odel-L\"ob axiom in \textsf{DGL}}
  \label{fig:ax_w}
\end{figure}

Finally, in \textsf{DT} and \textsf{DS4}, the term \[
  \mathsf{ax_{T}} \defeq
    \lambda x : \Box A. \; \letbox{y}{x}{y}
\] has type $\Box A \rightarrow A$, i.e. inhabits axiom \textsf{T}.

With all that  we can show: 
\begin{thm}[Hilbert to Dual]
  If $\Gamma$ is a well-defined context and $\hat{\Gamma}
  \vdash_\mathcal{L} A$, then there exists a term $M$ such that
  $\ctxt{\cdot}{\Gamma} \vdash_{\textsf{D}\mathcal{L}} M : A$.
\end{thm}
\begin{proof} 
  By induction on the derivation of $\hat{\Gamma}
  \vdash_\mathcal{L} A$. In the case of the assumption rule, we
  use $(\textsf{var})$ to type the associated variable in
  $\hat{\Gamma}$. The cases for axioms of $(\textsf{IPL}_\Box)$
  are easy. For the modal axioms, we use the terms derived above.
  For modus ponens  we use application.

  This leaves the case of necessitation. Suppose $\hat{\Gamma}
  \vdash_\mathcal{L} A$ by it; then $A \equiv \Box A'$, and
  $\vdash_\mathcal{L} A'$. By the IH, there is a term $M'$ such
  that $\ctxt{\cdot}{\cdot} \vdash_{\textsf{D}\mathcal{L}} M' :
  A'$. We then use the appropriate introduction rule for
  box---e.g. $(\Box\mathcal{I}_\mathsf{K})$, and so on---to obtain
  $\ctxt{\cdot}{\Gamma} \vdash_{\textsf{D}\mathcal{L}} \ibox{M'} :
  \Box A'$.
\end{proof}

The essence of the opposite direction lies in showing that the
rules of the dual-context calculus are admissible in the
corresponding Hilbert system---that is, after erasing the proof
terms. We have done most of the required work in \S
\ref{sec:admissible}.

\begin{thm}[Dual to Hilbert]
  If $\ctxt{\Delta}{\Gamma} \vdash_{\textsf{D}\mathcal{L}} M : A$
  then $\Box\hat{\Delta}, \hat{\Gamma} \vdash_\mathcal{L} A$.
\end{thm}
\begin{proof} By induction on the derivation of
$\ctxt{\Delta}{\Gamma} \vdash_{\textsf{D}\mathcal{L}} M : A$.

If the premise holds by $(\textsf{var})$, then we use the
assumption rule of the Hilbert system. If the last step in the
derivation of the premise is the rule $(\rightarrow\mathcal{I})$,
we use the IH followed by the Deduction Theorem (Theorem
\ref{thm:deduction}). If the last step is by
$(\rightarrow\mathcal{E})$, we use modus ponens. It is simple to
translate the rules that pertain to the product, namely
$(\times\mathcal{I})$ and $(\times\mathcal{E}_i)$ to uses of the
\textsf{IPL} axioms pertaining to the product along with modus
ponens. It is also not hard to see that, under the given
translation, $(\Box\mathcal{E})$ can also be matched by a use of
the IH along with an invocation of the admissibility of cut for
Hilbert systems (Theorem \ref{thm:hilbscut}). Uses of the modal
variable rule $(\Box\mathsf{var})$ can be imitated by a use of the
assumption rule, modus ponens, and an instance of the \textsf{T}
axiom.

This leaves the introduction rules for the box. The rule
$(\Box\mathcal{I}_\mathsf{K})$ is matched with Scott's rule
(Theorem \ref{thm:scott}). The rule
$(\Box\mathcal{I}_\mathsf{K4})$ is matched with the Four rule
(Theorem \ref{thm:four}). The rule $(\Box\mathcal{I}_\mathsf{GL})$
is matched with the generalized L\"ob rule (Theorem
\ref{thm:lob}). Finally, the rule $(\Box\mathcal{I}_\mathsf{S4})$
is matched with the corollary to the Four rule (Corollary
\ref{thm:four_cor}).
\end{proof}

\section{Reduction}
  \label{sec:reduction}

We will now show that it is possible to eliminate cuts from proofs
in our systems. This elimination of cuts will be complete, in the
sense that normal proofs will satisfy the \emph{subformula
property}, i.e. they will not reference any external logical
formul\ae~that are unrelated to their assumptions or conclusion.
We will achieve this with the traditional technique of a confluent
and strongly normalizing small-step reduction. Rather strikingly,
this reduction will be the same across all systems---with the
exception of \textsf{DGL}, whose term former for the introduction
of the modality has a very different shape. This avoids repeating
work to deal with different systems, as most of our proofs are by
induction on the typing judgments, and most rules are shared
between all systems. We discuss \textsf{GL} separately in
\S\ref{sec:case-gl}.

In this paper we stop short of deciding equality of proofs, which
is a much more challenging problem. There are many reasons that
make it so. The first one is related to the known problematic
behaviour of $\eta$-contraction. The second arises from the fact
that our reduction does not eliminate all redundacy from our
proofs.  For example, the terms
\begin{align*}
  \ctxt{\cdot}{x : \Box A, y : B} &\vdash y : B &
  \ctxt{\cdot}{x : \Box A, y : B} &\vdash \letbox{u}{x}{y} : B
\end{align*} 
will be equal in the equational theory of \S\ref{sec:eqth}, but
will also be normal forms with respect to reduction. In a sense,
there have to be additional commuting conversions that---amongst
other things---`garbage collect' unnecessary eliminations, and
which we only discover when we consider the categorical semantics
\S\ref{sec:catsem}. The third problem is deeper: it arises because
$\Box$ essentially behaves as \emph{positive connective}. Deciding
equality in the presence of such connectives requires either
advanced rewriting or categorical techniques. For example, the
case of $\beta\eta$-equivalence in the presence of sums and an
empty type was open until Scherer resolved it in 2017
\cite{Scherer2017}; see \emph{op.  cit.} for an extensive
bibliography.

Normalization of proofs for this kind of system has not been
extensively studied before. Pfenning and Davies \cite{Davies2001}
hint at our notion, and use a strict subset of it as operational
semantics \cite{Davies2001a}. A similar notion was studied in the
context of \textsf{DILL} by Ohta and Hasegawa \cite{Ohta2006},
including $\eta$-contractions and the full set of commuting
conversions.

\subsection{The reduction}

\begin{figure}
  \centering
  \renewcommand{\arraystretch}{2.5}

\begin{small}
\begin{tabular}{cccc}


  $
  \begin{prooftree}
      \justifies
    (\lambda x{:}A.\ M)N \red{} M[N/x]
  \end{prooftree}
  $

  & 

  $
  \begin{prooftree}
    \justifies
      \pi_i(\langle M_1, M_2 \rangle) \red{} M_i
  \end{prooftree}
  $

  &


  $
  \begin{prooftree}
      M \red{} N
    \justifies
      \pi_i(M) \red{} \pi_i(N)
  \end{prooftree}
  $

  &

  $
  \begin{prooftree}
      M_i \red{} N_i
    \quad
      M_{1-i} \equiv N_{1-i}
    \justifies
      \langle M_0, M_1 \rangle \red{} \langle N_0, N_1 \rangle
  \end{prooftree}
  $

  \\


  $
  \begin{prooftree}
      M \red{} N
    \justifies
      \lambda x{:}A.\ M \red{} \lambda x{:}A.\ N
  \end{prooftree}
  $

  &

  $
  \begin{prooftree}
      M \red{} N
    \justifies
      \ibox{M} \red{} \ibox{N}
  \end{prooftree}
  $

  &


  $
  \begin{prooftree}
      M \red{} N
    \justifies
      MP \red{} NP
  \end{prooftree}
  $

  &

  $ 
  \begin{prooftree}
      P \red{} Q
    \justifies
      MP \red{} MQ
  \end{prooftree}
  $
      
  \\


  \multicolumn{2}{c}{
  $
  \begin{prooftree}
      M \red{} N
    \justifies
      \letbox{u}{M}{P} \red{} \letbox{u}{N}{P}
  \end{prooftree}
  $
  }

  &

  \multicolumn{2}{c}{
  $ 
    \begin{prooftree}
        P \red{} Q
      \justifies
        \letbox{u}{M}{P} \red{} \letbox{u}{M}{Q}
    \end{prooftree}
  $
  }

  \\

  
  \multicolumn{4}{c}{
  $
  \begin{prooftree}
    \justifies
      \pi_i\left(\letbox{u}{M}{N}\right) 
        \red{}
      \letbox{u}{M}{\pi_i(N)}
  \end{prooftree}
  $
  }

  \\

  \multicolumn{4}{c}{
  $
  \begin{prooftree}
    \justifies
      \left(\letbox{u}{M}{P}\right) Q
        \red{}
      \letbox{u}{M}{PQ}
  \end{prooftree}
  $
  }

  \\

  \multicolumn{4}{c}{
  $
  \begin{prooftree}
    \justifies
      \letbox{v}{(\letbox{u}{M}{N})}{P}
        \red{}
      \letbox{u}{M}{\letbox{v}{N}{P}}
  \end{prooftree}
  $
  }
\end{tabular}
\end{small}


  \caption{Reduction}
  \label{fig:reduction}
\end{figure}

The notion of reduction $\red{}$ is defined as the least relation
satisfying the rules of Figure \ref{fig:reduction}. This includes
the usual $\beta$-reduction, plus the modal $\beta$-reduction
\[
  \letbox{u}{\ibox{M}}{N} \red{} N[M/u]
\] 
which is suggested by Theorem \ref{thm:mcut}. It also includes
congruences---so that reductions can happen anywhere in a
term---and, finally, three \emph{commuting conversions}, which are
required for the subformula property to hold. 

We begin with the following lemma, which shall also prove useful
in \S\ref{sec:sn}. Recall the definition and discussion of
complementary variables from \S\ref{sec:complementary}.

\begin{lem}[Complement reduction]
  \label{lem:redcomp}
  If $\ctxt{\Delta}{\Delta^\bot} \vdash_{\mathsf{DK4}} M^\bot : A$
  then $M \red N$ implies $M^\bot \red N^\bot$.
\end{lem}

\begin{proof}
  By induction on $M \red N$. We only prove the cases that are not
  straightforward.
  \begin{indproof}
    \case{$\beta$ for $\rightarrow$} Suppose
    $\ctxt{\Delta}{\Delta^\bot, x : B} \vdash M : C$ for some $B,
    C$. Then 
      \[
        \left((\lambda x : B.\ M)N\right)^\bot
          \equiv
        (\lambda x^\bot  : B.\ M^\bot)N^\bot
          \red{}
        M^\bot[N^\bot/x^\bot] 
      \] 
      We want to show that the latter is just
      $\left(M[N/x]\right)^\bot$. We can rename $x$ so that it
      does not clash with any variable nor its complement in the
      context of $M$. We will show that (a) $x^\bot \not\in
      \fv{M}$, and that (b) $x \not\in \fv{M^\bot}$. By (a) we may
      apply Theorem \ref{thm:bot} and then use (b) to infer that
      \[
        \left(M[N/x]\right)^\bot
          \equiv
        M^\bot[N, N^\bot/x, x^\bot]
          \equiv
        M^\bot[N^\bot/x^\bot]
      \] 
      Both (a) and (b) follow from Theorem \ref{thm:freevar}. For
      (a): by (3) it suffices to show that $x^\bot \not\in
      \ufv{M}$ and $x^\bot \not\in \bfv{M}$. But $x^\bot$ does not
      occur in either context, so we use (6). For (b): we know
      that $x^\bot \not\in \ufv{M}$, so this implies by (4) that
      $x \not\in \ufv{M^\bot}$. Hence, it suffices by (3) to also
      show that $x \not\in \bfv{M^\bot}$.  But by (5) the latter
      is equal to $\bfv{M}$, and by well-formedness of contexts we
      know that know that $x \not\in \vars{\Delta}$, so by (6) it
      is not in $\bfv{M}$ either.

    \case{$\beta$ for $\ibox{(-)}$}
      It is easy to see that
      \begin{align*}
        \left(\letbox{u}{\ibox{M}}{N}\right)^\bot
          &\equiv
        \letbox{u}{(\ibox{M})^\bot}{N^\bot} \\
          &\equiv
        \letbox{u}{\ibox{M}}{N^\bot} \\
          &\red{}
        N^\bot[M/u]
      \end{align*}
      It now suffices to show that (a) $u^\bot \not\in \fv{N}$,
      and that (b) $u^\bot \not\in \fv{N^\bot}$. For then Theorem
      \ref{thm:bot} applies, and $(N[M/u])^\bot \equiv N^\bot[M,
      M^\bot/u, u^\bot] \equiv N^\bot[M/u]$. We will use Theorem
      \ref{thm:freevar} again, recalling that $\ctxt{\Delta, u :
      A}{\Delta^\bot} \vdash N^\bot : A$. 

      For (a): by (3) it suffices to show that $u^\bot \not\in
      \ufv{N}$ and $u^\bot \not\in \bfv{N}$. By (4) and (5) it
      suffices to show that $u \not\in \ufv{N^\bot}$ and $u^\bot
      \not\in \bfv{N^\bot}$. Both follow by (6), as neither $u$
      nor $u^\bot$ are allowed to occur anywhere in $\Delta$ and
      $\Delta^\bot$.

      For (b): we know that $u^\bot \not\in \bfv{N^\bot}$, so by
      (3) it suffices to show that $u^\bot \not\in \ufv{N^\bot}$.
      But we can use (6): $u$ cannot be in $\Delta$, so $u^\bot$
      cannot be in $\Delta^\bot$.
  \end{indproof}
\end{proof}

We then show that

\begin{thm}[Subject reduction]
  If $\ctxt{\Delta}{\Gamma} \vdash M : A$ and $M \red N$, then
  $\ctxt{\Delta}{\Gamma} \vdash N : A$.
\end{thm}
\begin{proof}
  By induction on $M \red N$. Most cases follow straightforwardly
  from the IH. The cases for the $\beta$ rules follow from
  Theorems \ref{thm:scut} and \ref{thm:mcut}.
\end{proof}

\subsection{Subformula property}
  \label{sec:subformula}

A calculus satisfies the subformula property when any
\emph{normal} proof (i.e. one that has no reducts) of a formula
$A$ from assumptions $\Gamma$ only involves formul\ae~that are
either (a) subformul\ae~of the conclusion of $A$, or (b)
subformul\ae~of some premise in $\Gamma$. This is tantamount to
saying that the proof has a very specific structure: it proceeds
by eliminating logical symbols of assumptions in $\Gamma$, and
then uses the results to construct a proof of $A$ using only
introduction rules. In short, \emph{the proof has no detours}, and
proceeds as quickly as possible from assumptions to conclusion:
see \cite{Prawitz1965} and \cite{Girard1989}.

Without the commuting conversions of Figure \ref{fig:reduction},
our systems do not satisfy the subformula property. The reason is
the presence of the elimination rule
\[
  \begin{prooftree}
      \ctxt{\Delta}{\Gamma} \vdash M : \Box A
    \quad\quad
      \ctxt{\Delta, u : A}{\Gamma} \vdash N : C
    \justifies
      \ctxt{\Delta}{\Gamma} \vdash \letbox{u}{M}{N} : C
    \using
      {(\Box\mathcal{E})}
  \end{prooftree}
\] Notice that the conclusion $C$ is given to us by the
\emph{minor premise} $\ctxt{\Delta, u : A}{\Gamma} \vdash N : C$,
and it is structurally unrelated to $\Box A$, the \emph{major
premise} that is being eliminated: in Girard's terminology, it is
\emph{parasitic}. This is so because---as we discussed in
\S\ref{sec:deriving}---the elimination rule is secretly a kind of
\emph{cut rule}, or a rule in the style of Schroeder-Heister
\cite{Schroeder-Heister1984}.

It is not so easy to see where the actual trouble lies at first.
The point is that the $\letbox{u}{(-)}{(-)}$ construct may `hide
redexes.' Once we introduce the extra reductions that are needed,
and prove the subformula property, this will become quite clear.
But---in the meantime---let us consider three examples.

Suppose that $\ctxt{\Delta, u : A}{\Gamma} \vdash \langle N_1, N_2
\rangle : A_1 \times A_2$ and $\ctxt{\Delta}{\Gamma} \vdash M :
\Box A$ are normal forms, and that the latter is not of the form
$\ibox{(-)}$. We may use $(\Box\mathcal{E})$ to obtain
\[
  \ctxt{\Delta}{\Gamma} \vdash
    \letbox{u}{M}{\langle N_1, N_2 \rangle} : A_1 \times A_2
\] 
This is indeed---and should be!---a normal form. But what if we
just want to prove $A_1$? We may apply one of the elimination
rules for products to get \[
  \ctxt{\Delta}{\Gamma} \vdash
    \pi_1\left(\letbox{u}{M}{\langle N_1, N_2 \rangle}\right) 
      : A_1
\] 
This is now a proof of $A_1$, but it surreptitiously contains a
proof $N_2$ of $A_2$, which is entirely unrelated to $A_1$
(neither need be a subexpression of the other). The problem is
that the eliminator $\letbox{u}{(-)}{(-)}$ obstructs the meeting
of the destructor $\pi_1(-)$ with the constructor $\langle N_1,
N_2 \rangle$. The solution is to allow a commuting conversion that
allows the two to meet by pulling the $\mathsf{let}$ construct
outside:
\[
  \pi_1\left(\letbox{u}{M}{\langle N_1, N_2 \rangle}\right) 
    \red{}
  \letbox{u}{M}{\pi_1(\langle N_1, N_2 \rangle)}
\] A similar situation occurs when $\ctxt{\Delta, u : A}{\Gamma}
\vdash \lambda x : B.\ P : B \rightarrow C$. We may form \[
  \ctxt{\Delta}{\Gamma} \vdash
    \letbox{u}{M}{\lambda x : B.\ P} : B \rightarrow C
\] which is a perfectly reasonable normal form. But if
$\ctxt{\Delta}{\Gamma} \vdash Q : B$ then \[
  \ctxt{\Delta}{\Gamma} \vdash
    \left(\letbox{u}{M}{\lambda x : B.\ P}\right) Q : C
\] is not: we should be able to reduce \[
  \left(\letbox{u}{M}{\lambda x : B.\ P}\right) Q
    \red{}
  \letbox{u}{M}{(\lambda x : B.\ P)Q}
\] Finally, there is third, less visible case of this phenomenon.
If we understand $(\Box\mathcal{E})$ to be a `bad' elimination, we
have considered the cases of `good' elimination ($\pi_i(-)$,
application) following `bad' elimination.  The final case is that
of `bad' elimination following another `bad' elimination. To give
an example, let us consider an elimination after a $\ibox{(-)}$
introduction: 
\[
  \ctxt{\Delta}{\Gamma} \vdash
    \letbox{u}{M}{\ibox{N}} : \Box A
\] 
We can then plug this into a term $\ctxt{\Delta, v : A}{\Gamma}
\vdash P : C$ by eliminating the box: \[
  \ctxt{\Delta}{\Gamma} \vdash
    \letbox{v}{(\letbox{u}{M}{\ibox{N}})}{P} : C
\] 
Now things are clear: the second let-construct is obstructing the
meeting of the first let-construct with the introduction form
$\ibox{N}$. We need to convert
\[
  \letbox{v}{(\letbox{u}{M}{\ibox{N}})}{P}
    \red{}
  \letbox{u}{M}{\letbox{v}{\ibox{N}}{P}}
\] 
whilst taking care to not confuse our bound variables.

With these commuting conversions in place we can now prove the
property. We only need to slightly strengthen the induction
hypothesis, using the notion of \emph{principal branch}.

\begin{thm}[Subformula Property]
  Let $\ctxt{\Delta}{\Gamma} \vdash M : A$, and suppose $M$ is a
  normal form. 
  \begin{enumerate}
    \item
      Every type occuring in the derivation of
      $\ctxt{\Delta}{\Gamma} \vdash M : A$ is either a
      subexpression of the type $A$, or a subexpression of a type
      in $\Delta$ or $\Gamma$.
    \item
      If $M$ is an elimination form that is \emph{not} form
      $\letbox{u}{P}{Q}$---i.e. if it is a projection $\pi_i(N)$
      or an application $PQ$---then it entirely consists of a
      sequence of eliminations: that is, there is a sequence of
      types $A_0, \dots, A_n$ for which
      \begin{itemize}
      \item[-]
        $A_0$ occurs in either $\Delta$ or $\Gamma$,
      \item[-]
        $A_n$ is $A$, and
      \item[-]
        $A_i$ is the \emph{major premise} of an elimination whose
        conclusion is $A_{i+1}$ for $0 \leq i < n$. 
      \end{itemize} 
      This sequence is called a \emph{principal branch}. In
      particular, $A_n$ is a subexpression of $A_0$.
  \end{enumerate}
\end{thm}

\begin{proof}
  By induction on the derivation of $\ctxt{\Delta}{\Gamma} \vdash
  M : A$. We omit some cases, as they are similar to others: the
  case of $(\Box\textsf{var})$ is like that of an ordinary
  variable, the other introduction forms ($\langle M, N \rangle$,
  $\ibox{M}$) are similar to the case for $\lambda$, and the case
  for $\pi_i(M)$ is similar to the one for $MN$.
  \begin{indproof}
    \case{$x$}
      Then $\ctxt{\Delta}{\Gamma} \vdash x : A$ and hence $(x : A)
      \in \Gamma$. This is the complete derivation, and satisfies
      both desiderata.

    \case{$\lambda x : A.\ M$}
      Then the immediate premise is of the form
      $\ctxt{\Delta}{\Gamma, x : A} \vdash M : B$. By the IH, all
      types that occur in that are either subexpressions of
      types in $\Delta$ or $\Gamma$, subexpressions of $A$, or
      subexpressions of $B$. All of these are subexpressions of
      types in $\Delta$, $\Gamma$, or $A \rightarrow B$.

    \case{$MN$}
      Then the major premise is $\ctxt{\Delta}{\Gamma} \vdash M :
      B \rightarrow A$ and the minor premise is
      $\ctxt{\Delta}{\Gamma} \vdash N : B$ for some type $B$.

      Look at $M$: it cannot be a $\lambda$-abstraction, for that
      would make $MN$ a redex.  It also cannot be any other
      introduction rule, for it introduce some other type (e.g. $A
      \times B$ or $\Box A$). Hence, it must be an elimination. Of
      the eliminations, it cannot be a let-expression, for a
      commuting conversion would make that a redex too.

      It follows that $M$ is a `good' elimination: either
      $\pi_i(-)$ or $PQ$. We can thus apply (2) from the IH thesis
      to conclude that there is a principal branch beginning with
      an assumption in $\Delta$ or $\Gamma$, and ending with $B
      \rightarrow A$. We can extend that to a principal branch for
      $M$, ending with $A$. This proves (2), and furthermore
      implies that $B \rightarrow A$ is a subexpression of some
      premise in either $\Delta$ or $\Gamma$.

      Over to (1): applying the IH to the major premise we know
      that every type that occurs in the derivation of
      $\ctxt{\Delta}{\Gamma} \vdash M : B \rightarrow A$ is either
      a subexpression of a type in $\Delta$ or $\Gamma$, or a
      subexpression of $B \rightarrow A$. But we have already
      deduced that $B \rightarrow A$ is a subexpression of some
      premise in either $\Delta$ or $\Gamma$, so that all types
      occuring in the derivation of the major premise satisfy the
      desideratum. 

      Applying the IH to the minor premise, every type that occurs
      in the derivation of $\ctxt{\Delta}{\Gamma} \vdash N : B$ is
      either a subexpression of some type in $\Delta$ or $\Gamma$,
      or a subexpression of $B$. But $B$ is a subexpression of $B
      \rightarrow A$, which in turn is a subexpression of
      a premise in one of the contexts. Hence all types occurring
      in that branch also occur in either $\Delta$ or $\Gamma$.
      This concludes the proof of this case, for we have examined
      all types appearing in the derivation.

    \case{$\letbox{u}{M}{N}$}
      The major premise is then $\ctxt{\Delta}{\Gamma} \vdash M :
      \Box B$ and the minor premise is $\ctxt{\Delta,
      u : B}{\Gamma} \vdash N : A$ for some $B$. (2) does not
      apply to let-constructs, so we only need to show (1).

      Consider $M$. It cannot be a $\ibox{(-)}$, for that would
      make the entire term a redex. It also cannot be any other
      introduction form, because it would introduce a type of a
      different shape. Hence, it must hence be an elimination
      form, but not another let-construct, for that would be a
      redex too due to our commuting conversion. Therefore it must
      be either of the form $\pi_i(M')$ or of the form $PQ$. It
      follows that (2) of the IH applies: there is a principal
      branch beginning with a premise and ending with $\Box B$. In
      particular, $\Box B$ is a subexpression of some type in
      $\Delta$ or $\Gamma$.
      
      By the IH, any type that occurs in the derivation of the
      major premise is either a subexpression of a type in
      $\Delta$ or $\Gamma$, or a subexpression of $\Box B$. But
      $\Box B$ is a subexpression of some type in one of those two
      contexts, so every type that occurs in the derivation of the
      major premise is actually a subexpression of a type in
      $\Delta$ or $\Gamma$.

      As for the minor premise, any type that occurs in it is
      either a subexpression of a type in $\Delta$ or $\Gamma$, or
      a subexpression of the types $B$ or $A$. But $B$ is a
      subexpression of $\Box B$, which by our previous reasoning
      is in turn a subexpression of some type in either $\Delta$
      or $\Gamma$. Thus all types occuring in it are either
      subexpressions of some type in $\Delta$ or $\Gamma$, or
      subexpressions of $A$. This concludes the proof.
      \qedhere
  \end{indproof}
\end{proof}

\subsection{Confluence}
  \label{sec:conf}

We will prove that

\begin{thm}
  \label{thm:conf}
  The reduction relation $\red{}$ is confluent.
\end{thm}

One can show this result in many ways. We will use the method of
\emph{parallel reduction}, which was discovered by Tait and
Martin-L\"of. The history of the method and a few variations of it
are discussed by Takahashi \cite{Takahashi1995}. The idea is
simple: we introduce a second notion of reduction, $\redp$, which
we will `sandwich' between reduction proper and its transitive
closure, so that $\red{} \subseteq\ \redp\ \subseteq\ \redt$. We
will then show that $\redp$ has the diamond property. By the
above inclusions, the transitive closure $\redp^\ast$ of $\redp$
is then equal to $\redt$, and hence $\red$ is confluent. In fact,
we will follow \cite{Takahashi1995} in doing something better: we
will define for each term $M$ its \emph{complete development}
$M^\star$. The complete development is intuitively defined by
`unrolling' all the redexes of $M$ at once.  We will then show
that if $M \redp N$, then $N \redp M^\star$.  $M^\star$ will then
suffice to close the diamond:
\[
  \begin{tikzcd}[row sep=small, column sep=small]
    & M
	\arrow[dr, Rightarrow]
	\arrow[dl, Rightarrow]
    &  \\
    P
      \arrow[dr, Rightarrow, dotted]
    & 
    & Q 
      \arrow[dl, Rightarrow, dotted] \\
    & M^\star
  \end{tikzcd}
\]

\begin{figure}
  \centering
  \renewcommand{\arraystretch}{2.7}

\begin{small}
\begin{tabular}{cccc}


  $
  \begin{prooftree}
    \justifies
      x \redp x
  \end{prooftree}
  $

  &

  $
  \begin{prooftree}
      M \redp N
    \quad
      P \redp Q
    \justifies
      (\lambda x : A.\ M)P \redp N[Q/x]
  \end{prooftree}
  $

  & 

  $
  \begin{prooftree}
      M_i \redp N
    \justifies
      \pi_i(\langle M_1, M_2 \rangle) \redp N
  \end{prooftree}
  $

  &

  $ 
  \begin{prooftree}
      M \redp N
    \justifies
      \pi_i(M) \redp \pi_i(N)
  \end{prooftree}
  $

  \\

  $
  \begin{prooftree}
      M_1 \redp N_1
    \quad
      M_2 \redp N_2
    \justifies
      \langle M_1, M_2 \rangle \redp \langle N_1, N_2 \rangle
  \end{prooftree}
  $

  &


  $
  \begin{prooftree}
      M \redp N
    \justifies
      \lambda x : A.\ M \redp \lambda x : A.\ N
  \end{prooftree}
  $

  &

  $
  \begin{prooftree}
      M \redp N
    \justifies
      \ibox{M} \redp \ibox{N}
  \end{prooftree}
  $

  &


  $ 
  \begin{prooftree}
      M \redp N
    \quad
      P \redp Q
    \justifies
      MP \redp NQ
	\end{prooftree}
  $

  \\

  \multicolumn{2}{c}{
  $
  \begin{prooftree}
      M \redp N
	  \quad
      P \redp Q
    \justifies
      \letbox{u}{M}{P} \redp \letbox{u}{N}{Q}
  \end{prooftree}
  $
  }

  &

  \multicolumn{2}{c}{
  $
  \begin{prooftree}
      M \redp N
    \quad
      P \redp Q
    \justifies
      \letbox{u}{\ibox{P}}{M} \redp N[Q/u]
  \end{prooftree}
  $
  }

  \\

  \multicolumn{4}{c}{
  $
  \begin{prooftree}
      M \redp N
    \quad
      P \redp Q
    \justifies
      \pi_i\left(\letbox{u}{M}{P}\right) \redp \letbox{u}{N}{\pi_i(Q)}
  \end{prooftree}
  $
  }

  \\

  \multicolumn{4}{c}{
  $
  \begin{prooftree}
      P \redp Q
    \quad
      M \redp N
    \quad
      S \redp T
    \justifies
      \left(\letbox{u}{P}{M}\right)S \redp \letbox{u}{Q}{NT}
  \end{prooftree}
  $
  }

  \\

  \multicolumn{4}{c}{
  $
  \begin{prooftree}
      P \redp Q
    \quad
      M \redp N
    \quad
      S \redp T
    \justifies
      \left(\letbox{u}{(\letbox{v}{P}{M})}{S}\right)
        \redp
      \letbox{v}{Q}{\letbox{u}{N}{T}}
  \end{prooftree}
  $
  }

\end{tabular}
\end{small}


  \caption{Parallel Reduction}
  \label{fig:parallel}
\end{figure}

The parallel reduction $\redp$ is defined in Figure
\ref{fig:parallel}. It is immediate that
\begin{lem}
  \label{lem:redprefl}
  $\redp$ is reflexive.
\end{lem}

\begin{defi}[Complete development]
  The \emph{complete development} $M^\star$ of a term $M$ is
  defined by the following clauses:
  \begin{small}
  \begin{align*}
    x^\star &\defeq x &
    (\langle M, N \rangle)^\star
      &\defeq \langle M^\star, N^\star \rangle \\
    \left(\pi_i(\langle M_1, M_2 \rangle)\right)^\star
      &\defeq M_i^\star &
    \left(\pi_i(\letbox{u}{M}{N})\right)^\star
      &\defeq \letbox{u}{M^\star}{\pi_i(N^\star)} \\
    \left(\left(\lambda x : A.\ M\right)N\right)^\star
      &\defeq M^\star[N^\star/x] & 
    \left(\left(\letbox{u}{P}{M}\right)N\right)^\star
      &\defeq \letbox{u}{P}{M^\star N^\star} \\
    \left(\pi_i(M)\right)^\star
      &\defeq \pi_i(M^\star) &
    \left(\lambda x : A.\ M\right)^\star
      &\defeq \lambda x : A.\ M^\star \\
    \left(\ibox{M}\right)^\star
      &\defeq \ibox{M^\star} &
    \left(\letbox{u}{\ibox{M}}{N}\right)^\star
      &\defeq N^\star[M^\star/u] \\
    \left(MN\right)^\star
      &\defeq M^\star N^\star &
    \left(\letbox{u}{M}{N}\right)^\star
      &\defeq \letbox{u}{M^\star}{N^\star}
  \end{align*} \[
    \left(\letbox{u}{(\letbox{v}{P}{M})}{N}\right)^\star
      \defeq \letbox{v}{P^\star}{\letbox{u}{M^\star}{N^\star}}
  \]
  \end{small}
\end{defi}

First, a little lemma capturing the essence of parallel reduction:
\begin{lem} 
  \label{lem:redp}
  If $M \redp N$ and $P \redp Q$, then $M[P/x] \redp N[Q/x]$.
\end{lem}

\begin{proof}
  Straightforward induction on $M \redp N$. 
\end{proof}

And here is the main result:

\begin{thm}
  If $M \redp P$, then $P \redp M^\star$.
\end{thm}

\begin{proof}
  By induction on $M \redp P$. The case for variables is trivial,
  the case for the congruence rules follows from the IH, and
  $\beta$ for function types is as usual. We show the rest.
  \begin{indproof}
    \case{$\beta$ for $\times$}
      Then we have $\pi_i(\langle M_1, M_2 \rangle) \redp M'_i$,
      with $M_i \redp M'_i$. By the IH, $M'_i \redp M^\star_i
      \equiv \left(\pi_i(\langle M_1, M_2 \rangle)\right)^\star$.

    \case{$\beta$ for $\Box$}
      Then we have $\letbox{u}{\ibox{M}}{N} \redp N'[M'/u]$ where
      $M \redp M'$ and $N \redp N'$. By the IH, $M' \redp M^\star$
      and $N' \redp N^\star$. It follows that $ N'[M'/u] \redp
      N^\star[M^\star/u] \equiv \left(
      \letbox{u}{\ibox{M}}{N}\right)^\star $ by Lemma
      \ref{lem:redp}.

    \case{comm. conv. for $\times$}
      Then we have $\pi_i(\letbox{u}{M}{P}) \redp
      \letbox{u}{N}{\pi_i(Q)}$ where $M \redp N$ and $P \redp Q$.
      By the IH, $N \redp M^\star$ and $Q \redp P^\star$, whence
      $\letbox{u}{N}{\pi_i(Q)} \redp
      \letbox{u}{M^\star}{\pi_i(P^\star)}$, which is
      $\alpha$-equivalent to
      $\left(\pi_i(\letbox{u}{M}{P})\right)^\star$.
  \end{indproof} 
  The cases for the other commuting conversions are
  similar.
\end{proof}

\subsection{Strong normalization}
  \label{sec:sn}

In this section, we will prove that

\begin{thm}
  \label{thm:sn}
  The reduction relation $\red{}$ is strongly normalizing.
\end{thm}

There is a very orderly way of doing so for all systems, save for
\textsf{GL}. The idea is to \emph{embed} the modal proofs in the
simply-typed $\lambda$-calculus, for which strong normalization is
a known result \cite{Prawitz1971, deGroote2002}.  This strategy is
used by \cite{Martini1996, Tsukada2010}, and hinted at for the
dual-context \textsf{S4} system by \cite{Davies2001a}.  Because of
the binding structure of $\letbox{u}{(-)}{(-)}$, one cannot do so
by simply erasing the modalities: we would then map modal
$\beta$-reductions to syntactic equality in the simply-typed
$\lambda$-calculus, which would not provide enough leverage to
lift strong normalization to the modal calculi.

Instead, we use a strategy inspired by the proof of strong
normalization for Moggi's monadic metalanguage \cite{Benton1998}:
we will interpret $\Box$ as the \emph{product-with-the-unit
comonad}.\footnote{\cite{Benton1998} translate the monadic type
$TA$ to the exception monad $\mathbf{1} + (-)$. They simulate the
commuting conversions for `bind' using the commuting conversions
for coproducts, which leads to a more direct proof.}
More specifically, we define a translation $(-)^\maltese$ of modal
types to simple types: 
\begin{align*}
  (p_i)^\maltese &\defeq p_i &
  (A \times B)^\maltese &\defeq A^\maltese \times B^\maltese \\
  (A \rightarrow B)^\maltese &\defeq A^\maltese \rightarrow B^\maltese &
  (\Box A)^\maltese &\defeq \mathbf{1} \times A^\maltese
\end{align*} Next, we extend $(-)^\maltese$ to terms:
\begin{gather*}
  \begin{align*}
    (x)^\maltese &\defeq x &
    \left(\langle M, N \rangle\right)^\maltese 
      &\defeq \langle M^\maltese, N^\maltese \rangle \\
    \left(\pi_i(M)\right)^\maltese &\defeq \pi_i(M^\maltese) &
    (\lambda x : A.\ M)^\maltese &\defeq \lambda x : A^\maltese.\ M^\maltese \\
    (MN)^\maltese &\defeq M^\maltese N^\maltese &
    (\ibox{M})^\maltese &\defeq \langle \ast, M^\maltese\rangle
  \end{align*} \\
  (\letbox{u}{M}{N})^\maltese \defeq
    N^\maltese[\pi_2(M^\maltese)/u]
\end{gather*} 
where $\ast : \mathbf{1}$ is the introduction form for the unit
type. We can then show that 
\begin{thm}[Simulation] \hfill
  \label{thm:sim}
  \begin{enumerate}
    \item
      $\left(M[N/x]\right)^\maltese \equiv M^\maltese[N^\maltese/x]$
    \item
      If $\ctxt{\Delta}{\Gamma} \vdash_{\textsf{D}\mathcal{L}} M :
      A$ then $\Delta^\maltese, \Gamma^\maltese \vdash M^\maltese
      : A^\maltese$ in the simply-typed $\lambda$-calculus.
    \item
      If $M \red N$ then $M^\maltese \red_{\beta}^\ast N^\maltese$
      in the simply-typed $\lambda$-calculus.
  \end{enumerate}
\end{thm}
\begin{proof} \hfill
  \begin{enumerate}
    \item By induction on $M$. 

    \item By induction on $\ctxt{\Delta}{\Gamma} \vdash M : A$. We
    show the two most difficult cases, namely the elimination
    rule, and that of $\textsf{K4}$.
    \begin{indproof}

      \case{$\ctxt{\Delta}{\Gamma} \vdash \letbox{u}{P}{Q} : C$}

      By the IH, the premises imply that $\Delta^\maltese,
      \Gamma^\maltese \vdash P^\maltese : \mathbf{1} \times
      A^\maltese$, and $\Delta^\maltese, u : A^\maltese,
      \Gamma^\maltese \vdash Q^\maltese : C^\maltese$.  Applying
      exchange multiple times, we obtain $\Delta^\maltese,
      \Gamma^\maltese, u : A^\maltese \vdash Q^\maltese :
      C^\maltese$, and using cut yields $\Delta^\maltese,
      \Gamma^\maltese \vdash Q^\maltese[\pi_2(P^\maltese)/u] :
      C^\maltese$.

      \case{$\ctxt{\Delta}{\Gamma} \vdash_\textsf{DK4} \ibox{M} : \Box A$}

      By the IH, we have $\Delta^\maltese,
      \left(\Delta^\bot\right)^\maltese \vdash
      \left(M^\bot\right)^\maltese : A^\maltese$. Notice that,
      when acting on contexts, $(-)^\bot$ only acts on variables
      and $(-)^\maltese$ only on types, so
      $\left(\Delta^\bot\right)^\maltese \equiv
      \left(\Delta^\maltese\right)^\bot$. Substitute all the
      variables in $\left(\Delta^\maltese\right)^\bot$ with the
      corresponding ones in $\Delta^\maltese$. We thus obtain a
      term $\Delta^\maltese \vdash
      (M^\bot)^\maltese[\Delta^\maltese/\left(\Delta^\maltese\right)^\bot]
      : A^\maltese$. But we have by (1) that \[
        (M^\bot)^\maltese[\Delta^\maltese/(\Delta^\maltese)^\bot]
          \equiv
        (M^\bot[\Delta/(\Delta^\maltese)^\bot])^\maltese
          \equiv
        (M^\bot[\Delta/\Delta^\bot])^\maltese
          \equiv
        M^\maltese
      \] We thus apply weakening and the introduction rules for
      $\ast$ and products to obtain $\Delta^\maltese,
      \Gamma^\maltese \vdash \langle \ast, M^\maltese \rangle :
      \mathbf{1} \times A^\maltese$.
      
    \end{indproof}

    \item By induction on $M \red{} N$. The only cases that are
    not immediate are those involving modal constructors. For the
    modal $\beta$, we notice that by the congruence rules
    \[
        (\letbox{u}{\ibox{M}}{N})^\maltese
      \equiv
        N^\maltese[\pi_2(\langle \ast, M^\maltese \rangle)/u]
      \red{}
        N^\maltese[M^\maltese/x]
      \equiv
        \left(N[M/x]\right)^\maltese
    \] 
    where the last step follows by (1). For the first commuting
    conversion, we have 
    \[
      \left(\pi_i(\letbox{u}{M}{N})\right)^\maltese
        \equiv
      \pi_i(N^\maltese[\pi_2(M^\maltese)/u])
        \equiv
      \left(\letbox{u}{M}{\pi_i(N)}\right)^\maltese
    \] 
    so we use the fact $\red_\beta^\ast$ is reflexive. The cases
    for the other commuting conversions are similar, and they
    translate to syntactic equalities.
    \qedhere
  \end{enumerate}
\end{proof}

We can now almost obtain Theorem \ref{thm:sn} as a corollary: an
infinite sequence of reductions $N_1 \red{} N_2 \red \dots$ would
lead to an infinite sequence of reductions $N_1^\maltese
\red_\beta^\ast N^\maltese_2 \red_\beta^\ast \dots$ in the
simply-typed $\lambda$-calculus, which would contradict strong
normalization.  This is not yet a proof, because there is no
guarantee that each reduction $N_i^\maltese \red_\beta^\ast
N_{i+1}^\maltese$ is a non-trivial $\beta$-reduction. However,
scrutinising the proof of Theorem \ref{thm:sim} leads to the
conclusion that the only trivial reductions arise when $N_i \red{}
N_{i+1}$ due to a commuting conversion. It thus remains to prove
that $N_i \red{} N_{i+1}$ is infinitely often a proper
$\beta$-reduction.

For that, we employ a trick used by de Groote \cite{deGroote2002}:
we assign a \emph{permutation degree} to every term. We define it
by the following clauses. 
\begin{gather*}
  \begin{align*}
  |x| &\defeq 1 &
  |\lambda x : A.\ M| &\defeq |M| \\
  |MN| &\defeq |M| + \#(M) \cdot |N| &
  |\langle M, N \rangle| &\defeq |M| + |N| \\
  |\pi_i(M)| &\defeq |M| + \#(M) &
  |\ibox{M}| &\defeq |M|
  \end{align*} \\
  |\letbox{u}{M}{N}| \defeq |M| + \#(M) \cdot |N|
\end{gather*} 
where 
\begin{gather*}
  \begin{align*}
  \#(x) &\defeq 1 &
  \#(\lambda x : A.\ M) &\defeq 1 \\
  \#(MN) &\defeq \#(M) &
  \#(\langle M, N \rangle) &\defeq 1 \\
  \#(\pi_i(M)) &\defeq \#(M) &
  \#(\ibox{M}) &\defeq 1
  \end{align*} \\
  \#(\letbox{u}{M}{N}) \defeq 2 \cdot \#(M) \cdot \#(N)
\end{gather*}
Briefly, $\#(M) = 2^n$, where $n$ is the number of
$\letbox{u}{(-)}{(-)}$ constructs in $M$ that are either at an
`outermost' position, or can be commuted to be so. It follows that
\begin{lem} 
  If $M \red{} N$ by a commuting conversion, then $\#(M) = \#(N)$.
\end{lem} The metric $|M|$ uses $\#(M)$ to weigh the appearance of
$\letbox{u}{(-)}{(-)}$ constructs, with the weight being higher the
more deeply they appear in a term. It is easy to show that
\begin{lem}
  If $M \red{} N$ by a commuting conversion, then $|M| > |N|$.
\end{lem} It follows that in an infinite sequence of reductions it
must be the case that $\beta$-reductions occur infinitely often,
as the metric $|-|$ strictly reduces for each commuting
conversion. This completes the proof of strong normalization.

\subsection{The case of \textsf{GL}}
  \label{sec:case-gl}

It remains to prove normalization for \textsf{GL}, which---because
of the presence of a certain amount of self-reference---behaves in
an unusual way. 

Recall that $\fixbox{z}{M}$ is the term corresponding to an
application of the L\"ob rule to the proof $M$. One might first
think that this term must reduce in the manner of a fixpoint, for
example to $M[\fixbox{z}{M}/z]$ where the diagonal variable $z :
\Box A$ has been replaced by the entire term. Notice, however,
that this does not preserve subject reduction.  Instead, we follow
the statement of modal cut admissibility (Theorem \ref{thm:mcut}),
and replace the modal $\beta$-reduction used in the other systems
by
\[
  \letbox{u}{(\fixbox{z}{M})}{N}
    \red{}
  N[M[\fixbox{z}{M}/z]/u]
\] 
which preserves subject reduction. We also include the congruence rule
\[
  \begin{prooftree}
      M \red{} N
    \justifies
      \fixbox{z}{M} \red{} \fixbox{z}{N}
  \end{prooftree}
\]

This leads to to a system with manifest \emph{coinductive
behaviour}: The introduction form $\fixbox{z}{M}$ is a largely
inactive term former which---contrary to expectation---does not
unfold infinitely. Instead, we are free to use the congruence rule
to eliminate cuts in its body $M$. On the other hand, when this
term meets the elimination form $\letbox{u}{(-)}{N}$, it unfolds
as much as necessary to fill the position of the variable $u$ in
$N$.  Thus, $\fixbox{z}{M}$ unfolds `on demand' whenever it is
deconstructed.

We can extend complement reduction (Lemma \ref{lem:redcomp}) to
this system: if $\ctxt{\Delta}{\Delta^\bot, q^\bot : \Box A}
\vdash_{\mathsf{DGL}} M^\bot : A$, then $M \red N$ implies $M^\bot
\red N^\bot$. The proof is the essentially the same, even in the
case of the the new $\beta$-reduction: we have 
\[
  \left(\letbox{u}{\fixbox{z}{Q}}{N}\right)^\bot
    \red{}
  N^\bot[Q_\ast/u]
\]
where $Q_\ast \defeq Q[\fixbox{z}{Q}/z]$. We want to show that the
RHS is $\left(N[Q_\ast/u]\right)^\bot$. It suffices to show that
$u^\bot \not\in \fv{N}$ and $u^\bot \not\in \fv{N^\bot}$, for then
by Theorem \ref{thm:bot} we get $\left(N[Q_\ast/u]\right)^\bot
\equiv N^\bot[Q_\ast, Q_\ast^\bot/u, u^\bot] \equiv
N^\bot[Q_\ast/u]$. Recalling that $\ctxt{\Delta, u^\bot :
B}{\Delta^\bot, q^\bot : \Box A} \vdash N^\bot : A$ for some $B$,
we see that the rest of the argument is essentially identical to
that for $\textsf{K4}$.

It is then easy to extend the proof of confluence and the
subformula property to this notion of reduction.  However, the
method used to prove strong normalization in \S\ref{sec:sn} no
longer applies.  Intuitively the reason is that, up to the
isomorphism $A \cong \mathbf{1} \times A$, the G\"odel-L\"ob axiom
would be translated to the type $(A \rightarrow A) \rightarrow A$,
which is the type of fixed point combinators at $A$. We would thus
have to augment the target language with such fixed point
combinators. This would make it essentially equivalent to PCF
\cite{Plotkin1977}, which is the archetypal language with
non-terminating terms!

Instead, we will prove normalization using the method of
candidates (candidats de reducibilit\'{e}). This method originates
in Girard's proof of strong normalization for System F
\cite{Girard1972}. Our variant is a combination of two
presentations. The main structure of the proof is due to by
Koletsos \cite{Koletsos1985}, as presented in simplified form by
Gallier \cite{Gallier1995}. However, the Koletsos-Gallier
presentation does not carry typing information in the proof,
whereas in our calculi typing is vital. Thus, we enhance their
method, insofar as our can candidates consist of typing judgments
$\ctxt{\Delta}{\Gamma} \vdash M : A$ rather than simply terms $M :
A$. Ideas on how this is done were drawn from another chapter by
Gallier \cite{Gallier1990}, which surveys multiple variants of the
\emph{candidats} method. The proof itself is rather long. We
abbreviate it by only presenting the cases that are relevant to
the modal fragment of the language. The interested reader can
obtain the full proof from the author's
website.\footnote{\url{https://www.lambdabetaeta.eu}}

The overall structure of the proof is the following. Suppose we
have a family of nonempty \emph{sets of typing judgments}, 
\[
  \mathcal{P} = \{P_A\}_A
\]
indexed by the type $A$ they assign to the term they carry.  If
$C \subseteq P_A$, we write $\ctxt{\Delta}{\Gamma} \vdash M \in C$
as a shorthand for $\left(\ctxt{\Delta}{\Gamma} \vdash M :
A\right) \in C$. We will state six properties,
\textsf{P0}--\textsf{P5}, that such a family should satisfy. In
case it does indeed satisfy them, we show that 
\[
  \ctxt{\Delta}{\Gamma} \vdash_\textsf{DGL} M : A\
    \Longrightarrow\
  \ctxt{\Delta}{\Gamma} \vdash M \in P_A
\]

In our case, we show that the family of typing judgments \[
  \mathcal{SN} \defeq \{SN_A\}_A
\] satisfies the properties \textsf{P0}-\textsf{P5}, where $SN_A$
consists of all the judgments $\ctxt{\Delta}{\Gamma} \vdash M : A$
for which $M$ is strongly normalizing.  Then $SN_A = \Lambda_A$,
and all typable terms are strongly normalizing.

Here is a brief summary of the proof. We begin by stating the
first four properties, namely \textsf{P0}--\textsf{P3}. We also
define what it means for a set $C$ of derivable judgments to be a
\emph{candidate}. Then, we define a subset $\sem{A}{} \subseteq
P_A$ for each type $A$. We call judgments in $\sem{A}{}$
\emph{reducible}, and we show $\sem{A}{}$ to be a candidate.
Finally, we introduce two further properties, \textsf{P4} and
\textsf{P5}. If these hold of $P_A$, then we show that $\sem{A}{}$
contains all derivable judgments.

We write $\Gamma \sqsubseteq \Gamma'$ to mean that the context
$\Gamma$ is a subsequence of $\Gamma'$ (in other words, $\Gamma'$
is obtained after a series of weakening steps on $\Gamma$). We
also write $\Lambda$ for the set of all terms. We will make use of
the following helpful definitions. 

\begin{defi} \hfill
  \begin{enumerate}
    \item A term is a \emph{intro term} just if it is an introduction
    form, i.e. of the form \[
        \lambda x : A.\ M,
      \quad
        \langle M, N \rangle,
      \quad
        \fixbox{z}{M}
    \]

    \item A term is a \emph{simple term} just if it is a variable
    or an elimination form, i.e. of the forms \[
      x, \quad MN, \quad \pi_i(M), \quad \letbox{u}{M}{N}
    \]

    \item A \emph{stubborn term} is a term that is either a normal
    form w.r.t. $\red{}$, or a term that does not reduce to an
    intro term, i.e. if $M \redt{} N$ then $N$ is not an intro
    term.
  \end{enumerate}
\end{defi}

\paragraph{Candidates and the first four properties}

The sets of judgments we will use will always satisfy the
following four important properties.

\begin{defi}[Properties \textsf{P0}-\textsf{P3}] 
  For a family $\mathcal{P}$ we define the following properties.
  \begin{enumerate}[widest={P0}]
    \item[(\textsf{P0})]
      \begin{enumerate}
        \item 
          $\ctxt{\Delta}{\Gamma} \vdash M \in P_A$ and $\Gamma
          \sqsubseteq \Gamma'$ imply $\ctxt{\Delta}{\Gamma'}
          \vdash M \in P_A$

        \item
          $\ctxt{\Delta}{\Gamma} \vdash M \in P_A$
            and
          $\Delta \sqsubseteq \Delta'$
            imply
          $\ctxt{\Delta'}{\Gamma} \vdash M \in P_A$
      \end{enumerate}

    \item[(\textsf{P1})]
      $\ctxt{\Delta}{\Gamma} \vdash x \in P_A$ for all $(x : A)
      \in \Gamma$.

    \item[(\textsf{P2})]
      $\ctxt{\Delta}{\Gamma} \vdash M \in P_A$ 
      and 
      $M \red N$
      imply
      $\ctxt{\Delta}{\Gamma} \vdash N \in P_A$.

    \item[(\textsf{P3})] For simple terms $M$,
      \begin{enumerate}
        \item 
          If
          \begin{itemize}
            \item[--]
              $\ctxt{\Delta}{\Gamma}
              \vdash M \in P_{A \rightarrow B}$,
            \item[--]
              $\ctxt{\Delta}{\Gamma} \vdash N \in P_A$, and
            \item[--]
              whenever $M \redt \lambda x : A.\ M'$ then
              $\ctxt{\Delta}{\Gamma} \vdash
                (\lambda x : A.\ M')N \in P_B$
          \end{itemize} then this implies that
            $\ctxt{\Delta}{\Gamma} \vdash MN \in P_B$.
        \item 
          If 
          \begin{itemize}
            \item[--]
              $\ctxt{\Delta}{\Gamma}
              \vdash M \in P_{A \times B}$, and
            \item[--]
              whenever
              $M \redt \langle M_1, M_2 \rangle$
            then
              $\ctxt{\Delta}{\Gamma} \vdash \pi_1(\langle M_1, M_2 \rangle) \in P_A$
            and
              $\ctxt{\Delta}{\Gamma} \vdash \pi_2(\langle M_1, M_2 \rangle) \in P_B$,
          \end{itemize}
          then this implies that
            $\ctxt{\Delta}{\Gamma} \vdash \pi_1(M) \in P_A$ and
            $\ctxt{\Delta}{\Gamma} \vdash \pi_2(M) \in P_B$.
      \end{enumerate}
  \end{enumerate}
\end{defi}

\begin{defi}[$\mathcal{P}$-candidate] 
A set $C_A \subseteq P_A$ is \emph{$\mathcal{P}$-candidate at $A$}
just if
  \begin{enumerate}[widest={P0}]
    \item[(\textsf{R0})]
      \begin{enumerate}
        \item
          $\ctxt{\Delta}{\Gamma} \vdash M \in C_A$
          and
          $\Gamma \sqsubseteq \Gamma'$
          imply
          $\ctxt{\Delta}{\Gamma'} \vdash M \in C_A$.
        \item
          $\ctxt{\Delta}{\Gamma} \vdash M \in C_A$
          and
          $\Delta \sqsubseteq \Delta'$
          imply
          $\ctxt{\Delta'}{\Gamma} \vdash M \in C_A$.
      \end{enumerate}

    \item[(\textsf{R1})]
      $\ctxt{\Delta}{\Gamma} \vdash M \in C_A$ and $M \red N$ imply
      $\ctxt{\Delta}{\Gamma} \vdash N \in C_A$.

    \item[(\textsf{R2})]
      If $\ctxt{\Delta}{\Gamma} \vdash M \in P_A$ is simple, and
      $M \redt N$ and for an intro term $N$ implies
      $\ctxt{\Delta}{\Gamma} \vdash N \in C_A$, then it follows
      that $\ctxt{\Delta}{\Gamma} \vdash M \in C_A$.
  \end{enumerate}
\end{defi}

Notice that $(\textsf{R0})$ is analogous to $(\textsf{P0})$, and
$(\textsf{R1})$ is analogous to $(\textsf{P2})$. Moreover, these
conditions in tandem imply an analogue of $(\textsf{P1})$:

\begin{lem}
  \label{lem:varcand}
  For any $\mathcal{P}$-candidate $C_A$, if $(x : A) \in \Gamma$
  then $\ctxt{\Delta}{\Gamma} \vdash x \in C$.
\end{lem}

\begin{proof}
  By (\textsf{P1}), we have that $\ctxt{\Delta}{\Gamma} \vdash x
  \in P_A$, and by definition $x$ is simple, and a normal form, so
  it cannot ever reduce to an intro term. The result follows by
  (\textsf{R2}). 
\end{proof}

\noindent A family $\mathcal{P}$ for which
\textsf{P0}--\textsf{P3} hold is almost a candidate. In fact, the
only condition that is not automatically satisfied is
$\textsf{R2}$. To remedy that situation, we define a particular
subfamily $\sem{A} \subseteq P_A$ of \emph{reducible judgments},
which---as we show---satisfies it. This definition has the
familiar flavour of logical predicates.

\begin{defi}[Reducible judgments]
  We define for each type $A$ a set of derivable judgments
  $\sem{A}{} \subseteq P_A$ by induction on $A$.
  \begin{align*}
    \sem{p_i}{} &\defeq P_{p_i} \\
    \sem{A \times B}{} &\defeq
      \setcomp{
        \ctxt{\Delta}{\Gamma} \vdash M \in P_{A \times B}
      }{
        \ctxt{\Delta}{\Gamma} \vdash \pi_1(M) \in \sem{A}
          \land\
        \ctxt{\Delta}{\Gamma} \vdash \pi_2(M) \in \sem{B}{}
      } \\
    \sem{A \rightarrow B}{} &\defeq
      \setcomp{
          \ctxt{\Delta}{\Gamma} \vdash M \in P_{A \rightarrow B}
      }{
        \forall 
          \Delta \sqsubseteq \Delta',\
          \Gamma \sqsubseteq \Gamma',
          \ctxt{\Delta'}{\Gamma'} \vdash N \in \sem{A}{}.\
          \ctxt{\Delta'}{\Gamma'} \vdash MN \in \sem{B}{}
      } \\
    \sem{\Box A}{} &\defeq
      \setcomp{
	      \ctxt{\Delta}{\Gamma} \vdash M \in P_{\Box A}\
      }{
          M \redt \fixbox{z}{Q}\
            \Longrightarrow\
          \ctxt{\Delta}{\Delta^\bot, z^\bot : \Box A} 
            \vdash Q^\bot \in \sem{A}
 	    }
  \end{align*}
\end{defi}

\noindent We can now show that

\begin{thm}
  \label{thm:redcand}
  If $\mathcal{P} = \{P_A\}$ satisfies properties
  \textsf{P0}-\textsf{P3}, then
  \begin{enumerate}
    \item For any $A$, $\sem{A}{}$ is a $\mathcal{P}$-candidate.
    \item For any $A$, $\sem{A}{}$ contains all the stubborn terms in $P_A$.
  \end{enumerate}
\end{thm}
\begin{proof} 
  By induction on types. 

  \begin{indproof}
    \case{$\Box A$} For (1):
      \begin{enumerate}
        \item[(\textsf{R0})]
          (a) trivially holds, for none of the judgments for $Q$
          in $\sem{\Box A}$ depend on $\Gamma$.

          For (b): we first use $(\textsf{P0})$ to ascertain that
          $\ctxt{\Delta'}{\Gamma} \vdash M \in P_{\Box A}$. If $M$
          reduces to $\fixbox{z}{Q}$, we use (\textsf{R0})(a, b)
          of the IH for $A$ to weaken $\Delta$ and $\Delta^\bot$
          respectively.

        \item[(\textsf{R1})]
          Let $\ctxt{\Delta}{\Gamma} \vdash M \in \sem{\Box A}{}$
          and suppose $M \red N$. By (\textsf{P2}) we have
          $\ctxt{\Delta}{\Gamma} \vdash N \in P_{\Box A}$.  It
          remains to show that whenever $N \redt \fixbox{z}{Q}$
          then $\ctxt{\Delta}{\Delta'^\bot, z^\bot : \Box A}
          \vdash Q^\bot \in \sem{A}{}$. But then $M \redt
          \fixbox{z}{Q}$ as well, so this follows from $M \in
          \sem{\Box A}$.

        \item[(\textsf{R2})]
          Suppose that $\ctxt{\Delta}{\Gamma} \vdash M \in P_{\Box
          A}$ is a simple term, and whenever $M \redt
          \fixbox{z}{Q}$ then $\fixbox{z}{Q} \in \sem{\Box A}$,
          i.e. $\fixbox{z}{Q} \redt \fixbox{z}{Q'}$ implies
          $\ctxt{\Delta}{\Delta^\bot, z^\bot : \Box A} \vdash
          Q'^\bot \in \sem{A}$. We need to show that, if $M
          \redt \fixbox{z}{Q}$, then $\ctxt{\Delta}{\Delta^\bot,
          z^\bot : \Box A} \vdash Q \in \sem{A}$. But this follows
          by the reflexivity of $\redt$.

      \end{enumerate}
      For (2): if $M \in P_{\Box A}$ is stubborn, then it never
      reduces to an intro term $\fixbox{z}{Q}$, so it is vacuously
      in $\sem{\Box A}{}$. 
      \qedhere
  \end{indproof}
\end{proof}

\paragraph{Closure under formation: the latter two properties}
Unfortunately, this is not enough to show that the candidates
$\sem{A}$ contain \emph{all} the provable judgments of
\textsf{DGL}. We will thus need the following two additional
conditions on $\mathcal{P}$.

\begin{defi}[Properties \textsf{P4}-\textsf{P5}] \hfill
  \begin{enumerate}[widest={P4}]
    \item[(\textsf{P4})]
      \begin{enumerate}
        \item
          If $\ctxt{\Delta}{\Gamma, x : A} \vdash M \in P_B$ then
          $\ctxt{\Delta}{\Gamma} \vdash \lambda x : A.\ M \in P_{A
          \rightarrow B}$.

        \item
          $\ctxt{\Delta}{\Gamma} \vdash M \in P_A$ and
          $\ctxt{\Delta}{\Gamma} \vdash N \in P_B$ imply
          $\ctxt{\Delta}{\Gamma} \vdash \langle M, N \rangle \in
          P_{A \times B}$.

        \item
          $\ctxt{\Delta}{\Delta^\bot, z^\bot : \Box A} \vdash
          Q^\bot \in P_A$ implies $\ctxt{\Delta}{\Gamma} \vdash
          \fixbox{z}{Q} \in P_{\Box A}$
      \end{enumerate}

    \item[(\textsf{P5})]
      \begin{enumerate}
        \item
          $\Delta \sqsubseteq \Delta'$, $\Gamma \sqsubseteq
          \Gamma'$, $\ctxt{\Delta'}{\Gamma'} \vdash N \in P_A$,
          and $\ctxt{\Delta'}{\Gamma'} \vdash M[N/x] \in P_B$
          imply $\ctxt{\Delta'}{\Gamma'} \vdash (\lambda x : A.\
          M)N \in P_B$.

        \item
          $\ctxt{\Delta}{\Gamma} \vdash M \in P_A$
          and $\ctxt{\Delta}{\Gamma} \vdash N \in P_B$
          imply $\ctxt{\Delta}{\Gamma}
            \vdash \pi_1(\langle M, N \rangle) \in P_A$
          and $\ctxt{\Delta}{\Gamma} 
            \vdash \pi_2(\langle M, N \rangle) \in P_B$.

        \item
          If $\ctxt{\Delta}{\Gamma} \vdash M \in P_{\Box A}$ and
          $\ctxt{\Delta, u : A}{\Gamma} \vdash N \in P_C$, and
          whenever $M \redt \fixbox{z}{Q}$ then
          $\ctxt{\Delta}{\Gamma} \vdash
          N\left[Q[\fixbox{z}{Q}/z]/u\right] \in P_C$, then
          $\ctxt{\Delta}{\Gamma} \vdash \letbox{u}{M}{N} \in P_C$.
      \end{enumerate}
  \end{enumerate}
\end{defi}

\noindent These conditions ensure that the candidates $\sem{A}$
also have the following closure properties.

\begin{thm}
  \label{thm:cand2}
  If $\mathcal{P} = \{P_A\}$ satisfies properties
  \textsf{P0}--\textsf{P5}, then
  \begin{enumerate}
    \item
      If whenever $\Gamma \sqsubseteq \Gamma'$, $\Delta
      \sqsubseteq \Delta'$ and $\ctxt{\Delta'}{\Gamma'} \vdash N \in
      \sem{A}{}$ we have $\ctxt{\Delta'}{\Gamma'} \vdash M[N/x] \in
      \sem{B}{}$, then $\ctxt{\Delta}{\Gamma} \vdash
  	  \lambda x : A.\ M \in \sem{A \rightarrow B}$.

    \item
      If $\ctxt{\Delta}{\Gamma} \vdash M \in \sem{A}{}$ and
      $\ctxt{\Delta}{\Gamma} \vdash N \in \sem{B}{}$ then
      $\ctxt{\Delta}{\Gamma} \vdash \langle M, N \rangle \in
      \sem{A \times B}$.

    \item
      If $\ctxt{\Delta}{\Gamma} \vdash M \in \sem{\Box A}$, and
      whenever $\Delta \sqsubseteq \Delta'$ and
      $\ctxt{\Delta'}{\Delta'^\bot, z^\bot : \Box A} \vdash Q^\bot
      \in \sem{A}{}$ then $\ctxt{\Delta'}{\Gamma} \vdash
      N\left[Q[\fixbox{z}{Q}/z]/u\right] \in \sem{C}{}$, then $
      \ctxt{\Delta}{\Gamma} \vdash \letbox{u}{M}{N} \in \sem{C}{}
      $.
  \end{enumerate}
\end{thm}

\begin{proof}
  We only show (3). Write $Q_\star \defeq Q[\fixbox{z}{Q}/z]$.

      First, we show that $\letbox{u}{M}{N} \in P_C$, and we
      invoke (\textsf{P5})(c) to do so. It suffices to show that
      $\ctxt{\Delta}{\Gamma} \vdash M \in P_{\Box A}$, that
      $\ctxt{\Delta, u : A}{\Gamma} \vdash N \in P_C$, and that
      whenever $M \redt \fixbox{z}{Q}$ then $\ctxt{\Delta}{\Gamma}
      \vdash N[Q_\star/u] \in P_C$.  The first of these is implied
      by the assumption that $\ctxt{\Delta}{\Gamma} \vdash M \in
      \sem{A}{} \subseteq P_A$. For the second, we infer by Lemma
      \ref{lem:varcand} and Theorem \ref{thm:redcand} that
      $\ctxt{\Delta, u : A}{\Delta^\bot, u^\bot : A, z^\bot : A}
      \vdash u^\bot \in \sem{A}{}$.  Hence, as $\Delta \sqsubseteq
      \Delta, u : A$, we have by the assumption that \[
        \ctxt{\Delta, u : A}{\Gamma} 
          \vdash N \equiv N[u[\fixbox{z}{u}/z]/u] \in \sem{C}{}
      \] The final desideratum also follows: if $M \redt
      \fixbox{z}{Q}$ then, by the definition of $\sem{\Box A}{}$,
      we have that
        $\ctxt{\Delta}{\Delta^\bot, z^\bot : \Box A} 
          \vdash Q^\bot \in \sem{A}$
      and hence---by the assumption---that
      $\ctxt{\Delta}{\Gamma} 
        \vdash N[Q_\star/u] \in \sem{C}{} \subseteq P_C$.
      
      For the rest, we note that $\letbox{u}{M}{N}$ is simple, so
      we use (\textsf{R2}): it suffices to show that whenever
      $\letbox{u}{M}{N} \redt Q$ and Q is an intro term, then $Q
      \in \sem{C}{}$. If $\letbox{u}{M}{N}$ is stubborn, then the
      desideratum is trivial. Otherwise, if $\letbox{u}{M}{N}
      \redt Q$ where $Q$ is a intro term, then the reduction must
      be of the form \begin{align*}
                 \letbox{u}{M}{N}
        &\redt &&\letbox{u}{\fixbox{z}{U}}{N'} \\
        &\red  &&N'[U^\star/u] \\
        &\redt &&Q
      \end{align*} where $M \redt \fixbox{z}{U}$ and $N \redt N'$:
      otherwise the let construct would persist. But, by
      assumption, $\ctxt{\Delta}{\Gamma} \vdash M \in \sem{\Box
      A}{}$, so by multiple applications of (\textsf{R1}) we infer
      that $\ctxt{\Delta}{\Gamma} \vdash \fixbox{z}{U} \in
      \sem{\Box A}{}$ and hence that $\ctxt{\Delta}{\Delta',
      z^\bot : \Box A} \vdash U^\bot \in \sem{A}$. By the
      assumption, we get $\ctxt{\Delta}{\Gamma} \vdash
      N[U_\star/u] \in \sem{C}$.  But $N[U_\star/u] \redt
      N'[U_\star/u] \redt Q$, so $Q \in \sem{C}$ by repeated
      applications of (\textsf{R1}).
\end{proof}

\paragraph{The main theorem}

\begin{defi}[Substitution] \hfill
  \begin{enumerate}
  \item
    A \emph{substitution} is a finite function $\sigma : \mathcal{V}
    \parfunc{} \Lambda$ from the set of all variables $\mathcal{V}$
    to the set of all possible untyped/raw terms $\Lambda$.
  \item
    A substitution $\sigma$ is \emph{type-preserving} from
    $\ctxt{\Delta'}{\Gamma'}$ to $\ctxt{\Delta}{\Gamma}$, written
    $\ctxt{\Delta'}{\Gamma'} \transit{\sigma}{}
    \ctxt{\Delta}{\Gamma}$, just if 
    \begin{enumerate}
      \item 
        $dom(\sigma) \subseteq \vars{\Delta} \cup \vars{\Gamma}$,
      \item
        $(x : B) \in \Gamma$ implies $\ctxt{\Delta'}{\Gamma'} \vdash
        \sigma(x) : B$, and
      \item 
        there exists $z$ such that $(u : B) \in \Delta$ implies
        $\ctxt{\Delta'}{\Delta'^\bot, z^\bot : \Box B} \vdash
        (\sigma(u))^\bot \in B$.
    \end{enumerate}
  \end{enumerate}
\end{defi}

\noindent We call the aforementioned $z$ the \emph{diagonal
variable} of the substitution. We have that

\begin{lem}
  If $\ctxt{\Delta'}{\Gamma'} \transit{\sigma}{}
  \ctxt{\Delta}{\Gamma}$ and $\Delta' \sqsubseteq \Delta''$ and
  $\Gamma' \sqsubseteq \Gamma'$ then $\ctxt{\Delta''}{\Gamma''}
  \transit{\sigma}{} \ctxt{\Delta}{\Gamma}$.
\end{lem}

We write $\sigma[x \mapsto N]$ to mean the substitution defined by
$\sigma[x \mapsto N](x) \defeq N$, and $ \sigma[x \mapsto N](y)
\defeq \sigma(y)$ if $y \not\equiv x$. Furthermore, if
$\ctxt{\Delta'}{\Gamma'} \transit{\sigma}{} \ctxt{\Delta}{\Gamma}$
with diagonal variable $z$, we define $\sigma^\dagger$ by \[
  \sigma^\dagger(y) \defeq 
    \begin{cases}
      z^\bot                      & \text{if $y \equiv z^\bot$} \\
      \left(\sigma(u)\right)^\bot & \text{if $y \equiv u^\bot \in \vars{\Delta^\bot}$} \\
      \sigma(y) & \text{otherwise}
    \end{cases}
\] The following technical fact is evident, but very convenient.

\begin{lem}[Modal Drop]
  \label{lem:drop}
  If $\ctxt{\Delta'}{\Gamma'} \transit{\sigma}{}
  \ctxt{\Delta}{\Gamma}$ with diagonal variable $z$, then \[
    \ctxt{\Delta'}{\Delta'^\bot, z^\bot : \Box A}
      \transit{\sigma^\dagger}{}
    \ctxt{\Delta}{\Delta^\bot, z^\bot : \Box A}
  \]
\end{lem}

\noindent We extend the action of substitutions on terms in the
usual capture-avoiding manner, e.g. \[
  \sigma(\fixbox{z}{M})
    \defeq \fixbox{z}{\sigma(M)} \\
\] 

\begin{lem}
  If $
    \ctxt{\Delta'}{\Gamma'}
      \transit{\sigma}{}
    \ctxt{\Delta}{\Gamma}
  $ and $
    \ctxt{\Delta}{\Gamma} \vdash M : C
  $ then $
    \ctxt{\Delta'}{\Gamma'} \vdash \sigma(M) : C
  $.
\end{lem}

\begin{proof}
  By induction on $M$. We only show the modal cases. That of
  $\letbox{u}{M}{N}$ is very similar to $\lambda$-abstraction.
  For $\fixbox{z}{M}$, we have that $\ctxt{\Delta}{\Delta^\bot,
  z^\bot : \Box A} \vdash M : A$ for some $A$ such that $C \equiv
  \Box A$. It follows by Lemma \ref{lem:drop} that \[
    \ctxt{\Delta'}{\Delta'^\bot, z^\bot : \Box A}
      \transit{\sigma^\dagger}{}
    \ctxt{\Delta}{\Delta^\bot, z^\bot : \Box A}
  \] Applying the IH then yields $\ctxt{\Delta'}{\Delta'^\bot,
  z^\bot : \Box A} \vdash \sigma^\dagger(M) : A$. But notice that
  $\sigma^\dagger(M) \equiv \left(\sigma(M)\right)^\bot$, so a
  single use of $(\Box\mathcal{I}_\textsf{GL})$ suffices.
\end{proof}

\begin{thm}[Candidats]
  \label{thm:candidats}
  Let $\mathcal{P} = \{P_A\}$ be a family satisfying properties
  \textsf{P1}--\textsf{P5}. Let $\ctxt{\Delta}{\Gamma}
  \vdash_{\textsf{DGL}} M : A$, and $\ctxt{\Delta'}{\Gamma'}
  \transit{\sigma}{} \ctxt{\Delta}{\Gamma}$ be a substitution with
  diagonal variable $z$ that respects the candidates, i.e. such
  that 
  \begin{enumerate}
    \item 
      $(x : B) \in \Gamma$ implies $\ctxt{\Delta'}{\Gamma'} \vdash
      \sigma(x) \in \sem{B}{}$, and 
    \item
      $(u : C) \in \Delta$ implies $\ctxt{\Delta'}{\Delta'^\bot,
      z^\bot : \Box C} \vdash (\sigma(u))^\bot \in \sem{C}{}$,
  \end{enumerate} Then $\ctxt{\Delta'}{\Gamma'} \vdash \sigma(M)
  \in \sem{A}{}$.
\end{thm}

\begin{proof}
  By induction on $M$. We only prove the modal cases.
  \begin{indproof}

    \case{$\fixbox{z}{M}$}

      Then $\ctxt{\Delta}{\Delta^\bot, z^\bot : \Box A} \vdash M :
      A$. By Lemma \ref{lem:drop}, we have that
      $\ctxt{\Delta'}{\Delta'^\bot, z^\bot : \Box A}
      \transit{\sigma^\dagger}{} \ctxt{\Delta}{\Delta^\bot, z^\bot
      : \Box A}$. Then, by the IH we have that
      $\ctxt{\Delta'}{\Delta'^\bot, z^\bot : \Box A} \vdash
      \sigma^\dagger(M) \equiv \sigma(M)^\bot \in \sem{A}{}$. So,
      by (\textsf{P4})(c), $\fixbox{z}{\sigma(M)} \in P_{\Box A}$.
      It now suffices---by the definition of $\sem{\Box A}{}$---to
      show that \[
        \fixbox{z}{\sigma(M)} \redt \fixbox{z}{M'}
      \] implies $\ctxt{\Delta'}{\Delta'^\bot, z^\bot : \Box A}
      \vdash M' \in \sem{A}{}$. But then we must have $\sigma(M)
      \redt M'$, so by repeated applications of (\textsf{R1}) we
      have $M' \in \sem{A}{}$.
  
    \case{$\letbox{u}{M}{N}$}

      We show the case for \textsf{K}.  We have
      $\ctxt{\Delta}{\Gamma} \vdash M : A$ and $\ctxt{\Delta,
      u : A}{\Gamma} \vdash N : C$. We use Theorem
      \ref{thm:cand2}(a): to show that \[
	\ctxt{\Delta'}{\Gamma'} \vdash \sigma(\letbox{u}{M}{N}) 
	  \equiv \letbox{u}{\sigma(M)}{\sigma(N)} \in \sem{C}{}
      \] It suffices to show that 
      $\ctxt{\Delta'}{\Gamma'} \vdash \sigma(M) \in \sem{\Box
      A}{}$---which we have by the IH---and that whenever
      $\Delta'' \sqsupseteq \Delta'$ and
      $\ctxt{\cdot}{\Delta''} \vdash Q \in \sem{A}{}$, then 
      $\ctxt{\Delta''}{\Gamma'} \vdash \sigma(N)[Q/u] \in \sem{C}{}$.

      Define \[
	\sigma' \defeq \sigma[u \mapsto Q]
      \] Then, by weakening the modal context in $\sigma$, we have \[
	\ctxt{\Delta'', u : A}{\Gamma'}
	  \transit{\sigma'}{}
	  \ctxt{\Delta, u : A}{\Gamma}
      \] By the IH, \[
	\ctxt{\Delta'', u : A}{\Gamma'}
	  \vdash \sigma'(N) \in \sem{C}{}
      \] But $\sigma'(N) \equiv \sigma(N)[Q/u]$.
  \qedhere
  \end{indproof}
\end{proof}

\begin{cor}
  \label{cor:cand}
  If $\mathcal{P} = \{P_A\}$ is a family satisfying
  properties \textsf{P0}--\textsf{P5}, then \[
    P_A = \Lambda_A
  \]
\end{cor}
\begin{proof} 
  By Theorem \ref{thm:candidats} we have that $\ctxt{\Delta}{\Gamma}
  \vdash M \in \sem{A}{}$ for every $\ctxt{\Delta}{\Gamma} \vdash M
  : A$. Hence $\Lambda_A \subseteq \sem{A}{} \subseteq P_A \subseteq
  \Lambda_A$.
\end{proof}

It is then reasonably straightforward to verify that the above
properties hold of the family $\mathcal{SN}$. In carrying out the
proof we shall often use induction on $d(M)$, the \emph{depth of
the term $M$}. We argue that this admissible as follows. First, we
construct the \emph{reduction tree} of $M$, which consists of $M$
and all its reducts, with an edge from reduct $M_1$ to reduct
$M_2$ just if $M_1 \red M_2$.  As $M$ has at most finite redexes,
the reduction tree is finitely branching: there can only be a
finite number of terms $M_i$ such that $N \red M_i$ for any term
$N$. Furthermore, if $M$ is strongly normalizing, then the
reduction tree has no infinite paths. By K\"onig's Lemma, the tree
is then finite, and $d(M)$ is the depth of the reduction tree of
$M$---i.e. the longest path in the tree that is rooted at $M$.
Because of this use of K\"onig's Lemma, it is not clear whether
this proof is constructive. We once more only prove the modal
cases.

\begin{description}

  \item[(P4)(c)]
    If $\fixbox{z}{Q} \red P$, then $P \equiv \fixbox{z}{Q'}$ for
    some $Q'$ with $Q \red Q'$. Hence $d(\ibox{M}) \leq d(M)$. But
    the last one is less than or equal to $d(M^\bot)$ by Lemma
    \ref{lem:redcomp}, which is finite by assumption.

  \item[(P5)(c)] 
        First, we note that by substituting $u$ for $Q$, the
        premise implies that $N$ is strongly normalizing, and
        thus that both $d(M)$ and $d(N)$ are finite.

        We now proceed by induction on $d(M) + d(N)$. If
        $\letbox{u}{M}{N} \red P$, then there are three
        possibilities:
        \begin{itemize}
          \item[--]
          $P \equiv \letbox{u}{M'}{N}$ and $M \red M'$. Then \[
            d(M') + d(N) < d(M) + d(N)
          \] and so, by the IH, $P$ is strongly normalizing.

          \item[--]
            Likewise for $N$.

          \item[--]
          $M \equiv \ibox{Q}$ and $P \equiv N[Q/u]$. Then, by
          assumption, $P$ is strongly normalizing.
        \end{itemize} In all cases, if $\letbox{u}{M}{N} \red{}
        P$, then $P$ is strongly normalizing. We conclude that
        the original term itself is strongly normalizing.
\end{description}

\section{Modal category theory}
  \label{sec:modalcats}

We will now introduce a modest amount of monoidal category theory
that is necessary for the formulation of a categorical semantics.
This is needed because we will model the modality by a
\emph{strong monoidal endofunctor}. In our case the monoidal
product will always be the cartesian product of a cartesian closed
category. We will show that this coincides with the notion of
\emph{product-preserving endofunctor}, and hence gives rise to the
isomorphism $\Box (A \times B) \cong \Box A \times \Box B$, which
is another way of stating the modal axiom \textsf{K}
(\S\ref{sec:axioms}).

We assume some familiarity with the relationship between
$\lambda$-calculi and cartesian closed categories (CCCs). A
category $\mathcal{C}$ with finite products is a CCC whenever for
each pair $A, B \in \mathcal{C}$ there is an object $B^A \in
\mathcal{C}$ and a morphism $\mathsf{ev}_{A, B} : B^A \times A
\rightarrow B$ such that for every $f : C \times A \rightarrow B$
there is a unique $\lambda(f) : C \rightarrow B^A$ such that
$\textsf{ev}_{A, B} \circ (\lambda(f) \times id_A) = f$.  For
further background on CCCs and the typed $\lambda$-calculus, we
refer the reader to \cite{Lambek1988, Crole1993, Awodey2010,
Abramsky2011a}. Some material on monoidal category theory is drawn
from the superbly lucid treatment by Melli\`{e}s \cite[\S
5]{Mellies2009}, which is specifically geared towards categorical
logic. Further material can be found in \cite[\S
XI.2]{MacLane1978}. 

\subsection{Lax and strong monoidal functors}

Let $\mathcal{C}$ and $\mathcal{D}$ be cartesian categories. We
regard them as monoidal categories $(\mathcal{C}, \times,
\mathbf{1})$ and $(\mathcal{D}, \times, \mathbf{1})$ respectively.

\begin{defi}
  A functor $F : \mathcal{C} \longrightarrow \mathcal{D}$ between
  two cartesian categories is \emph{lax monoidal} just if it is
  equipped with a natural transformation $m : F(-) \times F(-)
  \Rightarrow F(- \times -)$ as well as an arrow $m_0 : \mathbf{1}
  \rightarrow F(\mathbf{1})$ such that the following diagrams
  commute:
  \begin{equation}
    \label{eq:laxmon1}
    \begin{tikzcd}
      (FA \times FB) \times FC
       	\arrow[r, "\alpha"]
        \arrow[d, "m_{A, B} \times id_{FC}", swap]
      & FA \times (FB \times FC)
        \arrow[d, "id_{FA} \times m_{B, C}"] \\
      F(A \times B) \times FC
        \arrow[d, "m_{A \times B, C}", swap]
      & FA \times F(B \times C)
        \arrow[d, "m_{A, B \times C}"] \\
      F((A \times B) \times C)
        \arrow[r, "F\alpha", swap]
      & F(A \times (B \times C))
    \end{tikzcd}
  \end{equation} \begin{equation}
    \label{eq:laxmon2}
    \begin{tikzcd}
    FA \times \mathbf{1}
      \arrow[r, "\rho_A"]
      \arrow[d, "id_{FA} \times m_0", swap]
    & FA \\
    FA \times F\mathbf{1}
      \arrow[r, "m_{A, \mathbf{1}}", swap]
    & F(A \times \mathbf{1})
      \arrow[u, "F\rho_A", swap]
    \end{tikzcd} \quad
    \begin{tikzcd}
    \mathbf{1} \times FB
      \arrow[r, "\lambda_B"]
      \arrow[d, "m_0 \times id_{FB}", swap]
    & FB \\
    F\mathbf{1} \times FB
      \arrow[r, "m_{\mathbf{1}, B}", swap]
    & F(\mathbf{1} \times B)
      \arrow[u, "F\lambda_B", swap]
    \end{tikzcd}
  \end{equation}
\end{defi}

\begin{defi}
  A \emph{strong monoidal functor} between two cartesian
  categories is a lax monoidal functor where the components $m_{A,
  B} : FA \times FB \rightarrow F(A \times B)$ and the arrow $m_0:
  \mathbf{1} \rightarrow F\mathbf{1}$ are isomorphisms.
\end{defi}

These natural transformations can be extended to arbitrary
contexts. We write 
\[
  \prod_{i = 1}^n A_n 
    \defeq
  A_1 \times \dots \times A_n
\] 
where $\times$ associates to the left. Then, we define the
following morphisms by induction: 
\begin{align*}
  m^{(0)} &\defeq \mathbf{1} \xrightarrow{m_0} F\mathbf{1} \\
  m^{(1)} &\defeq FA_1 \xrightarrow{id_{FA_1}} FA_1 \\
  m^{(n+1)} &\defeq
    \prod_{i = 1}^{n+1} FA_i
      \xrightarrow{m^{(n)} \times id }
    F\left(\prod_{i = 1}^{n} A_i\right) \times FA_{n+1}
      \xrightarrow{m}
    F\left(\prod_{i = 1}^{n+1} A_i\right)
\end{align*} 
Then the $m^{(n)} : \prod_{i=1}^n FA_n \rightarrow
F\left(\prod_{i=1}^n A_i\right)$'s are a natural transformation,
i.e.
\[
  m^{(n)} \circ \prod_{i=1}^n Ff_i
  = F\left(\prod_{i=1}^n f_i\right) \circ m^{(n)}
\]

\noindent We also note that if $F : \mathcal{C} \longrightarrow
\mathcal{C}$ is a monoidal endofunctor, then so is $F^2 \defeq F
\circ F$, with \[
  n_{A, B} \defeq
    F^2 A \times F^2 B
      \xrightarrow{m_{A, B}}
    F(FA \times FB)
      \xrightarrow{Fm_{A, B}}
    F^2 (A \times B)
\] 
and $n_0 \defeq Fm_0 \circ m_0$. See \cite[\S 5.9]{Mellies2009}.

\subsection{Product-Preserving Functors}

Lax and strong monoidal functors are widely used as morphisms
between monoidal categories. However, our monoidal product will
always be the cartesian product, so it is worth examining how
these notions adapt to this particular setting. To begin, we should
compare them to another kind of morphism between cartesian
categories that `plays well with products,' namely that of
\emph{product-preserving functors}.

\begin{defi}
  A \emph{product-preserving functor} $F : \mathcal{C}
  \longrightarrow \mathcal{D}$ between two cartesian categories is a
  functor for which the canonical arrows 
  \begin{align*}
    p_{A, B} &\defeq \langle F\pi_1, F\pi_2 \rangle 
      : F(A \times B) \xrightarrow{\cong} FA \times FB 
      &
    p_0      &\defeq {!}_{F\mathbf{1}}
      : F\mathbf{1} \xrightarrow{\cong} \mathbf{1}
  \end{align*} 
  are isomorphisms. 
\end{defi}

This definition appears to be much stronger. Indeed,
product-preserving functors are strong monoidal. To show that, all
we need to do is show that the inverses 
\begin{align*}
  m_{A, B} &\defeq p^{-1}_{A, B} 
    : FA \times FB \xrightarrow{\cong} F(A \times B) &
  m_0      &\defeq p^{-1}_0 
    : \mathbf{1} \xrightarrow{\cong} F\mathbf{1}
\end{align*} 
satisfy the coherence conditions \eqref{eq:laxmon1} and
\eqref{eq:laxmon2}. Before we do that, we note two very useful
equations that product-preserving functors satisfy. The first is

\begin{prop}
  \label{prop:prodpresangle}
  If $F$ is product-preserving then $ m_{A,B} \circ \langle Ff, Fg
  \rangle = F\langle f, g \rangle$.
\end{prop}
\begin{proof}
  Calculate that $p_{A, B} \circ F\langle f, g \rangle = \langle
  Ff, Fg \rangle$ and notice $p^{-1}_{A, B} = m_{A, B}$.
\end{proof}

The second equation shows how $m_{A,B}$ may be used to relate
projections.

\begin{prop}
  \label{prop:prodpresproj}
  Let $F : \mathcal{C} \longrightarrow \mathcal{D}$ be
  product-preserving, and let $
    A \xleftarrow{\pi^{A, B}_1} 
      A \times B
    \xrightarrow{\pi^{A, B}_2} B
  $ and $
    FA \xleftarrow{\pi^{FA, FB}_1}
      FA \times FB
    \xrightarrow{\pi^{FA, FB}_2} FB
  $ be product diagrams in $\mathcal{C}$ and $\mathcal{D}$
  respectively. Then \[
    F\pi^{A, B}_i \circ m_{A, B} = \pi^{FA, FB}_i
  \]
\end{prop}

\begin{proof}
  Calculate that $\pi_i \circ m^{-1}_{A, B} = \pi_i \circ \langle F\pi_1,
  F\pi_2 \rangle = F\pi_i$ and notice $p^{-1}_{A, B} = m_{A, B}$.
\end{proof} 

\noindent We will often write such equations as $F\pi_1 \circ m =
\pi_1$ without further ado. Armed with these facts, we see that
the definitions of $m_{A, B}$ and $m_0$ given above are natural,
and that 
\begin{thm}
  Any product-preserving functor is strong monoidal.
\end{thm}

\opt{th}{
\begin{proof}
  For $f : C \rightarrow A$ and $g : D \rightarrow B$, we
  calculate that
    \begin{derivation}
      m_{A, B} \circ (Ff \times Fg)
	\since{definition}
      m_{A, B} \circ \langle Ff \circ \pi_1, Fg \circ \pi_2 \rangle
	\since{Proposition \ref{prop:prodpresproj}}
      m_{A, B} \circ \langle Ff \circ F\pi_1 \circ m_{C, D},
			     Fg \circ F\pi_2 \circ m_{C, D} \rangle
	\since{functoriality of F, naturality of product}
      m_{A, B} \circ \langle F(f \circ \pi_1),
			     F(g \circ \pi_2) \rangle \circ m_{C, D}
	\since{Proposition \ref{prop:prodpresangle}, definition}
      F(f \times g) \circ m_{C, D}
    \end{derivation}
  so that $ m : F(-) \times F(-) \Rightarrow F(- \times -)$ is a
  natural transformation. The associativity diagram commutes: the
  proof is a lengthy but simple calculation involving the
  naturality of the product arrow, the definition $\alpha \defeq
  \langle \pi_1 \pi_1, \langle \pi_2 \pi_1, \pi_2 \rangle
  \rangle$, and---more crucially---the invertibility of the
  $m_{A,B}$'s. Commutation of the other two diagrams follows from
  Proposition \ref{prop:prodpresproj} and the observation that
  $\rho_A \defeq \pi_1$ and $\lambda_B \defeq \pi_2$.
\end{proof}
}

\noindent Rather strikingly, the converse holds as well.

\begin{thm}
  Any functor that is strong monoidal with respect to the
  cartesian structure is product-preserving.
\end{thm}

\begin{proof}
  Note that $m_0^{-1} : F\mathbf{1} \longrightarrow \mathbf{1}$ is
  necessarily equal to the unique arrow ${!}_{F\mathbf{1}} :
  F\mathbf{1} \longrightarrow \mathbf{1}$ to the terminal object
  $\mathbf{1}$. Hence, it suffices to show that for any $A, B
  \in \mathcal{C}$, $m_{A,B}^{-1} = \langle F\pi_1, F\pi_2
  \rangle$. We will first show a particular case, viz. that 
  \[
    m^{-1}_{A, \mathbf{1}} = \langle F\pi_1, F\pi_2 \rangle
  \] 
  from which the general case will follow. Remembering that
  $\rho_A \defeq \pi_1 : A \times \mathbf{1} \rightarrow A$, we
  have that $\rho^{-1}_A = \langle id_{FA}, {!}_{FA} \rangle : FA
  \rightarrow FA \times \mathbf{1}$. Hence, by reversing $\rho_A$
  and $m_{A, \mathbf{1}}$ in the first diagram of
  (\ref{eq:laxmon2}) we obtain \[
    m_{A, \mathbf{1}}^{-1}
      =
    (id_A \times m_0) \circ 
        \langle id_{FA}, {!}_{FA} \rangle 
        \circ F\pi_1 
      =
    \langle F\pi_1,
            m_0 \circ {!}_{F(A \times \mathbf{1})}
    \rangle
    : F(A \times \mathbf{1}) \rightarrow FA \times F\mathbf{1}
  \] But, as $m_0 : \mathbf{1} \xrightarrow{\cong} F\mathbf{1}$,
  $F\mathbf{1}$ is also a terminal object, and any arrow into it
  is of the form $m_0 \circ {!}_A : A \rightarrow F\mathbf{1}$.
  This applies to $F\pi_2 : F(A \times \mathbf{1}) \rightarrow
  F\mathbf{1}$, so $
    m^{-1}_{A, \mathbf{1}} = \langle F\pi_1, F\pi_2 \rangle 
  $.

  Now for the general case. As $m_{A, B}$ is a natural
  isomorphism, its inverse is a natural transformation with
  components $m^{-1}_{A, B}$. The naturality square for $(id_A,
  {!}_B)$ is \[
    \begin{tikzcd}
      F(A \times B)
	\arrow[r, "m_{A, B}^{-1}"]
	\arrow[d, "F(id_A \times {!}_B)", swap]
      & FA \times FB
	\arrow[d, "id_{FA} \times F({!}_B)"]
	\\
      F(A \times \mathbf{1})
	\arrow[r, "m_{A, \mathbf{1}}^{-1}", swap]
      & FA \times F\mathbf{1}
    \end{tikzcd}
  \] Calculating down and across gives \[
    m^{-1}_{A, \mathbf{1}} \circ F(id_A \times {!}_B)
    = \langle F\pi_1, F\pi_2 \rangle \circ F(id_A \times {!}_B)
    = \langle F\pi_1, F({!}_B \circ \pi_2) \rangle
  \] whereas across and down gives \[
    (id_{FA} \times F({!}_B)) \circ m^{-1}_{A, B}
    = \langle \pi_1 \circ  m^{-1}_{A, B},  F({!}_B) \circ \pi_2
      \circ m^{-1}_{A, B} \rangle
  \] The first two components of these should be equal, therefore
  $\pi_1 \circ m^{-1}_{A, B} =  F\pi_1$. Similarly, $\pi_2 \circ
  m^{-1}_{A, B} =  F\pi_2$, and hence $m^{-1}_{A, B} = \langle
  F\pi_1, F\pi_2 \rangle$.
\end{proof}

\subsection{Monoidal natural transformations}

The standard notion of natural transformation between lax monoidal
functors is the following.

\begin{defi}
  Let $F, G : \mathcal{C} \longrightarrow \mathcal{D}$ be two lax
  monoidal functors between two cartesian categories.  A
  \emph{monoidal natural transformation} between $F$ and $G$ is a
  natural transformation $\alpha : F \Rightarrow G$ such that the
  following diagrams commute:
  \[
    \begin{tikzcd}
        FA \times FB
          \arrow[r, "\alpha_A \times \alpha_B"]
          \arrow[d, "m_{A, B}", swap]
      & GA \times GB
          \arrow[d, "n_{A, B}"] \\
      F(A \times B)
          \arrow[r, "\alpha_{A \times B}", swap]
      & G(A \times B)
    \end{tikzcd} \quad \begin{tikzcd}
      \mathbf{1}
        \arrow[rd, "n_0"]
        \arrow[d, "m_0", swap]
      & \\
      F\mathbf{1}
        \arrow[r, "\alpha_\mathbf{1}", swap]
      & G\mathbf{1}
    \end{tikzcd}
  \]
\end{defi}

\noindent Surprisingly, it is not hard to show that

\begin{thm}
  \label{thm:monoidal}
  If $F, G : \mathcal{C} \longrightarrow \mathcal{D}$ are
  product-preserving functors between two cartesian categories,
  then any natural transformation $\alpha : F \Rightarrow G$ is
  a monoidal natural transformation.
\end{thm}
\begin{proof}
  We trivially have ${!}_{G\mathbf{1}} \circ \alpha_\mathbf{1} =
  {!}_{F\mathbf{1}}$. But ${!}_{G\mathbf{1}}$ and
  ${!}_{F\mathbf{1}}$ are isomorphisms, so---by inverting
  them---we obtain $\alpha_\mathbf{1} \circ m_0 = n_0$.
  Furthermore, we have the following naturality diagram: \[
    \begin{tikzcd}
      F(A \times B)
        \arrow[r, "F\pi_1"]
        \arrow[d, "\alpha_{A \times B}", swap]
      & FA
        \arrow[d, "\alpha_A"] \\
      G(A \times B)
        \arrow[r, "G\pi_1", swap]
      & GA
    \end{tikzcd}
  \] and a similar one for $B$. Hence, $n_{A, B}^{-1} \circ
  \alpha_{A \times B} \defeq \langle G\pi_1, G\pi_2 \rangle \circ
  \alpha_{A \times B}$ is equal to \[
    \langle G\pi_1 \circ \alpha_{A \times B}, 
            G\pi_2 \circ \alpha_{A \times B} \rangle
      =
    \langle \alpha_A \circ F\pi_1, \alpha_B \circ F\pi_2 \rangle
      =
    (\alpha_A \times \alpha_B) \circ  \langle F\pi_1, F\pi_2 \rangle
  \] which is just $(\alpha_A \times \alpha_B) \circ m_{A,
  B}^{-1}$. It then suffices to invert $m_{A, B}$ and $n_{A, B}$.
\end{proof}

\subsection{The categorical interpretation of modal rules}

In this section we introduce the main structures needed to produce
categorical models for our calculi. We begin with the basic two
examples of Kripke categories (\textsf{K}), and Bierman--de Paiva
categories (\textsf{S4}). These are the most well-behaved, and
most commonly encountered cases. We then discuss the slightly more
obscure cases of Kripke-4 categories (\textsf{K4}), Kripke-T
categories (\textsf{T}), and finally G\"odel-L\"ob categories
(\textsf{GL}).

\subsubsection{Kripke categories}

The combination of a CCC with a product-preserving endofunctor is
the quintessential structure in our development, so we give it a
name.

\begin{defi}
 A \emph{Kripke category} $(\mathcal{C}, \times, \mathbf{1}, F)$
 is a cartesian closed category $\mathcal{C}$ along with a
 \emph{product-preserving endofunctor} $F : \mathcal{C}
 \longrightarrow \mathcal{C}$.
\end{defi}

\noindent Kripke categories are the minimal setting in which one
can model Scott's rule (see \S \ref{sec:admissible}), by defining
an operation
\[
  (-)^\bullet : \mathcal{C}\left(\prod_{i = 1}^n A_i, B\right)
    \rightarrow \mathcal{C}\left(\prod_{i = 1}^n FA_i, FB\right)
\] by \[
  f : \prod_{i = 1}^n A_i \rightarrow B\
    \quad \longmapsto \quad
  f^\bullet 
    \defeq
        \prod_{i = 1}^n FA_i
      \xrightarrow{m^{(n)}}
        F\left(\prod_{i = 1}^n A_i\right)
      \xrightarrow{Ff}
        FB
\]

\noindent The operation $(-)^\bullet$ satisfies the following
distribution/naturality laws.

\begin{prop} \hfill
  \label{prop:kripkedist}
  \begin{enumerate}
    \item
      Let $f : \prod_{i = 1}^n B_i \rightarrow C$ and $g_i : \prod_{j
      = 1}^k A_j \rightarrow B_i$ for $i = 1, \dots, n$.
      Then \[
        \left(f \circ \langle \vct{g_i} \rangle\right)^\bullet
        = f^\bullet \circ \left\langle \vct{g^\bullet_i} \right\rangle
      \]
    \item
      For $f : \prod_{i=1}^n A_i \rightarrow B$ and $\langle
      \vec{\pi_j} \rangle : \prod_{i=1}^n FA_i \rightarrow
      \prod_{j \in J} FA_j$ for $J$ a list drawn from $\{1, \dots,
      n\}$, \[
        \left(f \circ \langle \vec{\pi_j} \rangle\right)^\bullet
          =
        f^\bullet \circ \langle \vec{\pi_j} \rangle
      \]
  \end{enumerate}
\end{prop}

As product-preserving endofunctors abound in the literature on
category-theory, we do not see any use in giving examples of
Kripke categories. Rather, their omnipresence is an attestation of
the importance of our system as a candidate internal language.

\subsubsection{Bierman-de Paiva categories}

In order to model \textsf{S4}, we need a Kripke category whose
product-preserving functor is additionally a \emph{comonad}.

\begin{defi} 
  A comonad $(F, \epsilon, \delta)$ consists of an endofunctor $F
  : \mathcal{C} \longrightarrow \mathcal{C}$, and two natural
  transformations \[
    \epsilon : F \Rightarrow \mathsf{Id}, \qquad
    \delta : F \Rightarrow F^2
  \] such that the following diagrams commute: \[
    \begin{tikzcd}
      FA
        \arrow[r, "\delta_A"]
        \arrow[d, "\delta_A", swap]
      & F^2 A 
        \arrow[d, "\delta_{FA}"] \\
      F^2 A 
        \arrow[r, "F\delta_A", swap] 
      & F^3 A
    \end{tikzcd} \quad
    \begin{tikzcd}
      FA
        \arrow[r, "\delta_A"]
        \arrow[d, "\delta_A", swap]
        \arrow[rd, "id_{FA}"]
      & F^2 A 
        \arrow[d, "\epsilon_{FA}"] \\
      F^2 A
        \arrow[r, "F\epsilon_A", swap]
      & FA
    \end{tikzcd}
  \]
\end{defi}

\noindent In particular, we will require that the comonad used is
\emph{monoidal}, in that it satisfies some additional coherence
conditions with respect to the monoidality.

\begin{defi}
  A \emph{monoidal comonad} on a cartesian category $\mathcal{C}$
  is a comonad $(F, \epsilon, \delta)$ such that $F : \mathcal{C}
  \longrightarrow \mathcal{C}$ is a lax monoidal functor, and
  $\epsilon : F \Rightarrow \mathsf{Id}$ and $\delta : F
  \Rightarrow F^2$ are monoidal natural transformations.
  Concretely, this amounts to the commutation of
  \begin{equation}
    \label{eq:epsilonmonoidal}
    \begin{tikzcd}
        FA \times FB
        \arrow[r, "\epsilon_A \times \epsilon_B",]
        \arrow[d, "m_{A, B}", swap]
      & A \times B
        \arrow[d, equal] \\
      F(A \times B)
        \arrow[r, "\epsilon_{A \times B}", swap]
      & A \times B
    \end{tikzcd} \quad \begin{tikzcd}
      \mathbf{1}
        \arrow[rd, equal]
        \arrow[d, "m_0", swap]
      & \\
      F\mathbf{1}
        \arrow[r, "\epsilon_\mathbf{1}", swap]
      & \mathbf{1}
    \end{tikzcd}
  \end{equation} \begin{equation}
    \label{eq:deltamonoidal}
    \begin{tikzcd}
        FA \times FB
        \arrow[r, "\delta_A \times \delta_B",]
        \arrow[dd, "m_{A, B}", swap]
      & F^2 A \times F^2 B
        \arrow[d, "m_{FA, FB}"] \\
      & F(FA \times FB) 
        \arrow[d, "Fm_{A, B}"] \\
      F(A \times B)
        \arrow[r, "\delta_{A \times B}", swap]
      & F^2(A \times B)
    \end{tikzcd} \quad \begin{tikzcd}
      \mathbf{1}
        \arrow[rd, "m_0"]
        \arrow[dd, "m_0", swap]
      & & \\
      & F\mathbf{1}
        \arrow[rd, "Fm_0"] 
      & \\
      F\mathbf{1}
        \arrow[rr, "\delta_\mathbf{1}", swap]
      & & F^2\mathbf{1}
    \end{tikzcd}
  \end{equation}
\end{defi}

\noindent However, since the functors that we use are
product-preserving, or strong monoidal, it follows automatically
by Theorem \ref{thm:monoidal} that

\begin{cor}
  If $(F, \epsilon, \delta)$ is a comonad whose functor $F$ is
  product-preserving, then it is a monoidal comonad.
\end{cor}

We shall not hence explicitly worry about monoidality, neither in
this section nor in later ones, and we will use the above
equations without further ado.

\begin{defi}
  A \emph{Bierman-de Paiva category} $(\mathcal{C}, \times,
  \mathbf{1}, F, \epsilon, \delta)$ consists of a Kripke category
  $(\mathcal{C}, \times, \mathbf{1}, F)$ whose functor $F :
  \mathcal{C} \longrightarrow \mathcal{C}$ is part of a comonad
  $(F, \epsilon, \delta)$.
\end{defi}

\noindent Bierman-de Paiva categories (abbrv. BdP categories) are
the minimal setting in which both the Four and Veridicality rules
can be modelled. The Four rule is modelled by something already
well-known in category theory, namely the \emph{co-Kleisli
lifting}: 
\[
  (-)^\ast :
    \mathcal{C}\left(\prod_{i = 1}^n FA_i , B\right)
      \rightarrow
    \mathcal{C}\left(\prod_{i = 1}^n FA_i, FB\right)
\] which is defined as follows: \[
  \begin{prooftree}
    f : 
      \prod_{i = 1}^n FA_i
        \rightarrow
      B
      \justifies
    f^\ast \defeq
      \prod_{i = 1}^n FA_i
        \xrightarrow{\prod_{i=1}^n \delta_{A_i}}
      \prod_{i = 1}^n F^2 A_i
        \xrightarrow{m^{(n)}}
      F\left(\prod_{i = 1}^n FA_i\right)
        \xrightarrow{Ff}
      FB
  \end{prooftree}
\]

\noindent This operation interacts in a useful manner with the 
transformations $\delta$ and $\epsilon$.

\begin{prop} \hfill
  \label{prop:deltaepsilonast}
  \begin{enumerate}
    \item
      Let $f : \prod_{i = 1}^n FA_i \rightarrow B$. Then
      $\delta_B \circ f^\ast = \left(f^\ast\right)^\ast$.
    \item
      Let $f : \prod_{i=1}^n FA_i \rightarrow B$. Then
      $\epsilon_B \circ f^\ast = f$.
  \end{enumerate}
\end{prop}

\opt{th}{
\begin{proof} \hfill
  \begin{enumerate}
    \item
      Let $E \defeq  \prod_{i = 1}^n FA_i$.  Then 
      \begin{derivation}
          \delta_B \circ f^\ast
        \since{definition}
          \delta_B 
            \circ Ff
            \circ m^{(n)}
            \circ \prod_{i=1}^n \delta_{A_i}
        \since{$\delta$ natural}
          F^2 f
            \circ \delta_E 
            \circ m^{(n)}
            \circ \prod_{i=1}^n \delta_{A_i}
        \since{monoidal equation for $\delta$}
          F^2 f
            \circ Fm^{(n)}
            \circ m^{(n)}
            \circ \prod_{i=1}^n \delta_{FA_i}
            \circ \prod_{i=1}^n \delta_{A_i}
        \since{product is functorial}
          F^2 f 
            \circ Fm^{(n)}
            \circ m^{(n)}
            \circ \prod_{i=1}^n \delta_{FA_i}\delta_{A_i}
        \since{ equation of comonads }
          F^2 f
            \circ Fm^{(n)}
            \circ m^{(n)}
            \circ \prod_{i=1}^n F\delta_{A_i}\delta_{A_i}
        \since{product functorial, $F$ product-preserving}
          F^2 f 
            \circ Fm^{(n)}
            \circ F\left(\prod_{i=1}^n \delta_{A_i}\right)
            \circ m^{(n)}
            \circ \prod_{i=1}^n \delta_{A_i}
        \since{$F$ functor, definitions}
          \left(f^\ast\right)^\ast
      \end{derivation}
    \item
      Straightforward calculation involving---amongst other
      things---the naturality and monoidality of $\epsilon$.
  \end{enumerate}
\end{proof}
}

\noindent The co-Kleisli extension also satisfies a handful of
very useful naturality/distribution laws.

\begin{prop} \hfill
  \label{prop:astdist}
  \begin{enumerate}
    \item
      $id_{FA}^\ast = \delta_{FA}$
    \item
      $\epsilon_A^\ast = id_{FA}$
    \item
      Let $f : \prod_{i = 1}^n B_i \rightarrow C$ and $g_i : \prod_{j
      = 1}^k FA_j \rightarrow B_i$ for $i = 1, \dots, n$.  Then \[
        \left(f \circ \langle \vct{g_i} \rangle\right)^\ast
         =
        f^\bullet \circ \left\langle \vct{g^\ast_i} \right\rangle
      \]
    \item
      For $k_i : \prod_{j=1}^m FA_j \rightarrow B_j$ and $l :
      \prod_{i=1}^n FB_i \rightarrow C$, \[
        \left(
          l
          \circ
          \left\langle \vct{k_i^\ast}\right\rangle
        \right)^\ast
          =
        l^\ast \circ \left\langle \vct{k_i^\ast}\right\rangle
      \]
    \item
      For $f : \prod_{i=1}^n FA_i \rightarrow B$ and $\langle
      \vec{\pi_j} \rangle : \prod_{i=1}^n FA_i \rightarrow
      \prod_{j \in J} FA_j$ for $J$ a list with elements from
      $\{1, \dots, n\}$,
      \[
        \left(f \circ \langle \vec{\pi_j} \rangle\right)^\ast
          =
        f^\ast \circ \langle \vec{\pi_j} \rangle
      \]
  \end{enumerate}
\end{prop}
\begin{proof}
  (1) and (2) are standard comonad equations. (3) is a
  straightforward calculation, similar to Proposition
  \ref{prop:kripkedist}(1). (4) follows from (3), Proposition
  \ref{prop:deltaepsilonast}(1), and $f^\ast \defeq f^\bullet
  \circ \prod \delta$. (5) is a corollary of (3), once we notice
  that $\pi_i^\ast = \delta_{A_i} \circ \pi_i$, and use $f^\ast
  \defeq f^\bullet \circ \prod \delta$. 
\end{proof}

Product-preserving comonads are also often encountered in the
category-theoretic literature, so we refrain from sketching any
examples of Bierman-de Paiva categories.

\subsubsection*{Idempotence}

It is interesting to separately consider those BdP categories for
which the comonad $(F, \epsilon, \delta)$ is \emph{idempotent},
i.e. those for which $\delta : F \Rightarrow F^2$ is an
\emph{isomorphism}. There are many equivalent ways to define
idempotence: see \cite[\S 4.3.2]{Borceux1994}. One of them is that
$\delta_{FA} \circ \epsilon_{FA} = id_{F^2 A}$ for each object
$A$. Here, we will use the equation $F\epsilon_A = \epsilon_{FA}$
for each object $A$. Restated in our notation, it becomes \[
\epsilon_A^\bullet = \epsilon_{FA} : F^2 A \rightarrow FA \] 

\begin{prop}
  \label{prop:astidemepsilon}
  If $(F, \epsilon, \delta)$ is idempotent, then for $f :
  \prod_{i=1}^n FA_i \rightarrow FB$ we have
  \[
    (\epsilon_B \circ f)^\ast = f
  \] 
\end{prop}
\begin{proof}
  By Proposition \ref{prop:astdist}(3) and
  \ref{prop:deltaepsilonast}(2),
  $
    (\epsilon \circ f)^\ast = \epsilon^\bullet \circ f^\ast =
    \epsilon \circ f^\ast = f
  $ 
\end{proof}

\noindent This situation creates an additional sort of naturality
for $(-)^\ast$, essentially by making Proposition
\ref{prop:astdist}(4) universally applicable.

\begin{prop}
  \label{prop:astidemsubst}
  For any $f : \prod_{j=1}^m FB_j \rightarrow FC$, and any $k_j :
  \prod_{i=1}^n FA_i \rightarrow FB_j$, we have \[
    \left(f \circ \left\langle \vct{k_i} \right\rangle
    \right)^\ast = f^\ast \circ \left\langle \vct{k_i} \right\rangle
  \]
\end{prop}
\begin{proof}
  Using Propositions \ref{prop:astidemepsilon} and
  \ref{prop:astdist}(4), \[
      \left(f \circ \left\langle \vct{k_j} \right\rangle
      \right)^\ast
    =
      \left(f \circ \left\langle \vct{\left(\epsilon_{B_j} \circ
      k_j\right)^\ast} \right\rangle \right)^\ast
    =
      f^\ast \circ \left\langle \vct{\left(\epsilon_{B_j} \circ
      k_j\right)^\ast} \right\rangle
    =
      f^\ast \circ \left\langle \vct{k_j} \right\rangle
      \tag*{\qedhere}
  \]
\end{proof}

\noindent Thus, we have the following characterisation.

\begin{thm}[Idempotence]
  \label{thm:idem}
  $(F, \epsilon, \delta$) is idempotent if and only if the map \[
    (-)^\ast :
      \mathcal{C}\left(\prod_{i = 1}^n FA_i , B\right)
        \rightarrow
      \mathcal{C}\left(\prod_{i = 1}^n FA_i, FB\right)
  \] is an isomorphism, natural with respect to precomposition of
  any $k : \prod_{j = 1}^n FD_j \rightarrow \prod_{i = 1}^n FA_i$.
\end{thm}
\begin{proof}
  For the backwards direction, the inverse of $(-)^\ast$ is
  $\epsilon_B \circ -$. It is a left and right inverse by
  Propositions \ref{prop:deltaepsilonast}(2) and
  \ref{prop:astidemepsilon}. Naturality follows by Proposition
  \ref{prop:astidemsubst} once we write $k = \left\langle
  \vct{\pi_i \circ k} \right\rangle$. For the forwards direction,
  we calculate 
  \[
    \delta_{FA} \circ \epsilon_{FA}
      =
    id^\ast_{FA} \circ \epsilon_{FA}
      =
    (id_{FA} \circ \epsilon_{FA})^\ast
      =
    \epsilon_{FA}^\ast
      =
    id_{F^2 A}
  \] by Prop. \ref{prop:astdist}(1), naturality for $\epsilon_{FA}
  : F^2 A \rightarrow FA$, and Prop. \ref{prop:astdist}(2).
\end{proof}

\subsubsection*{Comparison with Bierman \& de Paiva}

Those familiar with previous literature on the categorical
semantics of \textsf{S4} modalities will no doubt ask about the
relationship of BdP categories with the models discussed by
Bierman and de Paiva \cite[\S 7]{Bierman2000a}. At their core, the
models of the system with delayed substitutions that we discussed
in \S\ref{sec:depaiva} also consist of a monoidal comonad.
However, the natural transformation $m : F(-) \times F(-)
\Rightarrow F(- \times -)$ is not required to be invertible, and
so $F$ is only lax monoidal. As a result, one must also explicitly
require the coherence equations \eqref{eq:epsilonmonoidal} and
\eqref{eq:deltamonoidal}, which---as we mentioned
above---hold automatically whenever $F$ preserves products.

In the penultimate section of their paper, Bierman and de Paiva
\cite[\S 10]{Bierman2000a} briefly discuss a slightly stronger
notion of model, which they attribute to Schalk. Therein the
morphisms $\langle id_{FA}, id_{FA} \rangle : FA \rightarrow FA
\times FA$ and ${!}_{FA} : FA \rightarrow \mathbf{1}$ are
homomorphisms for the coalgebras $\delta_A : FA \rightarrow F^2 A$
and $m \circ (\delta_A \times \delta_A) : FA \times FA \rightarrow
F^2 A \times F^2 A \rightarrow F(FA \times FA)$. These equations
allow one to prove the soundness of two additional commuting
conversions, which embody certain structural rules for the delayed
substitutions: they allow one to \emph{weaken} by `garbage
collecting' a delayed substitution for a variable that does not
occur, and to \emph{contract} two identical delayed substitutions.
Indeed, these requirements also reappear in models of linear logic
known as \emph{linear categories} \cite[\S 7.4]{Mellies2009}.

Curiously, our notion of model is even stronger: it is easy to
calculate (using monoidality and naturality of $\delta$, product
preservation, and the invertibility of $Fm$) that the
aforementioned morphisms are automatically coalgebra morphisms in
the product-preserving setting. Indeed, two commuting conversions
similar to the ones mentioned above, here called
$(\textsf{commweak})$ and $(\textsf{commcontr})$, will be
necessary to prove completeness.

\subsubsection{Kripke-4 categories}

Kripke-4 categories model \textsf{K4}. They are essentially `half a
comonad,' and only come with a comultiplication $\delta$. We still
require that one of the comonadic equations, viz. the one that only
refers to $\delta$, holds.

\begin{defi}
  \label{def:kripke-4}
  A \emph{Kripke-4} category $(\mathcal{C}, \times, \mathbf{1}, F,
  \delta)$ is a Kripke category $(\mathcal{C}, \times, \mathbf{1},
  F)$ along with a natural transformation $\delta : F \Rightarrow
  F^2$ such that the following diagram commutes: \[
    \begin{tikzcd}
      FA
        \arrow[r, "\delta_A"]
        \arrow[d, "\delta_A", swap]
      & F^2 A 
        \arrow[d, "\delta_{FA}"] \\
      F^2 A 
        \arrow[r, "F\delta_A", swap] 
      & F^3 A
    \end{tikzcd}
  \]
\end{defi}

\noindent We know by Theorem \ref{thm:monoidal} that $\delta : F
\Rightarrow F^2$ is a monoidal natural transformation.

We can model the general version of Four rule
(\ref{sec:admissible}) in Kripke-4 categories, but in a way that
is slightly more involved than the simple co-Kleisli lifting of
Bierman--de Paiva categories.  To see this, let $(\mathcal{C},
\times, \mathbf{1}, F, \delta)$ be a Kripke-4 category, and write
\[
  \prod_{i = 1}^n A_i \times_l \prod_{j = 1}^m B_j
\] to mean the left-associating product
  $A_1 \times \dots \times A_n \times B_1 \times \dots \times
  B_m$.
Also, write \[
  \langle \vct{f_i}, \vct{g_i}, \vct{h_j} \rangle
\] to mean the left-associating mediating morphism $\langle f_1,
\dots, f_n, g_1, \dots, g_m, h_1, \dots, g_p \rangle$. With this
notation we can now define a map of hom-sets \[
  (-)^\# :
    \mathcal{C}\left(\prod_{i = 1}^n FA_i
      \times_l \prod_{i = 1}^n A_i , B\right)
      \rightarrow
    \mathcal{C}\left(\prod_{i = 1}^n FA_i, FB\right)
\] as follows: \[
  \begin{prooftree}
    f : 
        \prod_{i = 1}^n FA_i
          \times_l
        \prod_{i=1}^n A_i
      \rightarrow
        B
    \justifies
        f^\# \defeq
          \prod_{i = 1}^n FA_i
      \xrightarrow{\langle \vct{\delta_{A_i} \pi_i}, \vct{\pi_i} \rangle} 
        \prod_{i = 1}^n F^2 A_i \times_l \prod_{i = 1} A_i
      \xrightarrow{m^{(2n)}}
        F\left(\prod_{i = 1}^n FA_i \times_l \prod_{i = 1}^n A_i\right)
	\xrightarrow{Ff} B
  \end{prooftree}
\]

\noindent Even though it might seem slightly contrived at first
sight, we can show that the $(-)^\#$ operation satisfies some
naturality equations similar to the ones encountered before.

\begin{prop} \hfill
  \label{prop:hashdist}
  \begin{enumerate}
    \item
      Let $f : \prod_{i = 1}^n B_i \rightarrow C$ and $g_i :
      \prod_{j = 1}^k FA_j \times_l \prod_{j = 1}^k A_j
      \rightarrow B_i$ for $i = 1, \dots, n$. Then \[
        \left(f \circ \langle \vct{g_i} \rangle\right)^\#
        = f^\bullet \circ \left\langle \vct{g^\#_i} \right\rangle
      \]
    \item
      Let $J$ be a list with elements from $\{1, \dots, n\}$. Then
      we have \[
        \left(f \circ \langle \vct{\pi_{Fj}}, \vct{\pi_j} \rangle \right)^\# 
          =
        f^\# \circ \langle \vct{\pi_j} \rangle
      \] where $\langle \vct{\pi_{Fj}}, \vct{\pi_j} \rangle :
      \prod_{i=1}^n FA_i \times_l \prod_{i=1}^n A_i \rightarrow
      \prod_{j \in J} FA_j \times_l \prod_{j \in J} A_j$ is the
      projection that `follows $J$ in both contexts'
      $\prod_{i=1}^n FA_i$ and $\prod_{i=1}^n A_i$.
    \item
      For $k : \prod_{i=1}^n FA_i \times_l \prod_{i=1}^n A_i
      \rightarrow B$ and $l : FB \rightarrow C$, then \[
        (l \circ k^\#)^\ast = l^\ast \circ k^\#
      \]
  \end{enumerate}
\end{prop}

\begin{proof}
  Straightforward calculations, similar to Propositions
  \ref{prop:kripkedist} and \ref{prop:astdist}.
\end{proof}

\begin{prop}
  \label{prop:deltahash}
  Let $f : \prod_{i = 1}^n FA_i \times_l \prod_{i=1}^n A_i
  \rightarrow B$. Then \[
    \delta_B \circ f^\#
      =
    \left(f^\#\right)^\ast
  \]
\end{prop}

\opt{th}{
\begin{proof}
  Let
    $E \defeq  \prod_{i = 1}^n FA_i \times_l \prod_{i=1}^n A_i$.
  Then 
  \begin{derivation}
      \delta_B \circ f^\# 
    \since{definition}
      \delta_B \circ
        Ff \circ m^{(2n)} \circ 
        \langle \vct{\delta_{A_i}\pi_i}, \vct{\pi_i} \rangle
    \since{$\delta$ natural}
      F^2 f \circ
        \delta_E \circ m^{(2n)} \circ 
        \langle \vct{\delta_{A_i}\pi_i}, \vct{\pi_i} \rangle
    \since{$\delta$ monoidal}
      F^2 f \circ Fm^{(2n)} \circ m^{(2n)} \circ 
      \left(\prod_{i=1}^n \delta_{FA_i} \times_l \prod_{i=1}^n \delta_{A_i}\right) \circ 
        \langle \vct{\delta_{A_i}\pi_i}, \vct{\pi_i} \rangle
    \since{product after bracket law}
      F^2 f \circ Fm^{(2n)} \circ m^{(2n)} \circ 
        \left\langle
          \vct{\delta_{FA_i}\delta_{A_i}\pi_i},
          \vct{\delta_{A_i}\pi_i}
        \right\rangle
    \since{law pertaining to $\delta$}
      F^2 f \circ Fm^{(2n)} \circ m^{(2n)} \circ 
        \left\langle
          \vct{F\delta_{A_i}\delta_{A_i}\pi_i},
          \vct{\delta_{A_i}\pi_i}
        \right\rangle
    \since{naturality of product morphisms, projections}
      F^2 f \circ Fm^{(2n)} \circ m^{(2n)} \circ 
        \left\langle
          \vct{F\delta_{A_i}\pi_i},
          \vct{\pi_i}
        \right\rangle
	    \circ
        \left\langle
          \vct{\delta_{A_i}\pi_i}
        \right\rangle
    \since{Proposition \ref{prop:prodpresproj}}
      F^2 f \circ Fm^{(2n)} \circ m^{(2n)}
        \circ 
          \left\langle
            \vct{F\delta_{A_i}F\pi_i \circ m^{(n)}},
            \vct{F\pi_i \circ m^{(n)}}
          \right\rangle
        \circ
          \left\langle
            \vct{\delta_{A_i}\pi_i}
          \right\rangle
    \since{naturality of product morphism, $F$ strong monoidal}
      F^2 f \circ Fm^{(2n)}
        \circ 
          F\left(
            \left\langle
              \vct{\delta_{A_i}\pi_i},
              \vct{\pi_i}
            \right\rangle
          \right)
	    \circ
        m^{(n)}
	    \circ
        \left\langle
          \vct{\delta_{A_i}\pi_i}
        \right\rangle
    \since{definitions}
      Ff^\# \circ m^{(n)} \circ \prod_{i=1}^n \delta_{A_i}
  \end{derivation}
\end{proof}
}

When the morphism of type $\prod_{i = 1}^n FA_i \times_l
\prod_{i=1}^n A_i \rightarrow B$ does not depend on the $A_i$,
then the operation $(-)^\#$ can be reduced to $(-)^\ast$.

\begin{prop}
  \label{prop:degen4}
  Let $f : \prod_{i = 1}^n FA_i \rightarrow B$. Then, writing $\pi
  : \prod_{i = 1}^n FA_i \times_l \prod_{i = 1}^n A_i \rightarrow
  \prod_{i = 1}^n FA_i$ for the projection, \[
    \left(f \circ \pi\right)^\# = f^\ast
  \]
\end{prop}

\opt{lmcs}{
\begin{proof}
  Straightforward calculation using Propositions
  \ref{prop:prodpresangle} and \ref{prop:prodpresproj}.
\end{proof}
}

\opt{th}{
\begin{proof}
  We calculate:
  \begin{derivation}
      \left(f \circ \pi \right)^\#
    \since{definition}
        F\left(f \circ \langle \vct{\pi_{FA_i}} \rangle\right) 
      \circ
        m^{(2n)}
      \circ
        \langle \vct{\delta_{A_i} \pi_i}, \vct{\pi_i} \rangle
    \since{functoriality, and Proposition \ref{prop:prodpresangle}}
      Ff \circ m^{(n)}
      \circ
        \langle \vct{F\pi_{FA_i}} \rangle
      \circ
        m^{(2n)}
      \circ
        \langle \vct{\delta_{A_i} \pi_i}, \vct{\pi_i} \rangle
    \since{naturality of product morphism, 
	    and Proposition \ref{prop:prodpresproj}}
        Ff
      \circ
        m^{(n)}
      \circ
        \langle \vct{\pi_{F^2 A_i}} \rangle 
      \circ
        \langle \vct{\delta_{A_i} \pi_i}, \vct{\pi_i} \rangle
    \since{naturality of product morphisms, projections}
        Ff \circ m^{(n)} 
      \circ
        \langle \vct{\delta_{A_i} \pi_i} \rangle
  \end{derivation}
\end{proof}
}

\noindent And, finally, we prove another crucial distribution
property of $(-)^\#$. Namely, if we substitute `the same thing'
for both contexts, the hash distributes as such:

\begin{prop}
  \label{prop:hashsubst}
  Let $f : \prod_{i=1}^n FB_i \times \prod_{i=1}^n B_i \rightarrow
  B$, and $g_i : \prod_{i=1}^n FA_i \times \prod_{i=1}^n A_i
  \rightarrow B_i$. Then, writing $\pi : \prod_{i=1}^n FA_i \times \prod_{i=1}^n A_i
  \rightarrow \prod_{i=1}^n FA_i$ for the projection, \[
    \left(f 
      \circ \langle 
              \vct{g_i^\# \circ \pi}, 
              \vct{g_i}
            \rangle
    \right)^\#
      =
    f^\# \circ \left\langle \vct{g_i^\#} \right\rangle
  \]
\end{prop}
\begin{proof}
  Using Propositions \ref{prop:hashdist}(1), \ref{prop:degen4},
  and \ref{prop:deltahash}, we calculate
  \[
      \left(f 
        \circ \left\langle 
                \vct{g_i^\# \circ \pi},
                \vct{g_i}
              \right\rangle
      \right)^\#
        =
      f^\bullet
        \circ \left\langle 
                \vct{\left(g_i^\# \circ \pi\right)^\#},
                \vct{g_i^\#}
              \right\rangle
        =
      f^\bullet
        \circ \left\langle 
                \vct{\left(g_i^\#\right)^\ast},
                \vct{g_i^\#}
              \right\rangle
        =
      f^\bullet
        \circ \left\langle 
                \vct{\delta \circ g_i^\#},
                \vct{g_i^\#}
              \right\rangle
  \]
\end{proof}

\begin{exa}[The topos of bifurcating trees, part 1]
  \label{exa:bifurcating-1}
  The \emph{topos of trees} \cite{Birkedal2012} is the category
  $\PSH{\omega}$ of presheaves over $\omega \defeq 1 < 2 < \dots$.
  Concretely, each object $X$ of $\PSH{\omega}$ is a diagram of
  sets and functions
  \[
    X_0 
      \xleftarrow{r_0} X_1 
      \xleftarrow{r_1} X_2 
      \xleftarrow{r_2} \dots
  \]
  The idea is that $X$ is a set that is computed over time:
  $X_{i+1}$ contains the elements that can be emitted after
  computing for $i+1$ steps, and $r_i : X_{i+1} \rightarrow X_i$
  \emph{trims} such elements to what they were at $i$ steps. The
  topos of trees is a synthetic model of step-indexed computation,
  as it provides a principled way of reasoning about infinite
  behaviours without ever constructing `completed' objects: the
  latter only appear as global sections $x : \mathbf{1}
  \Rightarrow X$ of an object $X$, i.e. families $(x_i \in X_i)_i$
  compatible under trimming.

  A variation on this idea is the following. Let
  $(\mathbb{B}^\ast, \sqsubseteq)$ be the prefix order on words
  over booleans $\mathbb{B} \defeq \{0, 1\}$. We can then
  construct the \emph{topos of bifurcating trees}, namely the
  presheaf topos $\PSH{\mathbb{B}^\ast}$ over the prefix order.
  Its objects are diagrams 
  \[
    \begin{tikzpicture}[node distance=3cm]
      \node (Xepsilon) {$X_\epsilon$};
      \node (X0) [above right = .5cm of Xepsilon] {$X_0$};
      \node (X1) [below right = .5cm of Xepsilon] {$X_1$};
      \node (X00) [above right = .5cm of X0] {$X_{00}$};
      \node (X01) [right = .5cm of X0] {$X_{01}$};
      \node (X10) [right = .5cm of X1] {$X_{10}$};
      \node (X11) [below right = .5cm of X1] {$X_{11}$};
      \node (dots00) [right = .5cm of X00] {$\ldots$};
      \node (dots01) [right = .5cm of X01] {$\ldots$};
      \node (dots10) [right = .5cm of X10] {$\ldots$};
      \node (dots11) [right = .5cm of X11] {$\ldots$};
      \path[->] (X0) edge node [above] {$l_\epsilon$} (Xepsilon);
      \path[->] (X1) edge node [below] {$r_\epsilon$} (Xepsilon);
      \path[->] (X00) edge node [above] {$l_0$} (X0);
      \path[->] (X01) edge node [below] {$r_0$} (X0);
      \path[->] (X10) edge node [above] {$l_1$} (X1);
      \path[->] (X11) edge node [below] {$r_1$} (X1);
    \end{tikzpicture}
  \]
  Intuitively, each element of $x_w \in X_{w}$ might, in one time
  step, evolve in two ways: to an element $x_{w0} \in X_{w0}$, or
  to an element $x_{w1} \in X_{w1}$, such that $x_w = l_w(x_{w0})
  = r_w(x_{w1})$. This encodes a certain degree of nondeterminism:
  a global section $x : \mathbf{1} \Rightarrow X$ now represents
  an infinite computation that may make a nondeterminsitic choice
  at every tick of the clock.

  We may then perform the following construction: given a presheaf
  $X$ over $\mathbb{B}^\ast$, we construct a presheaf $\Box X$ by
  letting
  \[
    (\Box X)_w \defeq \prod_{v \sqsubset w} X_v
  \]
  where $\sqsubset$ is the \emph{strict} order associated to the
  partial order $\sqsubseteq$. Thus, $(\Box X)_w$ consists of
  families $\{x_v\}_{v \sqsubset w}$ of an element for each word
  $v$ that is a strict prefix of $w$. Whether those families are
  \emph{matching}, i.e. whether for example $l_v(x_{v0}) = x_v$
  whenever $v0 \sqsubset w$ is immaterial: the constructions
  sketched here work whether we read $\prod$ as a categorical
  limit or as a dependent product. This may prove to be yet
  another way in which we may obtain `intensional' models of modal
  logic. The restriction maps
    $l_w : \prod_{v \sqsubset w1} X_{v} \rightarrow \prod_{v \sqsubset w} X_v$ 
  and
    $r_w : \prod_{v \sqsubset w1} X_{v} \rightarrow \prod_{v \sqsubset w} X_v$ 
  are defined by restricting the domain of $\prod$. 

  Furthermore, for each natural transformation $f : X \rightarrow
  Y$ we define
  $\Box f : \Box X \rightarrow \Box Y$ by
  \begin{alignat*}{2}
    (\Box f)_w : \prod_{v \sqsubset w} X_v &\rightarrow &&\prod_{v \sqsubset w} Y_v \\
                 p\                        &\mapsto\    &&\lambda v.\ f_v(p(v))
  \end{alignat*}

  $\PSH{\mathbb{B}^\ast}$ is a Kripke-4 category: we may define a
  $\delta : \Box \Rightarrow \Box^2$ at each $X$ by
  \begin{alignat*}{2}
    \delta_{X, w} : 
      \prod_{v \sqsubset w} X_v &\rightarrow &&\prod_{v \sqsubset w}\prod_{z \sqsubset v} Y_z \\
                            p\  &\mapsto\    &&\lambda v.\ \lambda z.\ p(z)
  \end{alignat*}
  This is well-typed, as $\sqsubset$ is transitive and hence $z
  \sqsubset w$. This bears a strong likeness to Kripke semantics:
  the transitivity of the `Kripke site' $(\mathbb{B}^\ast,
  \sqsubseteq)$ leads to a proof-relevant witness of the
  \textsf{4} axiom! In contrast, the `non-reflexive' flavour of
  $\Box$ means that there is no way natural $\epsilon_X : \Box X
  \rightarrow X$: for each $w \in \mathbb{B}^\ast$, the component
  $\epsilon_{X, w}$ would have type $\prod_{v \sqsubset w} X_v
  \rightarrow X_w$, and there in general no way to produce an
  element of $X_w$ given an element for each prefix of $w$.
\end{exa}

\subsubsection{Kripke-T categories}

The following structure will be used to interpret axiom \textsf{T}.

\begin{defi}
  A \emph{Kripke-T} category $(\mathcal{C}, \times, \mathbf{1}, F,
  \epsilon)$ consists of a Kripke category $(\mathcal{C}, \times,
  \mathbf{1}, F)$ along with a natural transformation \[
    \epsilon : F \Rightarrow \textsf{Id}
  \] 
\end{defi}

\noindent Using Theorem \ref{thm:monoidal} again we see $\epsilon
: F \Rightarrow \textsf{Id}$ is a monoidal natural transformation.
Modelling veridicality rule (\S\ref{sec:admissible}) amounts to
precomposition with the product of a bunch of components of
$\epsilon : F \Rightarrow \textsf{Id}$. This operation interacts
nicely with Scott's rule:

\begin{prop}
  \label{prop:epsilonkripke}
  Let $f : \prod_{i=1}^n A_i \rightarrow B$. Then
  $
    \epsilon_B \circ f^\bullet
      =
    f \circ \prod_{i=1}^n \epsilon_{A_i}
  $.
\end{prop}

\opt{th}{
\begin{proof}
  Let $E \defeq \prod_{i=1}^n A_i$. Then \[
    \epsilon_B \circ f^\bullet
      = 
    \epsilon_B \circ Ff \circ m^{(n)}
      =
    f \circ \epsilon_E \circ m^{(n)}
      =
    f \circ \prod_{i=1}^n \epsilon_{A_i}
  \] by the definition of $(-)^\bullet$, the naturality of
  $\epsilon$, and its being a monoidal transformation.
\end{proof}
}

\subsubsection{G\"odel-L\"ob categories}

We are looking for a setting where L\"ob's rule can be modelled.
This is not as natural as the previous ones: it is closely
modelled after the syntax of our calculus, and its definition will
involve all of the operations $(-)^\bullet$, $(-)^\#$, and
$(-)^\ast$.

We begin by defining the following central notion.

\begin{defi}[Modal Fixed Point]
  Let $(\mathcal{C}, \times, \mathbf{1}, F, \delta)$ be a
  Kripke-4 category.  
  \begin{enumerate}
    \item
    A \emph{modal fixed point} of $f : \prod_{i=1}^n FB_i \times
    \prod_{i=1}^n B_i \times FA \rightarrow A$ is an arrow
    \[
      f^\dagger : \prod_{i=1}^n FB_i \rightarrow FA
    \] 
    such that the following diagram commutes:
    \[
      \begin{tikzcd}[column sep=large]
        \prod_{i=1}^n FB_i
          \arrow[r, "\langle id^\#{,}
                             \left(f^\dagger\right)^\ast \rangle"]
          \arrow[d, "f^\dagger", swap]
        & F\left(\prod_{i=1}^n FB_i \times \prod_{i=1}^n B_i\right) \times F^2 A
          \arrow[dl, "f^\bullet"] \\
        FA
        & 
      \end{tikzcd}
    \]
  \item
   An object $A \in \mathcal{C}$ \emph{has modal fixed points}
   just if for any $B_i \in \mathcal{C}$ there is a homset map 
   \[
    (-)^\dagger_{\vct{B}} : 
      \mathcal{C}\left(\prod_{i=1}^n FB_i \times
                  \prod_{i=1}^n B_i \times FA, A\right)
        \rightarrow
      \mathcal{C}\left(\prod_{i=1}^n FB_i, FA\right)
    \] 
    such that $f^\dagger_{\vct{B}}$ is a modal fixed point of each
    $f : \prod_{i=1}^n FB_i \times \prod_{i=1}^n B_i \times FA
    \rightarrow A$.
  \end{enumerate}
\end{defi}

We will often write $f^\dagger$ for the modal fixed point of $f$,
dropping the subscript entirely.  This is an \emph{external}
specification of modal fixed points, in the sense that they are
given as a map on the homsets of the Kripke-4 category. We might
instead consider an \emph{internal} specification, i.e. through an
appropriate notion of a \emph{modal fixed point combinator}. This
will---unsurprisingly---be an arrow $F(A^{FA}) \rightarrow FA$,
which is the type of the G\"odel-L\"ob axiom $\Box(\Box A
\rightarrow A) \rightarrow \Box A$. This kind of combinator comes
in two varieties.

\begin{defi}
  Let $(\mathcal{C}, \times, \mathbf{1}, F, \delta)$ be a
  Kripke-4 category. 
  \begin{enumerate}
    \item
      A \emph{strong modal fixed point combinator at $A \in
      \mathcal{C}$} is an arrow \[
        Y_A : F(A^{FA}) \rightarrow FA
      \] such that the following diagram commutes: \[
        \begin{tikzcd}
          F(A^{FA})
            \arrow[r, "{\langle id, Y_A^\ast \rangle}"]
            \arrow[d, "Y_A", swap]
          & F(A^{FA}) \times F^2 A
            \arrow[ld, "\textsf{ev}^\bullet"] \\
          FA
        \end{tikzcd}
      \]
    \item 
      A \emph{weak modal fixed point combinator at $A \in
      \mathcal{C}$} is an arrow
      \[
        Y_A : F(A^{FA}) \rightarrow FA
      \] 
      such that for each $B$ and $f : \prod_{i=1}^n FB_i \times
      \prod_{i=1}^n B_i \times FA \rightarrow A$, the composite
      \[
        \prod_{i=1}^n FB_i
          \xrightarrow{\left(\lambda(f)\right)^\#}
        F(A^{FA})
          \xrightarrow{Y_A}
        FA
      \] 
      is a modal fixed point of $f$.
  \end{enumerate}
\end{defi}

We can prove that having a modal fixed point combinator at $A$ is
equivalent to having modal fixed points at $A$. But to do so we
will need a lemma concerning cartesian closure.

\begin{lem}
  \label{lem:evlob}
  If $f : \prod_{i=1}^n FA_i \times \prod_{i=1}^n A_i \times FB
  \rightarrow B$ and $a : \prod_{i=1}^n FA_i \rightarrow F^2 A$, then
  \[
    \textsf{ev}^\bullet \circ \langle \left(\lambda f\right)^\#,
    a \rangle
      =
    f^\bullet \circ \langle id^\#, a \rangle
  \]
\end{lem}

\opt{th}{
\begin{proof}
  Calculation:
  \begin{derivation}
      \textsf{ev}^\bullet \circ \langle \left(\lambda f\right)^\#,
          a \rangle
    \since{definitions}
      F\textsf{ev} \circ m
        \circ \left\langle
                F(\lambda f) \circ m \circ \langle \vct{\delta\pi_i}, \vct{\pi_i} \rangle,
                a
              \right\rangle
    \since{product equation, definition of $(-)^\ast$}
      F\textsf{ev} \circ m
        \circ \left(F(\lambda f) \times id \right)
        \circ \left\langle
                m \circ \langle \vct{\delta\pi_i}, \vct{\pi_i} \rangle,
                a
              \right\rangle
    \since{$m$ natural}
      F\textsf{ev} 
        \circ F\left(\lambda f \times id \right)
        \circ m
        \circ \left\langle
                m \circ \langle \vct{\delta\pi_i}, \vct{\pi_i} \rangle,
                a
              \right\rangle
    \since{cartesian closure, definition of $(-)^\bullet$ and $(-)^\#$}
      f^\bullet
        \circ \left\langle
                id^\#,
                a
              \right\rangle
  \end{derivation}
\end{proof}
}

\begin{thm}
  \label{thm:modalfix}
  Let $(\mathcal{C}, \times, \mathbf{1}, F, \delta)$ be a Kripke-4
  category. The following are equivalent: 
  \begin{enumerate}
    \item There is a strong modal fixed point combinator at $A$.
    \item There is a weak modal fixed point combinator at $A$.
    \item The object $A \in \mathcal{C}$ has modal fixed points.
  \end{enumerate}
\end{thm}

\begin{proof}
  To prove $(1) \Rightarrow (2)$, if we are given such a $Y$ we
  calculate
  \begin{derivation}
      Y \circ \lambda f^\#
    \since{definition of strong mfpc, naturality of product morphism}
      \textsf{ev}^\bullet
        \circ
      \langle
        \lambda f^\#,
        Y^\ast \circ \lambda f^\#
      \rangle
    \since{Proposition \ref{prop:hashdist}(3)}
      \textsf{ev}^\bullet
        \circ
      \langle
        \lambda f^\#,
        (Y \circ \lambda f^\#)^\ast
      \rangle
    \since{Proposition \ref{lem:evlob}}
      f^\bullet
        \circ
      \langle 
        id^\#, 
        (Y \circ \lambda f^\#)^\ast
      \rangle
  \end{derivation}
  so $Y$ yields modal fixed points. $(2) \Rightarrow (3)$ is
  trivial, so it remains to show $(3) \Rightarrow (1)$. Let
  \[
    g \defeq
      F(A^{FA}) \times A^{FA} \times FA
       	\xrightarrow{\langle \pi_2, \pi_3 \rangle}
      A^{FA} \times FA
        \xrightarrow{\textsf{ev}}
        A
  \] We show that $g^\dagger : F(A^{FA}) \rightarrow FA$ is a
  strong modal fixed point combinator at $A$. Indeed, it is not
  very hard to calculate that \[
    g^\dagger
      = \textsf{ev}^\bullet \circ \langle id, \left(g^\dagger\right)^\ast\rangle
      \tag*{\qedhere}
  \]
\end{proof}

\noindent We formulate the following naturality property of modal
fixed points, which is partly reminiscent of the ones of Simpson
and Plotkin \cite{Simpson2000}, but also resembles Proposition
\ref{prop:hashsubst}.

\begin{prop}
  \label{prop:mfpdist}
  If we define $(-)_B^\dagger$ by a weak modal fixed point
  combinator, then the resulting modal fixed points are
  \emph{natural}, in the sense that for any $f : \prod_{i=1}^n
  FA_i \times \prod_{i=1}^n A_i \times A \rightarrow A$ and any
  $g_i : \prod_{j=1}^m FB_j \times \prod_{j=1}^m B_j \rightarrow
  B_i$, then, writing $\pi : \prod_{j=1}^m FB_i \times
  \prod_{j=1}^m A_j \rightarrow \prod_{j=1}^m FB_i$ for the
  projection, \[
    \left(
      f
        \circ
      \left(\left\langle
        \vct{g_i^\# \circ \pi},
        \vct{g_i^\#}
      \right\rangle \times id_A\right)
    \right)^\dagger 
      =
    f^\dagger \circ \left\langle \vct{g_i^\#} \right\rangle
  \]
\end{prop}
\begin{proof}
  Recall that $f^\dagger \defeq Y \circ (\lambda(f))^\#$. The LHS
  is equal to 
  \[
    Y \circ \left(\lambda\left(f
        \circ
      \left(\left\langle
        \vct{g_i^\# \circ \pi},
        \vct{g_i^\#}
      \right\rangle \times id_A\right)\right)\right)^\#
  \] 
  which, using naturality of $\lambda(-)$ and Proposition
  \ref{prop:hashsubst}, is equal to $Y \circ
  \left(\lambda(f)\right)^\# \circ \left\langle \vct{g_i^\#}
  \right\rangle$.
\end{proof}

We are finally in a position to define the notion of model used
for \textsf{GL}.

\begin{defi}
  A \emph{G\"odel-L\"ob category} $\left(\mathcal{C}, \times,
  \mathbf{1}, F, \delta, (-)^\dagger\right)$ is a Kripke-4
  category $(\mathcal{C}, \times, \mathbf{1}, F, \delta)$ that has
  modal fixed points at all objects $A$, given by maps 
  \[
    (-)^\dagger_{\vct{B}, A} : 
      \mathcal{C}\left(\prod_{i=1}^n FB_i \times
                  \prod_{i=1}^n B_i \times FA, A\right)
        \rightarrow
      \mathcal{C}\left(\prod_{i=1}^n FB_i, FA\right)
  \]
  which, moreover, are \emph{natural}, in the sense that for any
  $f : \prod_{i=1}^n FB_i \times \prod_{i=1}^n B_i \times A
  \rightarrow A$ and $g_i : \prod_{j=1}^m FC_j \times
  \prod_{j=1}^m C_j \rightarrow B_i$, 
  \[
    \left(
      f
        \circ
      \left(\left\langle
        \vct{g_i^\# \circ \pi},
        \vct{g_i^\#}
      \right\rangle \times id_A\right)
    \right)^\dagger_{\vct{C}, A}
      =
    f^\dagger_{\vct{B}, A} \circ \left\langle \vct{g_i^\#}
    \right\rangle
  \]
\end{defi}

Combining the preceding theorem and proposition assures us that,
we see that it does not matter how modal fixed points are given,
as we can always turn them into a standard G\"odel-L\"ob category.

We also show that, whenever the modal fixed point has no
`diagonal' occurences, it deteriorates to the $(-)^\#$ operation.

\begin{prop}
  \label{prop:projgl}
  If $f : \prod_{i=1}^n FA_i \times \prod_{i=1}^n A_i \rightarrow
  B$ and $\pi : \prod_{i=1}^n FA_i \times \prod_{i=1}^n A_i \times
  A \rightarrow \prod_{i=1}^n FA_i \times \prod_{i=1}^n A_i$ is
  the obvious projection, then
  \[
    \left(f \circ \pi\right)^\dagger = f^\#
  \]
\end{prop}
\begin{proof}
  Writing $g \defeq \left(f \circ \pi\right)^\dagger$, we have \[
      g
    = (f \circ \pi)^\bullet \circ \left\langle id^\#, g^\ast \right\rangle
    = f^\bullet \circ \pi \circ \left\langle id^\#, g^\ast \right\rangle
    = f^\bullet \circ id^\#
    = (f \circ id)^\# = f^\#
  \] by the definition of modal fixed point, and Prop.
  \ref{prop:kripkedist}(2), \ref{prop:hashdist}(1).
\end{proof}

\begin{exa}[The topos of bifurcating trees, part 2]
  Recall Example \ref{exa:bifurcating-1}. The topos of trees is a
  model of \emph{guarded recursion}. Define the functor $\later :
  \PSH{\omega} \rightarrow \PSH{\omega}$ by having it `delay' a
  computation by one time step, i.e. by mapping $X_1
  \xleftarrow{r_1} X_2 \xleftarrow{r_2} \dots$ to $\mathbf{1}
  \xleftarrow{!} X_1 \xleftarrow{r_1} \dots$. This admits a
  natural transformation $\textsf{next}_X : X \rightarrow \later
  X$ which `delays' a computation by trimming $x_{n+1} \in
  X_{n+1}$ to $r_n(x_{n+1}) \in X_n = (\later X)_{n+1}$. We may
  perform guarded recursion: given $f : \later X \Rightarrow X$,
  we define a global section $x : \mathbf{1} \Rightarrow X$ by
  \begin{align*}
    x_1     &\defeq f_1(\ast) : X_1 &
    x_{n+1} &\defeq f_{n+1}(x_n) : X_{n+1}
  \end{align*}
  Essentially $f : \later X \rightarrow X$ provides both a `seed'
  value $f_1(\ast)$ as well as a `coinductive step function'
  $f_{n+1} : X_n \rightarrow X_{n+1}$ at each tick of the clock.
  We thus have \emph{guarded fixed points}, in that $x = f \circ
  \textsf{next}_X \circ x$. This can be done internally, so it
  defines a map $\textsf{fix}_X : (\later A \rightarrow A)
  \rightarrow A$ corresponding to the strong G\"odel-L\"ob logic
  axiom of the logic \textsf{SL}.

  A similar construction can be carried out for the topos of
  bifurcating trees. We can define a natural transformation
  $\textsf{erstwhile}_X : X \rightarrow \Box X$ by 
  \begin{alignat*}{2}
    \textsf{erstwhile}_{X, w}
              : X_w &\rightarrow &&\prod_{v \sqsubset w} X_v \\
                 x_w\ &\mapsto\    &&\lambda v.\ x_w\rvert_v
  \end{alignat*}
  where $(-)\rvert_v$ is the obvious presheaf action $X_w
  \rightarrow X_v$ that trims an element at `stage' $w$ to an
  element at `stage' $v$ for any $v \sqsubset w$. A natural
  transformation $f : \Box X \Rightarrow X$ is a collection of
  maps $f_w : \prod_{v \sqsubset w} X_v \rightarrow X_w$. In
  short, $f$ witnesses a proof-relevant \emph{strong induction
  hypothesis}: when given witnesses of $X$ at each stage that
  strictly precedes $w$, it returns a witness of $X$ at stage $w$.
  We can use this witness to define $x : \mathbf{1} \Rightarrow X$
  by
  \[
    x_w \defeq f_w(\lambda v \sqsubset w.\ x_v)
  \]
  This definition is admissible precisely because $f_w$ only
  depends on $x_v$ for a \emph{strict} prefix $v$, and because the
  prefix order is \emph{well-founded} \cite[\S 14.1]{Hrbacek1999}.
  This closely mirrors the situation from the classical Kripke
  semantics of \textsf{GL}, which is sound and complete for
  transitive frames for which the converse of the accessibility
  relation is \emph{well-founded} \cite[\S 4]{Boolos1994a}. We have
  $x = f \circ \textsf{erstwhile}_X \circ x$.
\end{exa}

By slightly generalising this construction we can show that it
furnishes something stronger than the G\"odel-L\"ob axiom: like
the topos of trees, it is a proof-relevant model of \textsf{SL},
i.e. a \emph{guarded fixpoint category} in the sense of Milius and
Litak \cite{Milius2013}:

\begin{defi}[Guarded Fixpoint Category]
  A \emph{guarded fixpoint category} $(\mathcal{C}, \later, p,
  (-)^\moo)$ consists of a category $\mathcal{C}$ with finite
  products, an endofunctor $\later : \mathcal{C} \rightarrow
  \mathcal{C}$, a natural transformation $p : \textsf{Id}
  \Rightarrow \later$, and a map of homsets
  \[
    (-)^\moo_{B, A} : 
      \mathcal{C}(B \times \later A, A) \rightarrow
      \mathcal{C}(B, A)
  \]
  such that for each $f : B \times \later A \rightarrow A$, the
  morphism $f^\moo_{B, A} : B \rightarrow A$ is a \emph{guarded
  fixpoint of $f$}, i.e. a morphism for which the following
  diagram commutes:
  \[
    \begin{tikzcd}
      B 
        \arrow[r, "f^\moo"]
        \arrow[d, "{\langle id, f^\moo \rangle}"{swap}]
      & A
      \\
      B \times A
        \arrow[r, "{id \times p_A}"{swap}]
      & B \times \later A
        \arrow[u, "f"]
    \end{tikzcd}
  \]
\end{defi}

Indeed, Milius and Litak mention that any presheaf topos over
\emph{any} well-founded order forms a guarded fixpoint category
\cite[Example 2.4(5)]{Milius2013}. In fact, any such category is
also a G\"odel-L\"ob category.

\begin{thm}
  Let $(\mathcal{C}, F, p, (-)^\moo)$ be a guarded fixpoint
  category, where $F$ preserves finite products. Then
  $(\mathcal{C}, \times, \mathbf{1}, F, Fp)$ is a Kripke-4
  category, which may be equipped with the structure of a
  G\"odel-L\"ob strategy by defining the modal fixed point of each
  $f : \prod_{i=1}^n FB_i \times \prod_{i=1}^n B_i \times FA
  \rightarrow A$ to be
  \[
    f^\dagger \defeq \left(f^\moo\right)^\# : 
      \prod_{i=1}^n FB_i \rightarrow FA
  \]
\end{thm}
\begin{proof}
  A quick calculation shows that $Fp : F \Rightarrow F^2$
  satisfies Definition \ref{def:kripke-4}. That $f^\dagger$ is a
  modal fixed point is a straightforward calculation using Props.
  \ref{prop:hashdist}(1) and \ref{prop:deltahash}.
\end{proof}

We still do not know whether there are any interesting
G\"odel-L\"ob categories that do not arise from guarded fixpoint
categories via the above.

\section{Categorical semantics}
  \label{sec:catsem}

In this section we use the modal category theory developed in \S
\ref{sec:modalcats} to formulate a categorical semantics for our
dual-context calculi. This completes the circle in terms of the
Curry-Howard-Lambek correspondence by establishing the following
associations:
\begin{center}
  \begin{tabular}{ccccc}
    \textsf{CK} &$\longleftrightarrow$ &\textsf{DK}
      &$\longleftrightarrow$ &Kripke categories \\
    \textsf{CK4} &$\longleftrightarrow$ &\textsf{DK4}
      &$\longleftrightarrow$ &Kripke-4 categories \\
    \textsf{CGL} &$\longleftrightarrow$ &\textsf{DGL}
      &$\longleftrightarrow$ &G\"odel-L\"ob categories \\
    \textsf{CT} &$\longleftrightarrow$ &\textsf{DT}
      &$\longleftrightarrow$ &Kripke-T categories \\
    \textsf{CS4} &$\longleftrightarrow$ &\textsf{DS4}
      &$\longleftrightarrow$ &Bierman-de Paiva categories
  \end{tabular}
\end{center} where the first bi-implication refers to provability,
and the second to soundness and completeness of the dual-context
calculus with respect to the categorical model on the right. 

We begin by endowing our calculi with an equational theory. We
then propose a categorical interpretation, and show that it is
sound. Finally, we discuss completeness.

\subsection{Equational theory}
  \label{sec:eqth}

Our equational theory of modal proofs should at the very least
contain the reductions used in \S\ref{sec:reduction}. It should
also come with $\eta$ rules, which we did not include in
\S\ref{sec:reduction} due to their usual problematic behaviour
under reduction.

\begin{figure}
  \centering
  \renewcommand{\arraystretch}{3}

\begin{small}
\begin{tabular}{c c}

  $ 
  \begin{prooftree}
      \ctxt{\Delta}{\Gamma, x : A} \vdash M : B
        \qquad
      \ctxt{\Delta}{\Gamma} \vdash N : A
        \justifies
      \ctxt{\Delta}{\Gamma} \vdash (\lambda x : A. M)\,N = M[N/x] : B
        \using
      {(\rightarrow\beta)}
  \end{prooftree}
  $

  &

  $
  \begin{prooftree}
    \ctxt{\Delta}{\Gamma} \vdash M : A \rightarrow B
      \qquad
    x \not\in \fv{M}
      \justifies
    \ctxt{\Delta}{\Gamma} \vdash M = \lambda x : A. Mx : A \rightarrow B
      \using
    {(\rightarrow\eta)}
  \end{prooftree}
  $

  \\ 

  \multicolumn{2}{c}{
  $
  \begin{prooftree}
    \ctxt{\Delta}{\Gamma} \vdash M : \Box A
      \justifies
    \ctxt{\Delta}{\Gamma} \vdash \letbox{u}{M}{\ibox{u}} = M : \Box A
      \using
    {(\Box\eta)}
  \end{prooftree}
  $
  }

  \\

  \multicolumn{2}{c}{
  $
  \begin{prooftree}
      \ctxt{\cdot}{\Delta} \vdash M : A
    \qquad
      \ctxt{\Delta, u : A}{\Gamma} \vdash N : C
    \justifies
      \ctxt{\Delta}{\Gamma} \vdash \letbox{u}{\ibox{M}}{N} = N[M/x] : C
    \using
      {(\Box\beta_\mathsf{K})}
  \end{prooftree}
  $
  }

  \\

  \multicolumn{2}{c}{
  $
  \begin{prooftree}
      \ctxt{\Delta}{\Delta^\bot} \vdash M^\bot : A
    \qquad
      \ctxt{\Delta, u : A}{\Gamma} \vdash N : C
    \justifies
      \ctxt{\Delta}{\Gamma} \vdash \letbox{u}{\ibox{M}}{N} = N[M/x] : C
    \using
      {(\Box\beta_\mathsf{K4})}
  \end{prooftree} 
  $
  }

  \\
  
  \multicolumn{2}{c}{
  $
  \begin{prooftree}
      \ctxt{\Delta}{\Delta^\bot, z^\bot : \Box A} \vdash M^\bot : A
    \qquad
      \ctxt{\Delta, u : A}{\Gamma} \vdash N : C
    \justifies
      \ctxt{\Delta}{\Gamma} \vdash 
        \letbox{u}{\fixbox{z}{M}}{N} 
          =
        N\left[M\left[\fixbox{z}{M}/z\right]/u\right] : C
    \using
      {(\Box\beta_\mathsf{GL})}
  \end{prooftree} 
  $
  }

  \\

  \multicolumn{2}{c}{
  $
  \begin{prooftree}
      \ctxt{\Delta}{\cdot} \vdash M : A
    \qquad
      \ctxt{\Delta, u : A}{\Gamma} \vdash N : C
    \justifies
      \ctxt{\Delta}{\Gamma} \vdash \letbox{u}{\ibox{M}}{N} = N[M/x] : C
    \using
      {(\Box\beta_\mathsf{S4})}
  \end{prooftree}
  $
  }

\end{tabular}
\end{small}


  \caption{Equational theory}
  \label{fig:eqth}
\end{figure}

A fragment of the equational theory may be found in Figure
\ref{fig:eqth}. To generate the theory for, say, \textsf{DGL}, we
take the first three rules ($\beta$/$\eta$ for function types,
$\eta$ for modal types), as well as the appropriate $\beta$ rule
for each system---in this case, $(\Box\beta_\mathsf{GL})$. To
these we must not forget to include (a) rules that ensure that
equality is an equivalence relation, and (b) the usual congruence
rules for all term formers. The congruence rules for $\ibox{(-)}$
must be typed with care. For example, the congruence rule for
\textsf{DK4} should be
\[
  \begin{prooftree}
      \ctxt{\Delta}{\Delta^\bot} \vdash M^\bot = N^\bot : A
    \justifies
      \ctxt{\Delta}{\Gamma} \vdash \ibox{M} = \ibox{N} : \Box A
  \end{prooftree}
\]

We need not include substitution rules:

\begin{thm}
  Structural rules of weakening, exchange and
  contraction for contexts are admissible in the equational theory.
  Furthermore, the following rules are derivable: \begin{enumerate}
      \item Substitution: \[
        \begin{prooftree}
          \ctxt{\Delta}{\Gamma, x : A} \vdash M = N : C
            \qquad
          \ctxt{\Delta}{\Gamma} \vdash P = Q : A
            \justifies
          \ctxt{\Delta}{\Gamma} \vdash M[P/x] = N[Q/x] : C
        \end{prooftree}
      \]
      \item Modal Substitution: for example, in the case of
      $\textsf{DK}$: \[
        \begin{prooftree}
            \ctxt{\Delta, u : A}{\Gamma} \vdash M = N : C
          \qquad
            \ctxt{\cdot}{\Delta} \vdash P = Q : \Box A
          \justifies
            \ctxt{\Delta}{\Gamma} \vdash M[P/u] = N[Q/u] : C
        \end{prooftree}
      \]
    \end{enumerate}
\end{thm}

\paragraph{\bfseries Commuting Conversions}

\begin{figure}
  \centering
  \renewcommand{\arraystretch}{3}

\begin{small}
\begin{tabular}{c}

  $
  \begin{prooftree}
      \ctxt{\Delta}{\Gamma} \vdash N : C
    \qquad
      \ctxt{\Delta}{\Gamma} \vdash M : \Box A
    \qquad
      u \not\in \fv{N}
    \justifies
      \ctxt{\Delta}{\Gamma} \vdash \letbox{u}{M}{N} = N : C
    \using
    (\textsf{commweak})
  \end{prooftree}
  $
  
  \\

  (\textsf{commcontr}):

  \\

  $
  \begin{prooftree}
      \ctxt{\Delta}{\Gamma} \vdash M : \Box A
    \qquad
      \ctxt{\Delta, u : A, v : A}{\Gamma} \vdash N : C
        \qquad
        u, v \not\in \fv{M}
    \justifies
      \ctxt{\Delta}{\Gamma} \vdash
        \letbox{u}{M}{\letbox{v}{M}{N}} = \letbox{w}{M}{N[w, w/u, v]} = N : C
  \end{prooftree}
  $
  
  \\

  $
  \begin{prooftree}
      \ctxt{\Delta}{\Gamma} \vdash C[\letbox{u}{M}{N}] : C
    \qquad
      C[-] \text{ is non-modal, does not bind $u$}
    \justifies
      \ctxt{\Delta}{\Gamma} \vdash \letbox{u}{M}{C[N]} = C[\letbox{u}{M}{N}] : C
    \using
      (\textsf{commlet})
    \end{prooftree}
  $

\end{tabular}
\end{small}


  \caption{Commuting conversions}
  \label{fig:commconv}
\end{figure}

The most interesting rules are the unavoidable commuting
conversions that we need if we want our categorical semantics to
be \emph{complete}. To state these we will need the notion of
\emph{term contexts}, i.e. terms with a single \emph{hole}.

\begin{defi}[Term Contexts] \hfill
\begin{enumerate}
  \item 

  Term contexts are generated by the grammar
  \begin{align*}
    C[-]\ ::=\
           [-] 
    &\;|\; \lambda x : A. \; C[-]
     \;|\; C[-]\;M
     \;|\; M\ C[-]
     \;|\; \langle C[-], M \rangle
     \;|\; \langle M, C[-] \rangle 
     \;|\; \pi_i(C[-]) \\
    &\;|\; \ibox{C[-]} \;|\; \letbox{u}{C[-]}{M}
     \;|\; \letbox{u}{M}{C[-]}
  \end{align*}

  \item 

  A term context $C[-]$ is \emph{non-modal} just if it is
  generated without the clause $\ibox{C[-]}$.

  \item 

  $C[-]$ does not bind $u$ just if its generation uses neither
  $\letbox{u}{C[-]}{M}$ nor $\lambda u : A.\, C[-]$.

\end{enumerate}
\end{defi}

\noindent We write $C[M]$ for the term that results from
capture-insensitive substitution of the term $M$ for the hole
$[-]$ of the term context $C[-]$.

Our systems share the same set of commuting conversions, which may
be found in Figure \ref{fig:commconv}.  The rule
$(\textsf{commweak})$ is a `weakening,' or `garbage collection'
rule that disposes of a delayed substitution that binds a
non-occurring variable. This rule has never been considered in the
study of dual-context systems, for \textsf{DILL} \cite{Barber1996}
was a linear system, and Davies and Pfenning \cite{Davies2001} did
not study reduction, equality, or categorical semantics. However,
a similar rule was proposed by Goubault-Larrecq
\cite{Goubault-Larrecq1996} in his study of Bierman and de Paiva's
calculus for \textsf{S4}. This rule was later included in
\cite{Bierman2000a}.

Similarly, $(\textsf{commcontr})$ is a `contraction' rule. This is
also unfamiliar in dual-context calculi---essentially for the same
reasons as $(\textsf{commweak})$---but is also well-known in
Bierman--de Paiva style calculi as a `garbage collection' rule:
see \cite{Goubault-Larrecq1996}, \cite{Bierman2000a} and
\cite{Kakutani2007a}.

`Exchange' is treated as part of the much more general rule
$(\textsf{commlet})$, which makes `let' constructs commute with
all term formers except $\ibox{(-)}$. The equality is \[
  C[\letbox{u}{M}{N}] = \letbox{u}{M}{C[N]}
\] for any context $C$ that does not bind $u$, and whose hole
$[-]$ is not included within a $\ibox{(-)}$. Read in one
direction, $(\textsf{commlet})$ allows one to `pull' a delayed
substitution to an outermost position, as long as nothing extra is
bound in the process. In the other direction, it allows one to
`push' a delayed substitution as deeply as one can without
creating any free occurrences. A variant of this rule for
\textsf{DILL} was considered by \cite{Barber1996}, and is also
mentioned by \cite{Kakutani2007a}. 

\paragraph{\bfseries The $\eta$ rule}

As is usual with \emph{positive} type formers, of which $\Box$ is
an example, there are two ways to express the $\eta$ rule. The
first is the straightforward way, viz. that introduction is
post-inverse to elimination: for any term $\ctxt{\Delta}{\Gamma}
\vdash M : \Box A$, 
\[
  \ctxt{\Delta}{\Gamma} \vdash
    M = \letbox{u}{M}{\ibox{u}} : \Box A
\] 
The second version of that is an \emph{extended $\eta$-rule},
which allows us to $\eta$-expand a term of modal type, no matter
where it is found in a well-typed term:
\[
  \begin{prooftree}
      \ctxt{\Delta}{\Gamma} \vdash M : \Box A
        \qquad
      \ctxt{\Delta}{\Gamma, x : \Box A} \vdash N : B
        \justifies
      \ctxt{\Delta}{\Gamma} \vdash
        N[M/x] = \letbox{v}{M}{N[\ibox{v}/x]} : B
  \end{prooftree}
\]

In fact,

\begin{thm}
  The $\eta$ rule and the extended $\eta$-rule for the modal type
  are equivalent in the presence of commuting conversions.
\end{thm}
\begin{proof}
  Certainly the $\eta$ rule is a special case of the extended
  $\eta$-rule. In the opposite direction, we proceed by induction
  on the derivation of the term $N$. Most cases are simple. For
  products we have that 
  \begin{derivation}
      \langle N_1, N_2 \rangle[M/x] 
    \since[\equiv]{ substitution }
      \langle N_1[M/x], N_2[M/x] \rangle
    \since{ IH, twice }
      \langle
        \letbox{u}{M} N_1[\ibox{u}/x], 
        \letbox{v}{M} N_2[\ibox{v}/x]
      \rangle
    \since{ $(\textsf{commlet})$, twice }
      \letbox{u}{M}{\letbox{v}{M}{
        \langle
          N_1[\ibox{u}/x],
          N_2[\ibox{v}/x]
        \rangle}
      }
    \since{ $(\textsf{commcontr})$ }
      \letbox{w}{M}{
        \langle
          N_1[\ibox{w}/x],
          N_2[\ibox{w}/x]
        \rangle
      }
    \since[\equiv]{ substitution }
      \letbox{w}{M}{N[\ibox{w}/x]}
  \end{derivation} A similar `collapsing step' is also needed in
  the case of $\textsf{let}$. The case for $\ibox{M}$ is simple,
  as in all of our type theories $x$ does not occur in $M$; the
  result hence follows by $(\textsf{commweak})$.
\end{proof}

\paragraph{\bfseries Idempotence in \textsf{DS4}}

The $(\textsf{commlet})$ rule avoided instances of commutation
between a $\textsf{let}$ and a $\textsf{box}$. If such
commutations were allowed we would have for example the
following equality in \textsf{DS4}: \[
  \ctxt{\Delta}{\Gamma} \vdash
    \ibox{(\letbox{u}{M}{N})} = \letbox{u}{M}{\ibox{N}} : C
\] for $\ctxt{\Delta}{\cdot} \vdash M : \Box A$ and $\ctxt{\Delta,
u : A}{\cdot} \vdash N : C$. We will later show that these rules
are sound for the categorical semantics of \textsf{DS4} if and
only if the comonad used to interpret $\Box$ is \emph{idempotent}.

There are three equivalent ways to present idempotence in
\textsf{DS4}. The first two roughly say that $\ibox{(-)}$ and
$\textsf{let}$ commute. The third is a strong form of the extended
$\eta$-rule, which this time applies to modal variables. Variants
of this rule are sometimes known as \emph{crisp induction}
\cite[\S 5]{Shulman2018}.

\begin{thm}
  The following rules are equivalent
  \begin{enumerate}
    \item
      \[ 
        \begin{prooftree}
          \ctxt{\Delta}{\cdot} \vdash M : \Box A
            \quad
          \ctxt{\Delta, u : A}{\cdot} \vdash N : B
            \justifies
              \ctxt{\Delta}{\Gamma} \vdash
                \ibox{(\letbox{u}{M}{N})} =
                \letbox{u}{M}{\ibox{N}} : \Box B
        \end{prooftree}
      \] 
    \item
      \[
        \begin{prooftree}
          \ctxt{\Delta}{\Gamma} 
              \vdash C[\letbox{u}{M}{N}] : B
            \qquad
          C[-] \text{ does not bind $u$}
            \justifies
          \ctxt{\Delta}{\Gamma}
              \vdash \letbox{u}{M}{C[N]} = C[\letbox{u}{M}{N}] : B
      \end{prooftree}
      \]
    \item
      \[ 
        \begin{prooftree}
            \ctxt{\Delta}{\cdot} \vdash M : \Box A
              \quad
            \ctxt{\Delta, u : \Box A}{\Gamma} \vdash N : B
              \justifies
            \ctxt{\Delta}{\Gamma} \vdash
              N[M/u] = \letbox{v}{M}{N[\ibox{v}/u]} : B
        \end{prooftree}
      \] 
  \end{enumerate}
\end{thm}
\begin{proof}
  (1) is a special case of (2). To prove (2) from (1), we proceed
  by induction on $C$: use the commuting conversion
  $(\textsf{commlet})$ for the non-modal cases, and then (1) for
  the modal case $C[-] \defeq \ibox{C'[-]}$. If we have the
  premises of (1), we can show that, by (3), \[
    \left(\ibox{(\letbox{u}{v}{N})}\right)\left[M/v\right] : \Box B
  \] is equal to \[
    \letbox{w}{M}{\left(\ibox{(\letbox{u}{v}{N})}\right)
                  \left[\ibox{w}/v\right]} : \Box B
  \] The first expression simplifies to
  $\ibox{(\letbox{u}{M}{N})}$, and the second to \[
      \letbox{w}{M}{\left(\ibox{(\letbox{u}{\ibox{v}}{N})}\right)}
  \] which, by one step of $\beta$-reduction and
  $\alpha$-conversion is equal to $\letbox{u}{M}{\ibox{N}}$. We
  can show (3) from (1) by induction on the derivation of $N$ as
  before, but using (1) for the crucial case of $\ibox{N'}$.
\end{proof}

\subsection{Categorical interpretation}

We are now fully equipped to define the categorical semantics of
our dual-context systems. For background on the categorical
semantics of simply-typed $\lambda$-calculus in cartesian closed
categories, we refer the reader to \cite{Lambek1988, Crole1993,
Abramsky2011a}.

We start by interpreting types and contexts. Given any Kripke
category $(\mathcal{C}, \times, \mathbf{1}, F)$, and a map
$\mathcal{I}(-)$ associating each base type $p_i$ with an object
$\mathcal{I}(p_i) \in \mathcal{C}$, we define an object $\sem{A}{}
\in \mathcal{C}$ for every type $A$ by induction:
\begin{align*}
  \sem{p_i} &\defeq \mathcal{I}(p_i) &
  \sem{A \times B} &\defeq \sem{A} \times \sem{B} \\
  \sem{A \rightarrow B} &\defeq \sem{B}^{\sem{A}} &
  \sem{\Box A} &\defeq F\sem{A}
\end{align*} Then, given a context $\ctxt{\Delta}{\Gamma}$ where
$\Delta = u_1 : B_1, \dots u_n : B_n$ and $\Gamma = x_1 : A_1,
\dots, x_m : A_m$, we let \[
  \sem{\ctxt{\Delta}{\Gamma}}{}
    \defeq 
      FB_1 \times \dots \times FB_n
	\times
      A_1 \times \dots \times A_m
\] where the product is, as ever, left-associating. We then extend
$\sem{-}{}$ to associate an arrow
\[
  \sem{\ctxt{\Delta}{\Gamma} \vdash M : A}{} :
  \sem{\ctxt{\Delta}{\Gamma}}{} \rightarrow \sem{A}{}
\] of the category $\mathcal{C}$ to each derivation
$\ctxt{\Delta}{\Gamma} \vdash M : A$. The definition for rules
common to all calculi are the same for all logics, but we use each
of the maps defined in \S \ref{sec:modalcats} to interpret the
different introduction rules for the modality. To do that we need
the corresponding structure we introduced in
\S\ref{sec:modalcats}, e.g. for \textsf{K4} we need a Kripke-4
category, and so on.

\begin{figure}
  \input{catsemdef}
  \caption{Categorical Semantics}
  \label{fig:catsemdef}
\end{figure}

The full definition is given in Figure \ref{fig:catsemdef}. The
morphism $
  \pi_\Delta^{\ctxt{\Delta}{\Gamma}} :
    \sem{\ctxt{\Delta}{\Gamma}}{} \rightarrow
    \sem{\ctxt{\Delta}{\cdot}}{}
$ is the obvious projection. Moreover, the notation $\langle
\vct{\pi_\Delta}, f, \vct{\pi_\Gamma} \rangle$ stands for \[
  \langle \vct{\pi_\Delta}, f, \vct{\pi_\Gamma} \rangle
    \defeq
      \langle
	\pi_1, \dots, \pi_n, 
  	  f,
	\pi_{n+1}, \dots, \pi_{n+m}
      \rangle
\]

\subsection{Soundness}

The main tools used in proving soundness are lemmas giving the
categorical interpretation of various admissible rules, and a
fundamental result relating substitution of terms to composition
in the category. In the sequel we often use informal vector
notation for contexts: for example, we write $\vec{u} : \vec{B}$
for the context $u_1 : B_1, \dots, u_m : B_m$.  We also write
$[\vec{N}/\vec{u}]$ for the simultaneous capture-avoiding
substitution $[N_1/u_1, \dots, N_m/u_m]$.

First, we interpret weakening and exchange.

\begin{lem}[Semantics of Weakening]
  \label{lem:semweak} \hfill
  \begin{enumerate}
    \item
      Let $\ctxt{\Delta}{\Gamma, x : C, \Gamma'} \vdash M : A$ with $x
      \not\in \fv{M}$. Then \[
        \sem{\ctxt{\Delta}{\Gamma, x : C, \Gamma'} \vdash M : A}{}
          = \sem{\ctxt{\Delta}{\Gamma, \Gamma'} \vdash M : A}{}
              \circ \pi
      \] where $\pi : \sem{\ctxt{\Delta}{\Gamma, x : C, \Gamma'}}{}
      \rightarrow \sem{\ctxt{\Delta}{\Gamma, \Gamma'}}{}$ is the
      obvious projection.
    \item
      Let $\ctxt{\Delta, u : B, \Delta'}{\Gamma} \vdash M : A$
      with $u \not\in \fv{M}$. Then \[
        \sem{\ctxt{\Delta, u : B, \Delta'}{\Gamma} \vdash M : A}{}
          = \sem{\ctxt{\Delta, \Delta'}{\Gamma} \vdash M : A}{}
              \circ \pi
      \] where $\pi : \sem{\ctxt{\Delta, u : B,
      \Delta'}{\Gamma}}{} \rightarrow \sem{\ctxt{\Delta,
      \Delta'}{\Gamma}}{}$ is the obvious projection.
  \end{enumerate}
\end{lem}
\begin{proof}
  By induction on the two derivations. All cases are
  straightforward. The modal one uses Propositions
  \ref{prop:kripkedist}(2), \ref{prop:astdist}(5),
  \ref{prop:hashdist}(2), and \ref{prop:mfpdist}.
\end{proof}

\begin{lem}[Semantics of Exchange]
  \label{lem:semexch} \hfill
  \begin{enumerate}
    \item
      Let $\ctxt{\Delta}{\Gamma, x : C, y : D, \Gamma'} \vdash M : A$.
      Then \[
        \sem{\ctxt{\Delta}{\Gamma, x : C, y : D, \Gamma'} \vdash M : A}{}
          =
        \sem{\ctxt{\Delta}{\Gamma, y : D, x : C, \Gamma'} \vdash M : A}{}
          \circ (\cong)
      \] where $(\cong) :
          \sem{\ctxt{\Delta}{\Gamma, x : C, y : D, \Gamma'}}{}
      \xrightarrow{\cong}
          \sem{\ctxt{\Delta}{\Gamma, y : D, x : C, \Gamma'}}{}$
      is the obvious isomorphism.
    \item
      Let $\ctxt{\Delta, u : C, v : D, \Delta'}{\Gamma} \vdash M : A$.
      Then \[
        \sem{\ctxt{\Delta, u : C, v : D, \Delta'}{\Gamma} \vdash M : A}{}
          =
        \sem{\ctxt{\Delta, v : D, u : C, \Delta'}{\Gamma} \vdash M : A}{}
          \circ (\cong)
      \] where $(\cong) :
          \sem{\ctxt{\Delta, u : C, v : D}{\Gamma}}{}
      \xrightarrow{\cong}
          \sem{\ctxt{\Delta, v : D, u : C}{\Gamma}}{}$
      is the obvious isomorphism.
  \end{enumerate}
\end{lem}

\begin{proof}
  By induction on the two derivations. All cases are
  straightforward.
\end{proof}

\noindent Then, we move on to something particular to the cases of
\textsf{T} and \textsf{S4}, namely the interpretation of the Modal
Dereliction rule---see Theorem \ref{thm:dereliction}.

\begin{lem}[Semantics of Dereliction]
  \label{lem:semder}
  Let $\ctxt{\Delta}{\Gamma, \Gamma'}
  \vdash_{\textsf{D}\mathcal{L}} M : A$ where $\mathcal{L} \in
  \{\mathsf{T}, \mathsf{S4}\}$ and $\Gamma =
  \vec{z} : \vec{C}$. Then \[
    \sem{\ctxt{\Delta, \Gamma}{\Gamma'} \vdash M : A}{}_\mathcal{L}
    = \sem{\ctxt{\Delta}{\Gamma, \Gamma'} \vdash M : A}{}_\mathcal{L}
	\circ
	  \left(
	    \vct{id_\Delta}
	      \times
	    \vct{\epsilon_{C_i}}
	      \times
	    \vct{id_{\Gamma'}}
	  \right)
  \]
\end{lem}

\begin{proof}
  By induction on the derivation of $\ctxt{\Delta}{\Gamma,
  \Gamma'} \vdash_{\textsf{D}\mathcal{L}} M : A$. All cases are
  straightforward. The case for ($\Box\mathcal{E}$) depends on the
  semantics of exchange lemma.
\end{proof}

\noindent We also need to know that `boxing' a variable results in
the obvious projection. This depends essentially on the fact our
functors are product-preserving (and not just lax monoidal).

\begin{lem}[Identity Lemma]
  \label{lem:identity}
  For $(u_i : B_i) \in \Delta$, and $\mathcal{L} \in
  \{\textsf{K}, \textsf{K4}, \textsf{T}, \textsf{S4}\}$, \[
    \sem{\ctxt{\Delta}{\Gamma} \vdash 
      \ibox{u_i} : \Box B_i}{}_\mathcal{L}
    = \pi^{\ctxt{\Delta}{\Gamma}}_{\Box B_i}
  \]
\end{lem}

\opt{th}{
\begin{proof} \hfill
  \begin{indproof}
    \case{\textsf{K}, \textsf{T}}
      \begin{derivation}
        \sem{\ctxt{\Delta}{\Gamma} \vdash \ibox{u_i} : \Box B_i}{}
          \since{definition}
        \sem{\ctxt{\cdot}{\Delta} \vdash u_i : B_i}{\bullet}
            \circ \pi^{\ctxt{\Delta}{\Gamma}}_{\Delta}
          \since{definition}
        \left(\pi^{\ctxt{\cdot}{\Delta}}_{B_i}\right)^\bullet
            \circ \pi^{\ctxt{\Delta}{\Gamma}}_{\Delta}
          \since{Proposition \ref{prop:prodpresproj}, or
          Proposition \ref{prop:kripkedist}(2) with $f = id$}
        \pi^{\ctxt{\Delta}{\cdot}}_{\Box B_i}
            \circ \pi^{\ctxt{\Delta}{\Gamma}}_{\Delta}
          \since{projections}
        \pi^{\ctxt{\Delta}{\Gamma}}_{\Box B_i}
      \end{derivation}

    \case{\textsf{K4}}
      \begin{derivation}
        \sem{\ctxt{\Delta}{\Gamma} \vdash \ibox{u_i} : \Box B_i}{}
          \since{definition}
        \sem{\ctxt{\Delta}{\Delta^\bot} \vdash u_i^\bot : B_i}{\#}
            \circ \pi^{\ctxt{\Delta}{\Gamma}}_{\Delta}
          \since{definition}
        \left(\pi^{\ctxt{\Delta}{\Delta^\bot}}_{B_i}\right)^\#
            \circ \pi^{\ctxt{\Delta}{\Gamma}}_{\Delta}
          \since{definition}
        F\pi^{\ctxt{\Delta}{\Delta^\bot}}_{B_i}
            \circ m^{(2n)}
            \circ \langle \vct{\delta_{B_i}\pi_i}, \vct{\pi_i} \rangle
            \circ \pi^{\ctxt{\Delta}{\Gamma}}_{\Delta}
          \since{Proposition \ref{prop:prodpresproj}}
        \pi^{\ctxt{\Box\Delta}{\Box\Delta^\bot}}_{\Box B_i}
            \circ \langle \vct{\delta_{B_i}\pi_i}, \vct{\pi_i} \rangle
            \circ \pi^{\ctxt{\Delta}{\Gamma}}_{\Delta}
          \since{projections}
        \pi^{\ctxt{\Delta}{\Gamma}}_{\Box B_i}
      \end{derivation}

    \case{\textsf{S4}}
      \begin{derivation}
          \sem{\ctxt{\Delta}{\Gamma} \vdash \ibox{u_i} : \Box B_i}{}
        \since{definition}
          \sem{\ctxt{\Delta}{\cdot} \vdash u_i : B_i}{\ast}
              \circ \pi^{\ctxt{\Delta}{\Gamma}}_{\Delta}
        \since{definition}
          \left(\epsilon_{B_i}
            \circ \pi^{\ctxt{\Delta}{\cdot}}_{\Box B_i}\right)^\ast
            \circ \pi^{\ctxt{\Delta}{\Gamma}}_{\Delta}
        \since{Proposition \ref{prop:astdist}}
          \epsilon_{B_i}^\ast
            \circ \pi^{\ctxt{\Delta}{\cdot}}_{\Box B_i}
            \circ \pi^{\ctxt{\Delta}{\Gamma}}_{\Delta}
          \since{Proposition \ref{prop:astdist}, projections}
            \pi^{\ctxt{\Delta}{\Gamma}}_{\Box B_i}
      \end{derivation}
  \end{indproof}
\end{proof}
}

We aim to show that substitution in the syntax corresponds to
composition in the semantics. To make this result work, we need to
introduce a $\ibox{(-)}$ construct for $\textsf{GL}$. We write
\[
  \ibox{M} \defeq \fixbox{w}{M}
\] 
with $w, w^\bot$ fresh. It is then not hard to see that the
introduction rule of \textsf{K4} is admissible for \textsf{GL}
when $M$ has no occurrences of $w^\bot$: we simply use weakening
followed by the introduction rule for \textsf{GL}. This derived
operation is reflected in the semantics by the equation 

\begin{prop}
  \label{prop:boxgl}
  $  
    \sem{\ctxt{\Delta}{\Gamma} \vdash_{\textsf{DGL}} \ibox{M} : \Box
      A}{}
    =
      \sem{\ctxt{\Delta}{\Delta^\bot}{}
        \vdash_{\textsf{DGL}} M^\bot : A}^{\#}
    \circ
      \pi^{\ctxt{\Delta}{\Gamma}}_{\Delta}
  $
\end{prop}
\begin{proof}
  By the semantics of weakening and Proposition \ref{prop:projgl}.
\end{proof}
\noindent In short: when the variable that is being `diagonalised
over' does \emph{not} occur freely, the interpretation degenerates
to that of \textsf{K4}.

\begin{lem}[Semantics of Substitution]
  \label{lem:semsubst}
  Suppose that $\ctxt{\vec{u} : \vec{B}}{\vec{x} : \vec{A}}
  \vdash_{\mathsf{D}\mathcal{L}} P : C$. Let $\ctxt{\Delta}{\Gamma}
  \vdash_{\mathsf{D}\mathcal{L}} M_i : A_i$ for $i = 1, \dots, n$,
  and let \[
    \alpha_i
      \defeq 
      \sem{\ctxt{\Delta}{\Gamma} \vdash M_i : A_i}{}_\mathcal{L}
  \] If either
  \begin{enumerate}
    \item
      $\mathcal{L} \in \{\textsf{K}, \textsf{T}\}$ and
      $\ctxt{\cdot}{\Delta} \vdash N_j : B_j$ for $j = 1, \dots,
      m$, or
    \item
      $\mathcal{L} \in \{\textsf{K4}, \textsf{GL}\}$ and
      $\ctxt{\Delta}{\Delta^\bot} \vdash N^\bot_j : B_j$ for $j =
      1, \dots, m$, or
    \item
      $\mathcal{L} = \textsf{S4}$ and
      $\ctxt{\Delta}{\cdot} \vdash N_j : B_j$ for $j = 1, \dots,
      m$,
  \end{enumerate}
  then, letting $
    \beta_j \defeq
      \sem{\ctxt{\Delta}{\Gamma}
        \vdash \ibox{N_j} : \Box B_j}{}_\mathcal{L}
  $ for $j \in \{1, \dots, m\}$, we have that \[
    \sem{\ctxt{\Delta}{\Gamma} 
      \vdash P[\vec{N}/\vec{u}, \vec{M}/\vec{x}] : C}{}_\mathcal{L}
    =
    \sem{\ctxt{\vec{u} : \vec{B}}{\vec{x} : \vec{A}} 
      \vdash P : C}{}_\mathcal{L}
    \circ 
      \langle
        \beta_1, \ldots, \beta_m,
        \alpha_1, \ldots, \alpha_n
      \rangle
  \]
\end{lem}

\begin{proof}
  By induction on the derivation of
  $\ctxt{\vec{u}:\vec{B}}{\vec{x}:\vec{A}} \vdash P : C$. Most
  cases are straightforward, and use a combination of standard
  equations that hold in cartesian closed categories in order to
  perform calculations very close the ones detailed in \cite[\S
  1.6.5]{Abramsky2011a}. Because of the precise definitions we
  have used, we also need to make use of Lemma \ref{lem:semweak}
  to interpret weakening whenever variables in the context do not
  occur freely in the term. For the modal rules we use many of the
  equations we showed in \S\ref{sec:modalcats}, e.g in
  Propositions \ref{prop:kripkedist}, \ref{prop:astdist},
  \ref{prop:hashdist}, and so on.
\opt{th}{
  We will now prove these modal cases in detail.
  \begin{indproof}
    \case{$\Box\textsf{var}$} We show the case \textsf{T} only. The
    case for \textsf{S4} is similar, but uses Proposition
    \ref{prop:deltaepsilonast}(2) instead of Proposition
    \ref{prop:epsilonkripke}. Then $P \equiv u_i$ for some $u_i$
    amongst the $\vec{u}$.  Hence, the LHS is
    $\ctxt{\Delta}{\Gamma} \vdash N_i : B_i$, whereas we calculate
    that the RHS is
    \begin{derivation}
      \sem{\ctxt{\vec{u}:\vec{B}}{\vec{x}:\vec{A}} \vdash P :
      C}{} \circ \langle \vec{\beta}, \vec{\alpha} \rangle
          \since{definition}
      \epsilon_{B_i} \circ
      \pi^{\ctxt{\vec{u}:\vec{B}}{\vec{x}:\vec{A}}}_{u_i:B_i}
      \circ \langle \vec{\beta}, \vec{\alpha} \rangle
          \since{projection}
      \epsilon_{B_i} \circ
        \sem{\ctxt{\Delta}{\Gamma} \vdash \ibox{N_i} : \Box B_i}{}
          \since{definition}
      \epsilon_{B_i} \circ
        \sem{\ctxt{\cdot}{\Delta} \vdash N_i :
        B_i}{\bullet} \circ
        \pi_\Delta^{\ctxt{\Delta}{\Gamma}}
          \since{Proposition \ref{prop:epsilonkripke}}
      \sem{\ctxt{\cdot}{\Delta} \vdash N_i : B_i}{} \circ
      \prod_{D \in \Delta} \epsilon_D \circ
      \pi_\Delta^{\ctxt{\Delta}{\Gamma}}
          \since{Semantics of Dereliction (Lemma \ref{lem:semder})}
      \sem{\ctxt{\Delta}{\cdot} \vdash N_i : B_i}{} \circ
        \pi_\Delta^{\ctxt{\Delta}{\Gamma}}
          \since{Semantics of Weakening (Lemma \ref{lem:semweak})}
      \sem{\ctxt{\Delta}{\Gamma} \vdash N_i : B_i}{}
    \end{derivation}

    \case{$\Box\mathcal{I}_\textsf{K}$}
      We have that $\ctxt{\vec{u}:\vec{B}}{\vec{x}:\vec{A}} \vdash
      \ibox{P} : \Box C$, so that $\ctxt{\cdot}{\vec{u}:\vec{B}}
      \vdash P : C$, with the result that none of the variables
      $\vec{x}$ occurs free in $P$. We use this fact and the
      definition of substitution to calculate:
      \begin{derivation}
          \sem{\ctxt{\Delta}{\Gamma} \vdash
            \ibox{(P[\vec{N}/\vec{u}, \vec{M}/\vec{x}])} : \Box C}{}
        \since{definition, and non-occurence of the $\vec{x}$}
          \sem{\ctxt{\cdot}{\Delta} \vdash
            P[\vec{N}/\vec{u}] : C}{\bullet}
            \circ
          \pi^{\ctxt{\Delta}{\Gamma}}_\Delta
        \since{IH}
          \left(\sem{\ctxt{\cdot}{\vec{u}:\vec{B}} \vdash P : C}{} 
                  \circ
                \left\langle 
                  \vct{\sem{\ctxt{\cdot}{\Delta} \vdash N_i : B_i}{}}
                \right\rangle
          \right)^\bullet
            \circ
          \pi^{\ctxt{\Delta}{\Gamma}}_\Delta
        \since{Proposition \ref{prop:kripkedist}}
          \sem{\ctxt{\cdot}{\vec{u}:\vec{B}} \vdash P : C}{\bullet}
            \circ
          \left\langle 
            \vct{\sem{\ctxt{\cdot}{\Delta} \vdash N_i : B_i}{\bullet}}
          \right\rangle 
            \circ
          \pi^{\ctxt{\Delta}{\Gamma}}_\Delta
        \since{naturality of product morphism, definition, projection}
          \sem{\ctxt{\cdot}{\vec{u}:\vec{B}} \vdash P : C}{\bullet}
            \circ
          \pi^{\ctxt{\vec{u}:\vec{B}}{\vec{x}:\vec{A}}}_{\vec{u}:\vec{B}}
            \circ
          \left\langle \vct{\beta}, \vct{\alpha} \right\rangle
        \since{definition}
          \sem{\ctxt{\vec{u}:\vec{B}}{\vec{x}:\vec{A}} \vdash
            \ibox{P} : \Box C}{}
          \circ
            \left\langle \vct{\beta}, \vct{\alpha} \right\rangle
      \end{derivation}

    \case{$\Box\mathcal{I}_\textsf{K4}$}
      We have that $\ctxt{\vec{u}:\vec{B}}{\vec{x}:\vec{A}} \vdash
      \ibox{P} : \Box C$, so that
      $\ctxt{\vec{u}:\vec{B}}{\vec{u^\bot}:\vec{B}} \vdash P : C$,
      with the result that none of the variables $\vec{x}$ or
      $\vct{x^\bot}$ occur free in $P$. Hence, $
        \left(P[\vec{N}/\vec{u}, \vec{M}/\vec{x}]\right)^\bot
        \equiv P[\vec{N}/\vec{u}, \vct{N^\bot}/\vct{u^\bot}]
      $ by Theorem \ref{thm:bot}. Now we calculate:
        \begin{derivation}
            \sem{\ctxt{\Delta}{\Gamma} \vdash
              \ibox{(P[\vec{N}/\vec{u}, \vec{M}/\vec{x}])} : \Box C}{}
          \since{definition, 
                 and non-occurence of the $\vec{x}$ and $\vct{x^\bot}$ }
            \left(\sem{\ctxt{\Delta}{\Delta^\bot} \vdash
              P[\vec{N}/\vec{u}, \vct{N^\bot}/\vct{u^\bot}] :
                C}{}\right)^\#
            \circ
            \pi^{\ctxt{\Delta}{\Gamma}}_\Delta
          \since{IH}
            \left(\sem{\ctxt{\vec{u}:\vec{B}}{\vec{u^\bot}:\vec{B}}
              \vdash P : C}{}
            \circ
            \left\langle 
              \vct{\sem{\ctxt{\Delta}{\Delta^\bot} \vdash
                \ibox{N_i} : \Box B_i}{}},
              \vct{\sem{\ctxt{\Delta}{\Delta^\bot} \vdash
                N_i : B_i}{}} 
            \right\rangle
            \right)^\# 
            \circ
              \pi^{\ctxt{\Delta}{\Gamma}}_\Delta
          \since{definition}
            \left(\sem{\ctxt{\vec{u}:\vec{B}}{\vec{u^\bot}:\vec{B}}
              \vdash P : C}{}
            \circ
            \left\langle 
              \vct{\sem{\ctxt{\Delta}{\Delta^\bot} \vdash
                N_i : B_i}{\#} 
                \circ 
                \pi^{\ctxt{\Delta}{\Delta^\bot}}_{\Delta}},
              \vct{\sem{\ctxt{\Delta}{\Delta^\bot} \vdash
                N_i : B_i}{}} 
            \right\rangle
            \right)^\# 
            \circ
              \pi^{\ctxt{\Delta}{\Gamma}}_\Delta
          \since{Proposition \ref{prop:hashsubst}}
            \sem{\ctxt{\vec{u}:\vec{B}}{\vec{u^\bot}:\vec{B}}
              \vdash P : C}{\#}
            \circ
            \left\langle 
              \vct{\sem{\ctxt{\Delta}{\Delta^\bot} \vdash
                N_i : B_i}{\#}}
            \right\rangle
            \circ
              \pi^{\ctxt{\Delta}{\Gamma}}_\Delta
          \since{naturality, projections, definitions}
            \sem{\ctxt{\vec{u}:\vec{B}}{\vec{x}:\vec{A}} \vdash
            \ibox{P} : \Box C}{} \circ
                \left\langle 
            \vct{\beta},
            \vct{\alpha}
                \right\rangle
        \end{derivation}

    \case{$\Box\mathcal{I}_\textsf{GL}$}
      We have that $\ctxt{\vec{u}:\vec{B}}{\vec{x}:\vec{A}} \vdash
      \fixbox{z}{P} : \Box C$, so that
      $\ctxt{\vec{u}:\vec{B}}{\vec{u^\bot}:\vec{B}, z^\bot : \Box
      C} \vdash P : C$, with the result that none of the variables
      $\vec{x}$ or $\vct{x^\bot}$ occur free in $P$. Hence, $
      \left(P[\vec{N}/\vec{u}, \vec{M}/\vec{x}]\right)^\bot \equiv
      P[\vec{N}/\vec{u}, \vct{N^\bot}/\vct{u^\bot}] $ by Theorem
      \ref{thm:bot}. Now we calculate:
        \begin{derivation}
            \sem{\ctxt{\Delta}{\Gamma} \vdash
              \fixbox{z}{(P[\vec{N}/\vec{u}, \vec{M}/\vec{x}])} : \Box C}{}
          \since{definition and previous argument}
            \left(\sem{\ctxt{\Delta}{\Delta^\bot, z^\bot : \Box C} \vdash
              P[\vec{N}/\vec{u}, \vct{N^\bot}/\vct{u^\bot}, z/z] :
                C}{}\right)^\dagger
            \circ
            \pi^{\ctxt{\Delta}{\Gamma}}_\Delta
          \since{IH, weakening,
              $b_i \defeq
                    \sem{\ctxt{\Delta}{\Delta^\bot} \vdash
                    \ibox{N_i} : \Box B_i}{}$,
              $n_i \defeq
                    \sem{\ctxt{\Delta}{\Delta^\bot} \vdash
                    N_i : B_i}{}$}
            \left(\sem{\ctxt{\vec{u}:\vec{B}}{\vec{u^\bot}:\vec{B},
            z^\bot : \Box C} \vdash P : C}{}
            \circ
              \left\langle 
                \vct{b_i \circ 
                  \pi^{\ctxt{\Delta}{\Delta^\bot, z^\bot}}_{
                    \ctxt{\Delta}{\Delta^\bot}}},
                \vct{n_i \circ \pi^{\ctxt{\Delta}{\Delta^\bot, z^\bot}}_{
                    \ctxt{\Delta}{\Delta^\bot}}},
                \pi^{\ctxt{\Delta}{\Delta^\bot, z^\bot}}_{z^\bot}
              \right\rangle
            \right)^\dagger
            \circ
              \pi^{\ctxt{\Delta}{\Gamma}}_\Delta
          \since{ definition of product morphism }
            \left(\sem{\ctxt{\vec{u}:\vec{B}}{\vec{u^\bot}:\vec{B},
            z^\bot : \Box C} \vdash P : C}{}
            \circ
              \left(
                \left\langle 
                  \vct{b_i},
                  \vct{n_i}
                \right\rangle \times id
              \right)
            \right)^\dagger
            \circ
              \pi^{\ctxt{\Delta}{\Gamma}}_\Delta
          \since{ by the remark preceding this theorem, $b_i = n_i^\# \circ
            \pi^{\ctxt{\Delta}{\Delta^\bot}}_{\ctxt{\Delta}{\cdot}}$}
            \left(\sem{\ctxt{\vec{u}:\vec{B}}{\vec{u^\bot}:\vec{B},
            z^\bot : \Box C} \vdash P : C}{}
            \circ
              \left(
                \left\langle 
                  \vct{n_i^\# \circ 
                    \pi^{\ctxt{\Delta}{\Delta^\bot}}_{\ctxt{\Delta}{\cdot}}},
                  \vct{n_i}
                \right\rangle \times id
              \right)
            \right)^\dagger
            \circ
              \pi^{\ctxt{\Delta}{\Gamma}}_\Delta
          \since{ naturality of modal fixed points }
            \sem{\ctxt{\vec{u}:\vec{B}}{\vec{u^\bot}:\vec{B},
            z^\bot : \Box C} \vdash P : C}{\dagger}
              \circ \left\langle\vct{n_i^\#}\right\rangle
            \circ
              \pi^{\ctxt{\Delta}{\Gamma}}_\Delta
          \since{ naturality of product morphism,
            remark preceding theorem, definitions }
            \sem{\ctxt{\vec{u}:\vec{B}}{} \vdash
            \fixbox{z}{P} : \Box C}{}
              \circ
                \left\langle
                  \vct{\sem{\ctxt{\Delta}{\Gamma} \vdash
                  \ibox{N_i} : \Box B_i}}
                \right\rangle
          \since{weakening}
            \sem{\ctxt{\vec{u}:\vec{B}}{\vec{x}:\vec{A}} \vdash
            \fixbox{z}{P} : \Box C}{} \circ
                \left\langle 
            \vct{\beta},
            \vct{\alpha}
                \right\rangle
        \end{derivation}
    \case{$\Box\mathcal{I}_\textsf{S4}$}
      We have that $\ctxt{\vct{u}:\vct{B}}{\vct{x}:\vct{A}} \vdash
      \ibox{P} : \Box C$, so that $\ctxt{\vct{u}:\vct{B}}{\cdot}
      \vdash P : C$, with the result that none of the variables
      $\vct{x}$ occur in $P$. Hence $P[\vec{N}/\vec{u},
      \vec{M}/\vec{x}] \equiv P[\vec{N}/\vec{u}]$, and we
      calculate:
      \begin{derivation}
          \sem{\ctxt{\Delta}{\Gamma} \vdash
            \ibox{(P[\vec{N}/\vec{u}, \vec{M}/\vec{x}])} : \Box C}{}
        \since{definition, and non-occurence of the $\vec{x}$ in $P$}
          \sem{\ctxt{\Delta}{\cdot} \vdash P[\vec{N}/\vec{u}] : C}{\ast}
          \circ
            \pi^{\ctxt{\Delta}{\Gamma}}_\Delta
        \since{IH}
          \left(
            \sem{\ctxt{\vec{u}:\vec{B}}{\cdot} \vdash P : C}{}
              \circ
            \left\langle 
              \vct{\sem{\ctxt{\Delta}{\cdot} \vdash \ibox{N_i} : \Box B_i}{}}
            \right\rangle
          \right)^\ast
            \circ
          \pi^{\ctxt{\Delta}{\Gamma}}_\Delta
        \since{definition}
          \left(
            \sem{\ctxt{\vec{u}:\vec{B}}{\cdot} \vdash P : C}{}
              \circ
            \left\langle 
              \vct{\sem{\ctxt{\Delta}{\cdot} \vdash N_i : \Box B_i}{\ast}}
            \right\rangle
          \right)^\ast
            \circ
          \pi^{\ctxt{\Delta}{\Gamma}}_\Delta
        \since{Proposition \ref{prop:astdist}(4)}
          \sem{\ctxt{\vec{u}:\vec{B}}{\cdot} \vdash P : C}{\ast}
            \circ
          \left\langle 
            \vct{\sem{\ctxt{\Delta}{\cdot} \vdash N_i : \Box B_i}{\ast}}
          \right\rangle
            \circ
          \pi^{\ctxt{\Delta}{\Gamma}}_\Delta
        \since{projections, definition of $(-)^\ast$ and $\sem{-}{}$}
          \sem{\ctxt{\vec{u}:\vec{B}}{\vec{x}:\vec{A}} 
            \vdash \ibox{P} : \Box C}{}
            \circ
          \left\langle \vct{\beta}, \vct{\alpha} \right\rangle
      \end{derivation}
  \end{indproof}
}
\end{proof}

\begin{thm}[Soundness]
  If $\ctxt{\Delta}{\Gamma} \vdash_{\mathsf{D}\mathcal{L}} M = N :
  A$, then we have that \[
    \sem{\ctxt{\Delta}{\Gamma} \vdash M : A}{}_{\mathcal{L}}
      = \sem{\ctxt{\Delta}{\Gamma} \vdash N : A}{}_{\mathcal{L}}
  \]
\end{thm}

\begin{proof} 
  By induction on the derivation of $\ctxt{\Delta}{\Gamma}
  \vdash_{\mathsf{D}\mathcal{L}} M = N : A$. The congruence cases
  are clear, as are the majority of the ordinary clauses---see
  \cite{Crole1993} and \cite{Abramsky2011a}. The rules that remain
  are $(\Box\eta)$, the many variants of $(\Box\beta)$, and the
  commuting conversions.
  
  First, we prove the modal $\beta$ and $\eta$ cases by direct
  calculation. To do so, we use Lemma \ref{lem:identity}, so
  product preservation is essential even to prove the soundness of
  $(\Box\beta)$.

  Let $\Delta = \vec{u} : \vec{B}$ and $\Gamma = \vec{x} :
  \vec{A}$. We then calculate: \begin{derivation}
      \sem{\ctxt{\Delta}{\Gamma} \vdash \letbox{u}{\ibox{M}}{N} : C}{}
        \since{definition}
      \sem{\ctxt{\Delta, u : A}{\Gamma} \vdash N : C}{}
        \circ
          \langle
            \vct{\pi_\Delta},
            \sem{\ctxt{\Delta}{\Gamma} \vdash \ibox{M} : \Box A}{},
            \vct{\pi_\Gamma}
          \rangle
        \since{Lemma \ref{lem:identity}}
      \sem{\ctxt{\Delta, u : A}{\Gamma} \vdash N : C}{}
        \circ
          \langle
            \vct{\sem{\ctxt{\Delta}{\Gamma} \vdash \ibox{u_i} : \Box B_i}{}},
            \sem{\ctxt{\Delta}{\Gamma} \vdash \ibox{M} : \Box A}{},
            \vct{\sem{\ctxt{\Delta}{\Gamma} \vdash x_i : A_i}{}}
          \rangle
        \since{Lemma \ref{lem:semsubst}}
      \sem{\ctxt{\Delta}{\Gamma} \vdash 
        N[\vec{u_i}/\vec{u_i}, M/u, \vec{x_i}/\vec{x_i}] : C}{}
    \end{derivation}

    This covers all cases save \textsf{GL}. For that, it suffices
    to show that \[
      \sem{\ctxt{\Delta}{\Gamma} 
        \vdash \fixbox{z}{M} : \Box A}{}
      =
      \sem{ \ctxt{\Delta}{\Gamma} 
        \vdash \ibox{M[\fixbox{z}{M}/z]} : \Box A }{}
    \] and then surreptitiously swap the first expression with the
    second in the above calculation just before using the
    substitution lemma. \opt{th}{To show that, we may let \[ 
      m \defeq \sem{\ctxt{\Delta}{\Delta^\bot, z^\bot : \Box A} 
      \vdash M^\bot : \Box A}{}
    \] and then calculate:
    \begin{derivation}
        \sem{\ctxt{\Delta}{\Gamma} \vdash \fixbox{z}{M} : \Box A}{}
      \since{definition}
        m^\dagger \circ \pi^{\ctxt{\Delta}{\Gamma}}_{\Delta}
      \since{definition of modal fixed point}
        m^\bullet \circ
          \left\langle
            id^\#,
            (m^\dagger)^\ast
          \right\rangle
      \since{Proposition \ref{prop:degen4}{}}
        m^\bullet \circ
          \left\langle
            id^\#,
            (m^\dagger \circ
            \pi^{\ctxt{\Delta}{\Delta^\bot}}_{\Delta})^\#
          \right\rangle
        \circ \pi^{\ctxt{\Delta}{\Gamma}}_{\Delta}
      \since{Proposition \ref{prop:hashdist}(1)}
        \left(m \circ
          \left\langle
            id,
            m^\dagger \circ
            \pi^{\ctxt{\Delta}{\Delta^\bot}}_{\Delta}
          \right\rangle
        \right)^\#
        \circ \pi^{\ctxt{\Delta}{\Gamma}}_{\Delta}
      \since{definition, weakening}
        \left(m \circ
          \left\langle
            id,
            \sem{
              \ctxt{\Delta}{\Delta^\bot} \vdash
                \fixbox{z}{M} : \Box A
            }{}
          \right\rangle
        \right)^\#
        \circ \pi^{\ctxt{\Delta}{\Gamma}}_{\Delta}
      \since{Lemma \ref{lem:semsubst}}
        \sem{
          \ctxt{\Delta}{\Delta^\bot} \vdash
            M^\bot[\fixbox{z}{M}/z^\bot] : A
        }{\#}
        \circ \pi^{\ctxt{\Delta}{\Gamma}}_{\Delta}
      \since{Proposition \ref{prop:boxgl}}
        \sem{
          \ctxt{\Delta}{\Gamma} \vdash
            \ibox{M[\fixbox{z}{M}/z]} : \Box A
        }{}
    \end{derivation} 
    }

    The case of $\eta$ is even simpler, as it follows immediately
    from Lemma \ref{lem:identity}.

    The commuting conversions for weakening and contraction are
    straightforward. $(\textsf{commlet})$ requires a nested
    induction on contexts $C[-]$, which follows from the
    naturality of the various operations of the CCC.
\end{proof}

\paragraph{\bfseries Idempotence}

If the comonad $(F, \epsilon, \delta)$ provided as part of a
Bierman-de Paiva category is idempotent, then more equations are
sound. We have shown the equivalence between three such equations
in \S\ref{sec:eqth}, so it suffices to prove soundness for only
one of them:
  \begin{derivation}
      \sem{\ctxt{\Delta}{\Gamma} 
        \vdash \ibox{(\letbox{u}{M}{N})} : \Box B}{}
    \since{ definitions }
      \left(
        \sem{\ctxt{\Delta, u : A}{\cdot} \vdash N : B}{}
          \circ
        \left\langle
          id,
          \sem{\ctxt{\Delta}{\cdot} \vdash M : \Box A}{}
        \right\rangle
      \right)^\ast
    \since{ Proposition Theorem \ref{prop:astidemsubst} }
      \sem{\ctxt{\Delta, u : A}{\cdot} \vdash N : B}^\ast
        \circ
      \left\langle
        id,
        \sem{\ctxt{\Delta}{\cdot} \vdash M : \Box A}{}
      \right\rangle
    \since{ definitions }
      \sem{\ctxt{\Delta}{\Gamma} \vdash \letbox{u}{M}{\ibox{N}} :
      \Box B}{}
  \end{derivation}

\subsection{A note on completeness}

It is possible to prove that the categorical semantics given in
this section are \emph{complete} in the Lindenbaum-Tarski sense.
For example, we can prove that if $\sem{\ctxt{\Delta}{\Gamma}
\vdash_\textsf{DK} M : A} = \sem{\ctxt{\Delta}{\Gamma}
\vdash_\textsf{DK} N : A}$ in every Kripke category, then
$\ctxt{\Delta}{\Gamma} \vdash_\textsf{DK} M = N : A$ is provable
in the equational theory. To do so we must construct a Kripke
category by quotienting the syntax of the $\lambda$-calculus; if
the equation is satisfied in all models, it is satisfied in this
one in particular, which implies equality in the theory.

Indeed, we have shown this for all our calculi, as documented in
the conference version of this article and its associated
technical report \cite{Kavvos2017b}. However, this result is of
limited interest from the point-of-view of categorical logic: it
merely shows that the correspondence between inference rules and
axioms (e.g. between Scott's rule and the \textsf{K} axiom)
extends to the categorical level, e.g. to a correspondence between
the operation $(-)^\bullet$ and the product-preserving structure
of a Kripke category.

A more important result is to show that the model constructed
through quotienting syntax is \emph{initial} in the category of
all such models. Unfortunately, it does not seem that the
constructions used in the conference version of this paper yield
initial models. We leave the solution of this problem to future
work.

\section{Conclusion \& further directions}

We have extended the full Curry-Howard-Lambek correspondence to a
handful of normal modal logics, spanning the logical aspect
(Hilbert systems and provability), the computational aspect (modal
$\lambda$-calculi), and the categorical aspect (proof-relevant
semantics). 

In order to achieve the connection at the first junction, i.e.  that
between logic and computation, we have employed a systematic pattern
based on translating sequent calculus rules to introduction rules for
dual-context systems.  This worked remarkably well: not only did it
lead to already known systems, like that for \textsf{S4}, but also to
a number of new ones, including one for \textsf{GL}. One would hope
that there is a deeper aspect to this pattern---perhaps even a theorem
to the effect that rules of cut-free sequent calculi rule can be
immediately turned into well-behaved dual-context systems. Of course,
this is quite a long way from our current grasp, but we believe it is
worth investigating.

The second junction, i.e. the one between modal $\lambda$-calculi and
their categorical models, is achieved by focusing on
product-preserving functors and their extensions.  This deviates from
previous approaches---such as that of Bierman and de Paiva which
concentrated on lax monoidal endofunctors---yet supports numerous
models of interest. We sketched in some detail the topos of
bifurcating trees as a motivating example of a model with a
product-preserving modality which does not extend to a comonad. The
assumption of product preservation seeps deep into our proofs: it is
in fact used even in proving soundness of $\beta$-convertibility.

The resulting dual-context calculi sport a simpler syntax,
which---as we argued in the introduction---makes them particularly
suited to computational applications. It is also our hope that
this work will help elucidate the computational behaviour of
necessity modalities. In fact, the author believes that modalities
can be used to control the `flow of data' in a programming
language, in the sense that they create regions of the language
whose intercommunication is restricted. For example, one can
handwavingly argue that \textsf{S4} guarantees that `only modal
variables flow into terms of modal type,' whereas \textsf{K}
additionally ensures that no modal data flows into a term of
non-modal type. A first result of this type is the free variables
theorem (Theorem \ref{thm:freevar}), but it is rather weak. The
author has recently used the second junction---namely that between
computation and categories---to prove \emph{noninterference
theorems} for modal $\lambda$-calculi \cite{Kavvos2019a}. These
theorems show that modal type systems indeed effect certain
restrictions on information flow. Amongst other things, the
present article is meant to lay a foundation that enables the
further study of categorical semantics of similar calculi.


\section*{Acknowledgments}

I gratefully acknowledge many interesting discussions with Daniel
Gratzer, Dan Licata, Kristina Sojakova, Mitchell Riley, and Ed
Morehouse.  Thanks are also due to Sam Staton and Luke Ong for
their encouragement, Samson Abramsky for his advice, and Geraint
Jones for his Eindhoven-style calculation macros.

The lion's share of this work was done during the author's time as
a doctoral student at Oxford, where he was supported by the EPSRC
(award reference 1354534). It was then revised at Wesleyan
University, as part of work supported by the Air Force Office of
Scientific Research under award number FA9550-16-1-0292. Any
opinions, finding, and conclusions or recommendations expressed in
this material are those of the author(s) and do not necessarily
reflect the views of the United States Air Force. Its completion
was supported in part by a research grant (12386, Guarded Homotopy
Type Theory) from the VILLUM Foundation. 


\bibliographystyle{alpha}
\bibliography{dualcalc-1}

\newcommand{\etalchar}[1]{$^{#1}$}
\begin{thebibliography}{BBdPH93}

\bibitem[AB04]{Awodey2004}
Steven Awodey and Andrej Bauer.
\newblock {Propositions as [Types]}.
\newblock {\em Journal of Logic and Computation}, 14(4):447--471, aug 2004.

\bibitem[ABHR99]{Abadi1999}
Mart{\'{i}}n Abadi, Anindya Banerjee, Nevin Heintze, and Jon~G Riecke.
\newblock {A core calculus of dependency}.
\newblock In {\em Proceedings of the 26th ACM SIGPLAN-SIGACT symposium on
  Principles of programming languages - POPL '99}, pages 147--160, New York,
  New York, USA, 1999. ACM Press.

\bibitem[AM13]{Atkey2013}
Robert Atkey and Conor McBride.
\newblock {Productive coprogramming with guarded recursion}.
\newblock {\em ACM SIGPLAN Notices}, 48(9):197--208, nov 2013.

\bibitem[And92]{Andreoli1992}
Jean-Marc Andreoli.
\newblock {Logic Programming with Focusing Proofs in Linear Logic}.
\newblock {\em Journal of Logic and Computation}, 2(3):297--347, 1992.

\bibitem[AT11]{Abramsky2011a}
Samson Abramsky and Nikos Tzevelekos.
\newblock {Introduction to Categories and Categorical Logic}.
\newblock In Bob Coecke, editor, {\em New Structures for Physics}, pages 3--94.
  Springer-Verlag, 2011.

\bibitem[Awo10]{Awodey2010}
Steve Awodey.
\newblock {\em {Category Theory}}.
\newblock Oxford Logic Guides. Oxford University Press, 2010.

\bibitem[Bar96]{Barber1996}
Andrew~Graham Barber.
\newblock {Dual Intuitionistic Linear Logic}.
\newblock Technical report, ECS-LFCS-96-347, Laboratory for Foundations of
  Computer Science, University of Edinburgh, 1996.

\bibitem[BBdP98]{Benton1998}
Nick Benton, Gavin~M. Bierman, and Valeria de~Paiva.
\newblock {Computational types from a logical perspective}.
\newblock {\em Journal of Functional Programming}, 8(2):177--193, 1998.

\bibitem[BBdPH93]{Benton1993b}
Nick Benton, Gavin Bierman, Valeria de~Paiva, and Martin Hyland.
\newblock {A term calculus for Intuitionistic Linear Logic}.
\newblock In Marc Bezem and Jan~Friso Groote, editors, {\em Typed Lambda
  Calculi and Applications}, pages 75--90, Berlin, Heidelberg, 1993. Springer
  Berlin Heidelberg.

\bibitem[BdP92]{Bierman1992a}
Gavin~M. Bierman and Valeria de~Paiva.
\newblock {Intuitionistic Necessity Revisited}.
\newblock In {\em Proceedings of the Logic at Work Conference}, 1992.

\bibitem[BdP96]{Bierman1996a}
Gavin~M. Bierman and Valeria de~Paiva.
\newblock {Intuitionistic Necessity Revisited}.
\newblock Technical report, University of Birmingham, 1996.

\bibitem[BdP00]{Bierman2000a}
Gavin~M. Bierman and Valeria de~Paiva.
\newblock {On an Intuitionistic Modal Logic}.
\newblock {\em Studia Logica}, 65(3):383--416, 2000.

\bibitem[BdPR01]{Bellin2001}
Gianluigi Bellin, Valeria de~Paiva, and Eike Ritter.
\newblock {Extended Curry-Howard correspondence for a basic constructive modal
  logic}.
\newblock In {\em Proceedings of Methods for Modalities}, 2001.

\bibitem[BdRV01]{Blackburn2001}
Patrick Blackburn, Maarten de~Rijke, and Yde Venema.
\newblock {\em {Modal Logic}}.
\newblock Cambridge University Press, 2001.

\bibitem[Bel85]{Bellin1985}
Gianluigi Bellin.
\newblock {A system of natural deduction for GL}.
\newblock {\em Theoria}, 51(2):89--114, 1985.

\bibitem[BGM19]{Bahr2019}
Patrick Bahr, Christian~Uldal Graulund, and Rasmus~Ejlers M{\o}gelberg.
\newblock {Simply RaTT: a fitch-style modal calculus for reactive programming
  without space leaks}.
\newblock {\em Proceedings of the ACM on Programming Languages}, 3(ICFP):1--27,
  2019.

\bibitem[BM13]{Birkedal2013}
Lars Birkedal and Rasmus~Ejlers M{\o}gelberg.
\newblock {Intensional Type Theory with Guarded Recursive Types qua Fixed
  Points on Universes}.
\newblock In {\em 2013 28th Annual ACM/IEEE Symposium on Logic in Computer
  Science}, pages 213--222. IEEE, jun 2013.

\bibitem[BMSS12]{Birkedal2012}
Lars Birkedal, Rasmus M{\o}gelberg, Jan Schwinghammer, and Kristian
  St{\o}vring.
\newblock {First steps in synthetic guarded domain theory: step-indexing in the
  topos of trees}.
\newblock {\em Logical Methods in Computer Science}, 8(4), oct 2012.

\bibitem[Boo94]{Boolos1994a}
George~S. Boolos.
\newblock {\em {The Logic of Provability}}.
\newblock Cambridge University Press, Cambridge, feb 1994.

\bibitem[Bor94]{Borceux1994}
Francis Borceux.
\newblock {\em {Handbook of Categorical Algebra}}.
\newblock Cambridge University Press, Cambridge, 1994.

\bibitem[CBBB16]{Clouston2016}
Ranald Clouston, Al{\v{e}}s Bizjak, Hans {Bugge Grathwohl}, and Lars Birkedal.
\newblock {The guarded lambda calculus: Programming and reasoning with guarded
  recursion for coinductive types}.
\newblock {\em Logical Methods in Computer Science}, 12(3):1--39, 2016.

\bibitem[CF58]{Curry1958}
Haskell~B. Curry and Robert Feys.
\newblock {\em {Combinatory Logic}}.
\newblock Studies in Logic and the Foundation of Mathematics. North-Holland,
  1958.

\bibitem[CFMM16]{Curien2016}
Pierre-Louis Curien, Marcelo Fiore, and Guillaume Munch-Maccagnoni.
\newblock {A theory of effects and resources: adjunction models and polarised
  calculi}.
\newblock In {\em Proceedings of the 43rd Annual ACM SIGPLAN-SIGACT Symposium
  on Principles of Programming Languages - POPL 2016}, pages 44--56, New York,
  New York, USA, 2016. ACM Press.

\bibitem[Clo18]{Clouston2018}
Ranald Clouston.
\newblock Fitch-style modal lambda calculi.
\newblock In Christel Baier and Ugo Dal~Lago, editors, {\em Foundations of
  Software Science and Computation Structures}, pages 258--275, Cham, 2018.
  Springer International Publishing.

\bibitem[Cro93]{Crole1993}
Roy~L. Crole.
\newblock {\em {Categories for Types}}.
\newblock Cambridge University Press, 1993.

\bibitem[Cur52]{Curry1952}
Haskell~B. Curry.
\newblock {The elimination theorem when modality is present}.
\newblock {\em The Journal of Symbolic Logic}, 17(04):249--265, dec 1952.

\bibitem[Dav17]{Davies2017}
Rowan Davies.
\newblock {A Temporal Logic Approach to Binding-Time Analysis}.
\newblock {\em Journal of the ACM}, 64(1):1--45, mar 2017.

\bibitem[dG02]{deGroote2002}
Philippe de~Groote.
\newblock {On the Strong Normalisation of Intuitionistic Natural Deduction with
  Permutation-Conversions}.
\newblock {\em Information and Computation}, 178(2):441--464, 2002.

\bibitem[DJ03]{Danos2003}
Vincent Danos and Jean~Baptiste Joinet.
\newblock {Linear logic and elementary time}.
\newblock {\em Information and Computation}, 183(1):123--137, 2003.

\bibitem[DP01]{Davies2001a}
Rowan Davies and Frank Pfenning.
\newblock {A modal analysis of staged computation}.
\newblock {\em Journal of the ACM}, 48(3):555--604, 2001.

\bibitem[dPR11]{dePaiva2011}
Valeria de~Paiva and Eike Ritter.
\newblock {Basic Constructive Modality}.
\newblock In Jean-Yves Beziau and Marcelo Coniglio, editors, {\em Logic without
  frontiers - Festschrift for Walter Alexandre Carnielli on the occasion of his
  60th birthday}, pages 411--428. College Publications, London, 2011.

\bibitem[Fit93]{Fitting1993}
Melvin Fitting.
\newblock {\em {Basic Modal Logic}}, pages 368--448.
\newblock Oxford University Press, Inc., USA, 1993.

\bibitem[Gal90]{Gallier1990}
Jean Gallier.
\newblock {On Girard's "Candidats de Reductibilite"}.
\newblock In Piergiorgio Odifreddi, editor, {\em Logic and Computer Science},
  pages 123--203. Academic Press, 1990.

\bibitem[Gal93]{Gallier1993}
Jean Gallier.
\newblock {Constructive logics Part I: A tutorial on proof systems and typed
  $\lambda$-calculi}.
\newblock {\em Theoretical Computer Science}, 110(2):249--339, 1993.

\bibitem[Gal95]{Gallier1995}
Jean Gallier.
\newblock {On the Correspondence Between Proofs and Lambda Terms}.
\newblock In Philippe de~Groote, editor, {\em The Curry-Howard Isomorphism},
  pages 55--138. Academia, Louvain-la-Neuve, 1995.

\bibitem[Gen35a]{Gentzen1935a}
Gerhard Gentzen.
\newblock {Untersuchungen {\"{u}}ber das logische Schlie{\ss}en. I}.
\newblock {\em Mathematische Zeitschrift}, 39(1):176--210, 1935.

\bibitem[Gen35b]{Gentzen1935b}
Gerhard Gentzen.
\newblock {Untersuchungen {\"{u}}ber das logische Schlie{\ss}en. II}.
\newblock {\em Mathematische Zeitschrift}, 39(1):405--431, 1935.

\bibitem[Gir72]{Girard1972}
Jean-Yves Girard.
\newblock {\em {Interpr{\'{e}}tation fonctionelle et {\'{e}}limination des
  coupures de l'arithm{\'{e}}tique d'ordre sup{\'{e}}rieur}}.
\newblock PhD thesis, Universit{\'{e}} Paris VII, 1972.

\bibitem[Gir93]{Girard1993}
Jean-Yves Girard.
\newblock {On the unity of logic}.
\newblock {\em Annals of Pure and Applied Logic}, 59(3):201--217, feb 1993.

\bibitem[GL96]{Goubault-Larrecq1996}
Jean Goubault-Larrecq.
\newblock {On Computational Interpretations of the Modal Logic S4 - I. Cut
  Elimination}.
\newblock Technical report, 1996-35. Institut f{\"{u}}r Logik,
  Komplexit{\"{a}}t und Deduktionssysteme, Universit{\"{a}}t Karlsruhe, 1996.

\bibitem[GLN{\etalchar{+}}17]{Gross2017}
Jacob~A Gross, Daniel~R Licata, Max~S New, Jennifer Paykin, Mitchell Riley,
  Michael Shulman, and Felix Wellen.
\newblock {Differential Cohesive Type Theory (Extended Abstract)}.
\newblock In {\em Extended abstracts for the Workshop "Homotopy Type Theory and
  Univalent Foundations"}, 2017.

\bibitem[GLT89]{Girard1989}
Jean-Yves Girard, Yves Lafont, and Paul Taylor.
\newblock {\em {Proofs and Types}}.
\newblock Cambridge University Press, 1989.

\bibitem[GR12]{Gore2012}
Rajeev Gor{\'{e}} and Revantha Ramanayke.
\newblock {Valentini's Cut-Elimination for Provability Logic Resolved}.
\newblock {\em The Review of Symbolic Logic}, 5(02):212--238, 2012.

\bibitem[Gua18]{Guatto2018}
Adrien Guatto.
\newblock {A Generalized Modality for Recursion}.
\newblock In {\em Proceedings of the 33rd Annual ACM/IEEE Symposium on Logic in
  Computer Science}, LICS '18, pages 482--491, New York, NY, USA, 2018.
  Association for Computing Machinery.

\bibitem[HC96]{Hughes1996}
G.~E. Hughes and M.~J. Creswell.
\newblock {\em {A New Introduction to Modal Logic}}.
\newblock Routledge, 1996.

\bibitem[HJ99]{Hrbacek1999}
K~Hrbacek and T~Jech.
\newblock {\em {Introduction to Set Theory, Third Edition, Revised and
  Expanded}}.
\newblock Chapman \& Hall/CRC Pure and Applied Mathematics. Taylor \& Francis,
  1999.

\bibitem[HN12]{Hakli2012}
Raul Hakli and Sara Negri.
\newblock {Does the deduction theorem fail for modal logic?}
\newblock {\em Synthese}, 187(3):849--867, 2012.

\bibitem[Hof99a]{Hofmann1999b}
Martin Hofmann.
\newblock {Semantical analysis of higher-order abstract syntax}.
\newblock In {\em Proceedings. 14th Symposium on Logic in Computer Science
  (Cat. No. PR00158)}, pages 204--213. IEEE Comput. Soc, 1999.

\bibitem[Hof99b]{Hofmann1999}
Martin Hofmann.
\newblock {\em {Type Systems for Polynomial-Time Computation}}.
\newblock Habilitation thesis, Technischen Universit{\"{a}}t Darmstadt, 1999.

\bibitem[Kak07]{Kakutani2007a}
Yoshihiko Kakutani.
\newblock {Calculi for Intuitionistic Normal Modal Logic}.
\newblock In {\em Proceedings of Programming and Programming Languages (PPL)
  2007}, 2007.

\bibitem[Kav16]{Kavvos2016b}
G.~A. Kavvos.
\newblock {The Many Worlds of Modal Lambda Calculi: I. Curry-Howard for
  Necessity, Possibility and Time}.
\newblock {\em CoRR}, 2016.

\bibitem[Kav17]{Kavvos2017b}
G.~A. Kavvos.
\newblock {Dual-context calculi for modal logic}.
\newblock In {\em 2017 32nd Annual ACM/IEEE Symposium on Logic in Computer
  Science (LICS)}. IEEE, 2017.

\bibitem[Kav19]{Kavvos2019a}
G.~A. Kavvos.
\newblock {Modalities, Cohesion, and Information Flow}.
\newblock {\em Proceedings of the ACM on Programming Languages}, 3(POPL), 2019.

\bibitem[KB11a]{Krishnaswami2011a}
Neelakantan~R. Krishnaswami and Nick Benton.
\newblock {A semantic model for graphical user interfaces}.
\newblock {\em ACM SIGPLAN Notices}, 46(9):45, sep 2011.

\bibitem[KB11b]{Krishnaswami2011}
Neelakantan~R. Krishnaswami and Nick Benton.
\newblock {Ultrametric Semantics of Reactive Programs}.
\newblock In {\em 2011 IEEE 26th Annual Symposium on Logic in Computer
  Science}, pages 257--266. IEEE, jun 2011.

\bibitem[Kol85]{Koletsos1985}
George Koletsos.
\newblock {Church-Rosser theorem for typed functional systems}.
\newblock {\em The Journal of Symbolic Logic}, 50(03):782--790, 1985.

\bibitem[Kri63]{Kripke1963}
Saul~A. Kripke.
\newblock {Semantical Analysis of Modal Logic I. Normal Modal Propositional
  Calculi}.
\newblock {\em Zeitschrift f{\"{u}}r Mathematische Logik und Grundlagen der
  Mathematik}, 9(5-6):67--96, 1963.

\bibitem[Kri13]{Krishnaswami2013}
Neelakantan~R. Krishnaswami.
\newblock {Higher-order functional reactive programming without spacetime
  leaks}.
\newblock In {\em Proceedings of the 18th ACM SIGPLAN international conference
  on Functional programming - ICFP '13}, page 221, New York, New York, USA,
  2013. ACM, ACM Press.

\bibitem[Lei81]{Leivant1981}
Daniel Leivant.
\newblock {On the proof theory of the modal logic for arithmetic provability}.
\newblock {\em The Journal of Symbolic Logic}, 46(03):531--538, sep 1981.

\bibitem[Lit14]{Litak2014}
Tadeusz Litak.
\newblock {Constructive Modalities with Provability Smack}.
\newblock In G.~Bezhanishvili, editor, {\em Leo Esakia on Duality in Modal and
  Intuitionistic Logics}, volume~4 of {\em Outstanding Contributions to Logic},
  pages 187--216. Springer, Dordrecht, 2014.

\bibitem[LOPS18]{Licata2018}
Daniel~R. Licata, Ian Orton, Andrew~M. Pitts, and Bas Spitters.
\newblock {Internal Universes in Models of Homotopy Type Theory}.
\newblock In H.~Kirchner, editor, {\em 3rd International Conference on Formal
  Structures for Computation and Deduction (FSCD 2018)}, Leibniz International
  Proceedings in Informatics (LIPIcs), pages 22:1--22:17. Schloss
  Dagstuhl-Leibniz-Zentrum fuer Informatik, 2018.

\bibitem[LS88]{Lambek1988}
Joachim Lambek and Philip~J. Scott.
\newblock {\em {Introduction to Higher-Order Categorical Logic}}.
\newblock Cambridge University Press, 1988.

\bibitem[{Mac}78]{MacLane1978}
Saunders {Mac Lane}.
\newblock {\em {Categories for the Working Mathematician}}, volume~5 of {\em
  Graduate Texts in Mathematics}.
\newblock Springer New York, New York, NY, 1978.

\bibitem[MCHP04]{MurphyVII2004}
T.~Murphy, Karl Crary, Robert Harper, and Frank Pfenning.
\newblock {A symmetric modal lambda calculus for distributed computing}.
\newblock In {\em Proceedings of the 19th Annual IEEE Symposium on Logic in
  Computer Science, 2004.}, pages 286--295. IEEE, 2004.

\bibitem[Mel09]{Mellies2009}
Paul-Andr{\'{e}} Melli{\`{e}}s.
\newblock {Categorical Semantics of Linear Logic}.
\newblock In Pierre-Louis Curien, Hugo Herbelin, Jean-Louis Krivine, and
  Paul-Andr{\'{e}} Melli{\`{e}}s, editors, {\em Panoramas et synth{\`{e}}ses
  27: Interactive models of computation and program behaviour}.
  Soci{\'{e}}t{\'{e}} Math{\'{e}}matique de France, 2009.

\bibitem[ML13]{Milius2013}
Stefan Milius and Tadeusz Litak.
\newblock {Guard Your Daggers and Traces: On The Equational Properties of
  Guarded (Co-)recursion}.
\newblock {\em Electronic Proceedings in Theoretical Computer Science},
  126(Informatik 8):72--86, aug 2013.

\bibitem[MM96]{Martini1996}
Simone Martini and Andrea Masini.
\newblock {A Computational Interpretation of Modal Proofs}.
\newblock In {\em Proof Theory of Modal Logic}, Applied Logic Series, pages
  213--241. Springer Netherlands, 1996.

\bibitem[Mog91]{Moggi1991}
Eugenio Moggi.
\newblock {Notions of computation and monads}.
\newblock {\em Information and Computation}, 93(1):55--92, 1991.

\bibitem[Nak00]{Nakano2000}
Hiroshi Nakano.
\newblock {A modality for recursion}.
\newblock {\em Proceedings Fifteenth Annual IEEE Symposium on Logic in Computer
  Science (Cat. No.99CB36332)}, 2000.

\bibitem[ND18]{Nuyts2018}
Andreas Nuyts and Dominique Devriese.
\newblock {Degrees of Relatedness}.
\newblock In {\em Proceedings of the 33rd Annual ACM/IEEE Symposium on Logic in
  Computer Science - LICS '18}, pages 779--788, New York, New York, USA, 2018.
  ACM Press.

\bibitem[Neg11]{Negri2011}
Sara Negri.
\newblock {Proof Theory for Modal Logic}.
\newblock {\em Philosophy Compass}, 6(8):523--538, 2011.

\bibitem[NVD17]{Nuyts2017}
Andreas Nuyts, Andrea Vezzosi, and Dominique Devriese.
\newblock {Parametric quantifiers for dependent type theory}.
\newblock {\em Proceedings of the ACM on Programming Languages}, 1(ICFP), 2017.

\bibitem[OH06]{Ohta2006}
Yo~Ohta and Masahito Hasegawa.
\newblock {A Terminating and Confluent Linear Lambda Calculus}.
\newblock In Frank Pfenning, editor, {\em Term Rewriting and Applications. RTA
  2006}, volume 4098 of {\em Lecture Notes in Computer Science}, pages
  166--180. Springer, Berlin, Heidelberg, 2006.

\bibitem[OM57]{Ohnisi1957}
Masao Ohnisi and Kazuo Matsumoto.
\newblock {Gentzen method in modal calculi}.
\newblock {\em Osaka Journal of Mathematics}, 11(2):113--130, 1957.

\bibitem[OM59]{Ohnisi1959}
Masao Ohnisi and Kazuo Matsumoto.
\newblock {Gentzen method in modal calculi. II}.
\newblock {\em Osaka Journal of Mathematics}, 11(2):115--120, 1959.

\bibitem[Ono98]{Ono1998}
Hiroakira Ono.
\newblock {Proof-theoretic methods in nonclassical logic{--}an introduction}.
\newblock In Masako Takahashi, Mitsuhiro Okada, and Marangiola
  Dezani-Ciancaglini, editors, {\em Theories of Types and Proofs}, MSJ Memoirs,
  pages 207--254. The Mathematical Society of Japan, Tokyo, 1998.

\bibitem[Orc14]{Orchard2014}
Dominic Orchard.
\newblock {\em {Programming contextual computations}}.
\newblock PhD thesis, University of Cambridge, 2014.

\bibitem[PD01]{Davies2001}
Frank Pfenning and Rowan Davies.
\newblock {A judgmental reconstruction of modal logic}.
\newblock {\em Mathematical Structures in Computer Science}, 11(4):511--540,
  2001.

\bibitem[Pfe01]{Pfenning2001}
Frank Pfenning.
\newblock {Intensionality, extensionality, and proof irrelevance in modal type
  theory}.
\newblock {\em Proceedings of the 16th Annual IEEE Symposium on Logic in
  Computer Science (LICS 2001)}, 2001.

\bibitem[Pfe10]{Pfenning2010}
Frank Pfenning.
\newblock {Lecture Notes on Combinatory Modal Logic}, 2010.

\bibitem[Pfe13]{Pfenning2013}
Frank Pfenning.
\newblock {Weather Report}, 2013.

\bibitem[Pfe15]{Pfenning2015}
Frank Pfenning.
\newblock {Decomposing Modalities}.
\newblock {\em Logical Frameworks and Meta-Languages: Theory and Practice
  (LFMTP'15)}, 2015.

\bibitem[Plo77]{Plotkin1977}
Gordon~D. Plotkin.
\newblock {LCF considered as a programming language}.
\newblock {\em Theoretical Computer Science}, 5(3):223--255, 1977.

\bibitem[Plo93]{Plotkin1993}
Gordon~D. Plotkin.
\newblock {Type theory and recursion}.
\newblock In {\em Proceedings Eighth Annual IEEE Symposium on Logic in Computer
  Science}, page 374. IEEE Comput. Soc. Press, 1993.

\bibitem[Pra65]{Prawitz1965}
Dag Prawitz.
\newblock {\em {Natural Deduction: a proof-theoretical study.}}
\newblock Almquist and Wiksell, 1965.

\bibitem[Pra71]{Prawitz1971}
Dag Prawitz.
\newblock {Ideas and Results in Proof Theory}.
\newblock In J.~E. Fenstad, editor, {\em Proceedings of the Second Scandinavian
  Logic Symposium}, volume~63 of {\em Studies in logic and the foundations of
  mathematics}. North-Holland, Amsterdam, 1971.

\bibitem[RSS20]{Rijke2020}
Egbert Rijke, Michael Shulman, and Bas Spitters.
\newblock {Modalities in homotopy type theory}.
\newblock {\em Logical Methods in Computer Science}, 16(1):2:1--2:79, 2020.

\bibitem[Sch17]{Scherer2017}
Gabriel Scherer.
\newblock {Deciding equivalence with sums and the empty type}.
\newblock In {\em Proceedings of the 44th ACM SIGPLAN Symposium on Principles
  of Programming Languages - POPL 2017}, pages 374--386, New York, New York,
  USA, 2017. ACM Press.

\bibitem[SDP01]{Schurmann2001}
Carsten Sch{\"{u}}rmann, Jo{\"{e}}lle Despeyroux, and Frank Pfenning.
\newblock {Primitive recursion for higher-order abstract syntax}.
\newblock {\em Theoretical Computer Science}, 266(1-2):1--57, 2001.

\bibitem[SdV12]{Severi2012}
Paula~G. Severi and Fer-Jan~J. de~Vries.
\newblock {Pure type systems with corecursion on streams}.
\newblock {\em ACM SIGPLAN Notices}, 47(9):141, 2012.

\bibitem[SH84]{Schroeder-Heister1984}
Peter Schroeder-Heister.
\newblock {A natural extension of natural deduction}.
\newblock {\em The Journal of Symbolic Logic}, 49(04):1284--1300, 1984.

\bibitem[Sha14]{Shamkanov2014}
D.~S. Shamkanov.
\newblock {Circular proofs for the G{\"{o}}del-L{\"{o}}b provability logic}.
\newblock {\em Mathematical Notes}, 96(3-4):575--585, sep 2014.

\bibitem[Shu18]{Shulman2018}
Michael Shulman.
\newblock {Brouwer's fixed-point theorem in real-cohesive homotopy type
  theory}.
\newblock {\em Mathematical Structures in Computer Science}, 28(6):856--941,
  2018.

\bibitem[SI08]{Shikuma2008}
Naokata Shikuma and Atsushi Igarashi.
\newblock {Proving Noninterference by a Fully Complete Translation to the
  Simply Typed lambda-calculus}.
\newblock {\em Logical Methods in Computer Science}, 4(3):10, 2008.

\bibitem[Sim94]{Simpson1994}
Alex~K. Simpson.
\newblock {\em {The Proof Theory and Semantics of Intuitionistic Modal Logic}}.
\newblock PhD thesis, The University of Edinburgh, 1994.

\bibitem[SP00]{Simpson2000}
Alex~K. Simpson and Gordon~D. Plotkin.
\newblock {Complete axioms for categorical fixed-point operators}.
\newblock In {\em Proceedings of the 15th Annual IEEE Symposium on Logic in
  Computer Science (LICS 2000)}, pages 30--41. IEEE Comput. Soc, 2000.

\bibitem[SS14]{Schreiber2014}
Urs Schreiber and Michael Shulman.
\newblock {Quantum Gauge Field Theory in Cohesive Homotopy Type Theory}.
\newblock {\em Electronic Proceedings in Theoretical Computer Science},
  158:109--126, 2014.

\bibitem[SU06]{Sorensen2006}
Morten~Heine S{\o}rensen and Pawel Urzyczyn.
\newblock {\em {Lectures on the Curry-Howard Isomorphism}}.
\newblock Elsevier, 2006.

\bibitem[SV80]{Sambin1980}
G.~Sambin and S.~Valentini.
\newblock {A modal sequent calculus for a fragment of arithmetic}.
\newblock {\em Studia Logica}, 39(2-3):245--256, 1980.

\bibitem[SV82]{Sambin1982}
Giovanni Sambin and Silvio Valentini.
\newblock {The modal logic of provability. The sequential approach}.
\newblock {\em Journal of Philosophical Logic}, 11(3):311--342, 1982.

\bibitem[Tak95]{Takahashi1995}
M.~Takahashi.
\newblock {Parallel Reductions in $\lambda$-Calculus}.
\newblock {\em Information and Computation}, 118(1):120--127, apr 1995.

\bibitem[TI10]{Tsukada2010}
Takeshi Tsukada and Atsushi Igarashi.
\newblock {A logical foundation for environment classifiers}.
\newblock {\em Logical Methods in Computer Science}, 6(4):1--43, 2010.

\bibitem[Val82]{Valentini1982}
Silvio Valentini.
\newblock {Cut-elimination in a modal sequent calculus for K}.
\newblock {\em Bolletino dell'Unione Mathematica Italiana}, 1B:119--130, 1982.

\bibitem[Val83]{Valentini1983}
Silvio Valentini.
\newblock {The Modal Logic of Provability: Cut-Elimination}.
\newblock {\em Journal of Philosophical Logic}, 12(4):471--476, 1983.

\bibitem[Wad93]{Wadler1993}
Philip Wadler.
\newblock {A taste of linear logic}.
\newblock In Andrzej~M Borzyszkowski and Stefan Soko{\l}owski, editors, {\em
  Mathematical Foundations of Computer Science 1993}, pages 185--210, Berlin,
  Heidelberg, 1993. Springer Berlin Heidelberg.

\bibitem[Wad94]{Wadler1994}
Philip Wadler.
\newblock {A syntax for linear logic}.
\newblock In S.~Brookes, M.~Main, A.~Melton, M.~Mislove, and D.~Schmidt,
  editors, {\em Mathematical Foundations of Programming Semantics: 9th
  International Conference, New Orleans, LA, USA, April 7 - 10, 1993.
  Proceedings}, pages 513--529. Springer-Verlag Berlin Heidelberg, 1994.

\bibitem[Wan02]{Wansing2002}
Heinrich Wansing.
\newblock {Sequent Systems for Modal Logics}.
\newblock In {\em Handbook of Philosophical Logic}, pages 61--145. Springer
  Netherlands, Dordrecht, 2002.

\bibitem[Wel17]{Wellen2017}
Felix Wellen.
\newblock {\em {Formalizing Cartan Geometry in Modal Homotopy Type Theory}}.
\newblock Phd thesis, Karlsruher Instituts f{\"{u}}r Technologie, 2017.

\bibitem[Wij90]{Wijesekera1990}
Duminda Wijesekera.
\newblock {Constructive modal logics I}.
\newblock {\em Annals of Pure and Applied Logic}, 50(3):271--301, dec 1990.

\bibitem[WLP98]{Wickline1998a}
Philip Wickline, Peter Lee, and Frank Pfenning.
\newblock {Run-Time Code Generation and Modal-ML}.
\newblock In {\em Proceedings of the ACM SIGPLAN 1998 Conference on Programming
  Language Design and Implementation}, PLDI '98, pages 224--235, New York, NY,
  USA, 1998. Association for Computing Machinery.

\end{thebibliography}

\end{document}